\documentclass[3p,12pt]{elsarticle}
\usepackage{graphicx}
\usepackage{enumerate}
\usepackage{amsmath}
\usepackage{amssymb}

\newcounter{mycounter}

\newcommand{\centerpic}[2]
{
  \begin{minipage}[c]{#1\textwidth}
    \includegraphics[width=0.95\textwidth]{#2}%
  \end{minipage}
}

\newcommand{\downpic}[2]
 {
   \begin{minipage}[t]{#1\textwidth}
    \vspace{-2ex}
    \includegraphics[width=0.95\textwidth]{#2}
   \end{minipage}\,
}

\newcommand{\uppic}[2]
{
  \begin{minipage}[b]{#1\textwidth}
   \includegraphics[width=0.95\textwidth]{#2}
   \vspace{-0.9ex}
  \end{minipage}
 \,
}

\newcommand{\AnfangEnde}[1]{\widetilde{\overline{#1}}}
\newcommand{\Anfang}[1]{\overline{#1}}
\newcommand{\Ende}[1]{\widetilde{#1}}

\journal{Annals of Physics}

\begin{document}

\begin{frontmatter}

\title{Weak localization with nonlinear bosonic matter waves}

\author[rgbg]{Timo Hartmann}
\author[rgbg]{Josef Michl}
\author[gre,lyon]{{Cyril Petitjean}}
\author[frei]{{Thomas Wellens}}
\author[rgbg]{Juan-Diego Urbina}
\author[rgbg]{Klaus Richter}
\author[ulg]{Peter Schlagheck\corref{cor1}\fnref{telfax}}
\ead{Peter.Schlagheck@ulg.ac.be}
\address[rgbg]{Institut f\"ur Theoretische Physik, Universit\"at Regensburg,
  93040 Regensburg, Germany}
\address[gre]{SPSMS, UMR-E 9001 CEA / UJF-Grenoble 1, INAC, Grenoble, 
  F-38054, France}
\address[lyon]{Laboratoire de Physique, CNRS UMR5672, Ecole Normale Superieure de Lyon, 46 All\'ee d'Italie, F-69364 Lyon Cedex 07, France}
\address[frei]{Institut f\"ur Physik, Albert-Ludwigs-Universit\"at Freiburg, 
Hermann-Herder-Str.~3, 79104 Freiburg, Germany}
\address[ulg]{D\'epartement de Physique,
Universit\'e de Li\`ege, 4000 Li\`ege, Belgium}
\cortext[cor1]{Corresponding author}
\fntext[telfax]{phone: +3243669043, fax: +3243663629}

\begin{abstract}

We investigate the coherent propagation of dilute atomic Bose-Einstein
condensates through irregularly shaped billiard geometries that are
attached to uniform incoming and outgoing waveguides.
Using the mean-field description based on the nonlinear Gross-Pita\-evskii
equation, we develop a diagrammatic theory for the self-consistent
stationary scattering state of the interacting condensate, which is
combined with the semiclassical representation of the single-particle Green
function in terms of chaotic classical trajectories within the billiard.
This analytical approach predicts a universal dephasing of weak localization
in the presence of a small interaction strength between the atoms, which is
found to be in good agreement with the numerically computed reflection and
transmission probabilities of the propagating condensate.
The numerical simulation of this quasi-stationary scattering process indicates
that this interaction-induced dephasing mechanism may give rise to a signature
of weak antilocalization, which we attribute to the influence of non-universal
short-path contributions.

\end{abstract}


\end{frontmatter}

\section{{Introduction}}

\label{sec:intro}

Recent technological advances in the manipulation of ultracold atoms on
microscopic length scales have paved the way toward the exploration of
scattering and transport phenomena with coherent interacting matter waves.
Key experiments in this context include the creation of flexible waveguide
geometries with optical dipole beams \cite{DumO02PRL} and on atom chips
\cite{FolO00PRL,ForZim07RMP}, the coherent propagation of Bose-Einstein
condensed atoms in such wave\-guides by means of guided atom lasers
\cite{GueO06PRL,CouO08EPL,FabO11PRL,GatO11PRL}, the realization of optical 
billiard confinements \cite{MilO01PRL,FriO01PRL,HenO09NJP} and microscopic 
scattering and disorder potentials for cold atoms \cite{BilO08N,RoaO08N}, 
as well as the detection of individual atoms within a condensate through 
photoionization on an atom chip \cite{StiO07PRA}.
Moreover, it was recently demonstrated \cite{LinO09PRL} that artificial gauge
potentials {can be induced for cold atoms, which lead to a breaking of 
time-reversal invariance in the same way as do magnetic fields for electrons.
Such artificial gauge potentials can, e.g., be implemented} by
means of Raman dressing with two laser beams that include a finite orbital
angular momentum \cite{JuzO05PRA,RusO05PRL,DalO11RMP}.
Together with the possibility of combining different atomic (bosonic and
fermionic) species and of manipulating their interaction through Feshbach
resonances, the combination of these tools gives rise to a number of possible
scattering and transport scenarios that are now ready for experimental
investigation.

A particularly prominent quantum transport phenomenon in mesoscopic physics is
\emph{weak localization} \cite{AltO80PRB,Ber84PR}.
This concept refers to an appreciable enhancement of the reflection (or,
in the solid-state context, of the electronic resistance) in the presence of a
two- or three-dimensional ballistic or disordered scattering region, as
compared to the expectation based on a classical, i.e.\ incoherent, transport
process.
This enhancement, which {in turn} implies a reduction of the transmission
(or of the electronic conductance) due to current conservation, is 
{in particular} caused by ``coherent backscattering'', i.e.\ by the
constructive interference between backscattered classical paths and their
time-reversed counterparts, which was first observed in experiments on the
scattering of laser light from disordered media
\cite{VanLag85PRL,WolMar85PRL}.
In the solid-state context, weak localization is most conveniently detected by
measuring the electronic conductance in dependence of a weak magnetic field
that is oriented perpendicular to the scattering region, such that it causes a
dephasing between backscattered paths and their time-reversed counterparts.
A characteristic peak structure at zero magnetic field is then typically
observed \cite{AroSha87RMP,ChaO94PRL}.

From the electronic point of view,
the presence of interaction between the particles that participate at this
scattering process is generally expected to give rise to an additional
dephasing mechanism of this subtle interference phenomenon
\cite{AltAroKhm82JPC,AltAro85EfrPol,WhiJacPet08PRB}.
In the context of ultracold bosonic atoms, this expectation is partly
confirmed by previous theoretical studies on the coherent propagation of an
interacting Bose-Einstein condensate through a two-dimensional disorder
potential \cite{HarO08PRL}, which employed numerical simulations as well as
diagrammatic representations based on the mean-field description of the
condensate in terms of the nonlinear Gross-Pitaevskii equation.
This study did indeed reveal a reduction of the height of the coherent
backscattering peak with increasing effective interaction strength between the
atoms.
It also predicted, however, that this coherent backscattering peak might
turn into a \emph{dip} at finite (but still rather small) interaction
strengths \cite{HarO08PRL}.
This scenario is reminiscent of weak antilocalization due to spin-orbit
interaction, which was observed in mesoscopic magnetotransport
\cite{ZumO02PRL}.

In order to gain a new perspective on this novel 
phenomenon, we investigate, in this work, the coherent propagation of
Bose-Einstein condensates through ballistic scattering geometries that exhibit
chaotic classical dynamics.
Such propagation processes {can} be experimentally realized by guided atom
lasers in which the optical waveguides are locally ``deformed'' by means of
additional optical potentials{, e.g.\ by focusing a red-detuned laser
from a different direction onto this waveguide as was done in the experiment
of Ref.~\cite{GatO11PRL}}.
Alternatively, atom chips \cite{FolO00PRL,ForZim07RMP} or atom-optical 
billiards \cite{MilO01PRL,FriO01PRL,HenO09NJP} could be used in
order to engineer chaotic scattering geometries for ultracold atoms.
From the theoretical point of view, the wave transport through such scattering
geometries can be described using the semiclassical representation of the
Green function in terms of classical trajectories.
The constructive interference of reflected trajectories with their
time-reversed counterparts gives then rise to coherent backscattering
{\cite{adist3}}, while {a complete understanding of weak localization,
in particular} the corresponding reduction of the transmitted current{,
requires additional, classically correlated trajectory} pairs 
\cite{RicSie02PRL,HeuO06PRL}.

In order to account for the presence of atom-atom interaction on the
mean-field level of the nonlinear Gross-Pitaevskii equation, we combine, in
this paper, the semiclassical approach with the framework of nonlinear
diagrammatic theory developed in Refs.~\cite{WelGre08PRL,WelGre09PRA,Wel09AP}.
For the sake of simplicity, we shall, as is described in Section
\ref{sec:setup}, restrict ourselves to ideal {chaotic} billiard dynamics
consisting of free motion that is confined by hard-wall boundaries.
Since such billiard geometries give rise to uniform average densities within
the scattering region, we can, as demonstrated in Sections \ref{sec:semi} and
\ref{sec:loop}, derive explicit analytical expressions for the
retro-reflection and transmission probabilities as a function of the effective
interaction strength.
As shown in Section \ref{sec:comp}, these expressions agree very well with the
numerically computed retro-reflection and transmission probabilities for two
exemplary billiard geometries as far as the deviation from the case of
noninteracting (single-particle) transport is concerned.
On the absolute scale, however, the height of the weak localization peak 
{is reduced in this noninteracting case} by the presence of short-path
contributions, in particular by self-retracing trajectories, which, as shown
in Section \ref{sec:comp}, consequently turn this peak into a finite dip in
the presence of a small interaction strength.
We shall therefore argue in Section \ref{sec:conc} that such short-path
contributions are at the origin of this weak antilocalization{-like} 
phenomenon.

\section{Setup {of the nonlinear scattering process}}

\label{sec:setup}

We consider the quasi-stationary transport of coherent bosonic matter waves
through two-dimensional waveguide structures that are perturbed by the 
presence of a wide quantum-dot-like scattering potential.
Such propagating matter waves can be generated by means of a guided atom laser
\cite{GueO06PRL,CouO08EPL} where ultracold atoms are coherently outcoupled 
from a trapping potential that contains a Bose-Einstein condensate.
The control of the outcoupling process, which, e.g., can be achieved by applying
a radiofrequency field that flips the spin of the atoms in the (magnetic) trap
\cite{GueO06PRL}, permits one, in principle, to generate an energetically 
well-defined beam of atoms that propagate along the (horizontally oriented) 
waveguide in its transverse ground mode {\cite{LebPav01PRA}}.
This waveguide, as well as the quantum-dot-like scattering potential, can be
engineered by means of focused red-detuned laser beams which provide an
attractive effective potential for the atoms that is proportional to their
intensity.
The restriction to two spatial dimensions can, furthermore, be realized by
applying, in addition, a tight one-dimensional optical lattice perpendicular 
to the waveguide (i.e.~oriented along the vertical direction).

The central object of study in this work is the phenomenon of weak 
localization.
In the context of electronic mesoscopic physics, this 
{quantum interference} phenomenon can be detected by measuring the
electronic conductance, which is directly related to the {quantum} 
transmission through the Landauer-B\"uttiker theory 
\cite{Lan57IBM,Lan70PhMg,BueO85PRB}, as a function of the strength of an
externally applied magnetic field which breaks  time-reversal
invariance within the scattering region.
Such a time-reversal breaking mechanism can also be induced for cold atoms
{\cite{LinO09PRL,JuzO05PRA,RusO05PRL,DalO11RMP}, e.g., by coherently coupling
two intra-atomic levels via} a STIRAP process, using two laser beams of which
one involves a nonvanishing orbital angular momentum \cite{JuzO05PRA}.
{This gives rise to an effective vector potential in the kinetic term of
the Schr\"odinger equation, which is assumed such that it generates an
effective ``magnetic field'' that is homogeneous} within the scattering region
and vanishes within the attached waveguides.

\begin{figure}
  \begin{center}
    \includegraphics[width=0.49\linewidth]{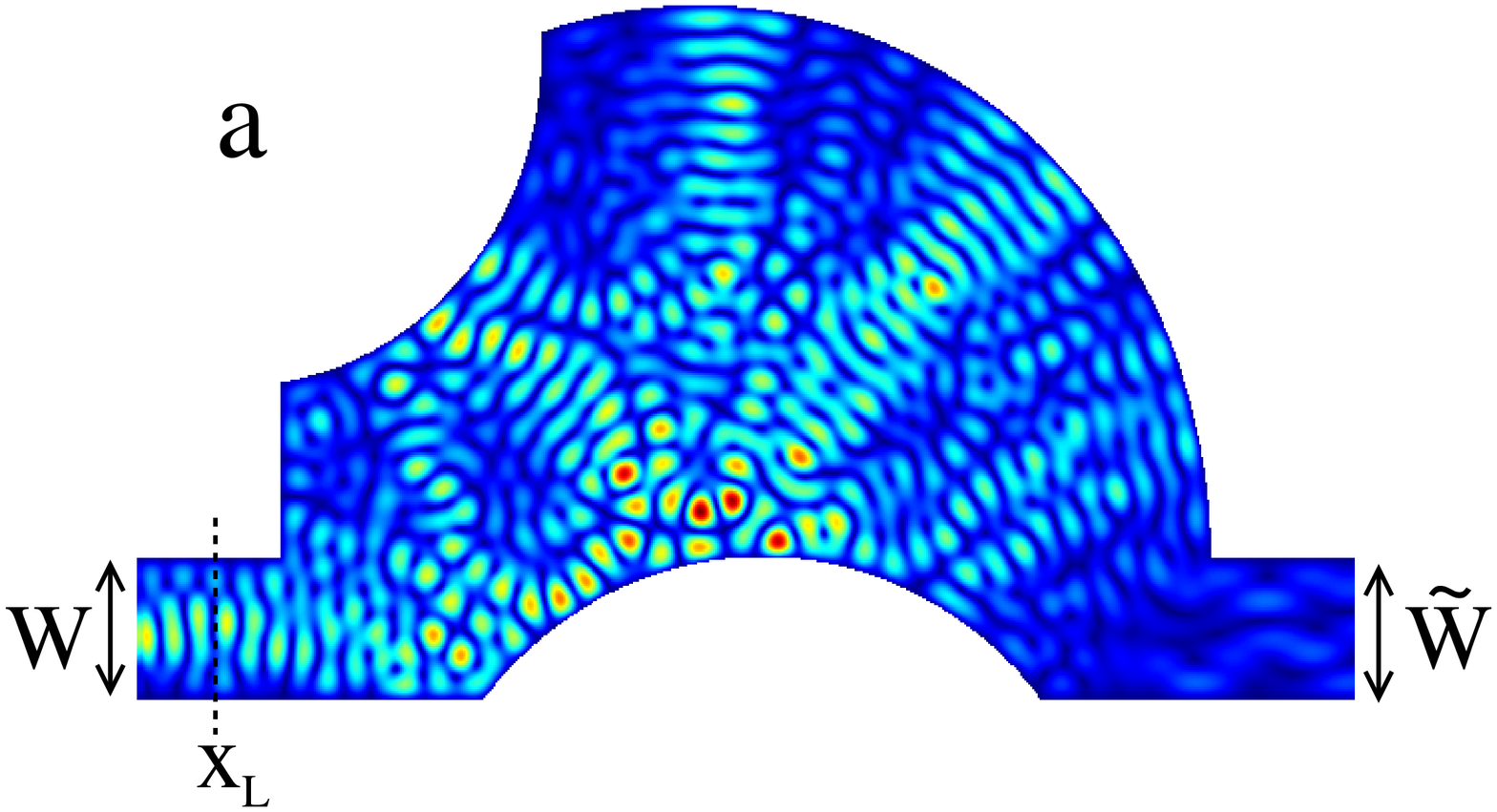}
    \hfill
    \includegraphics[width=0.49\linewidth]{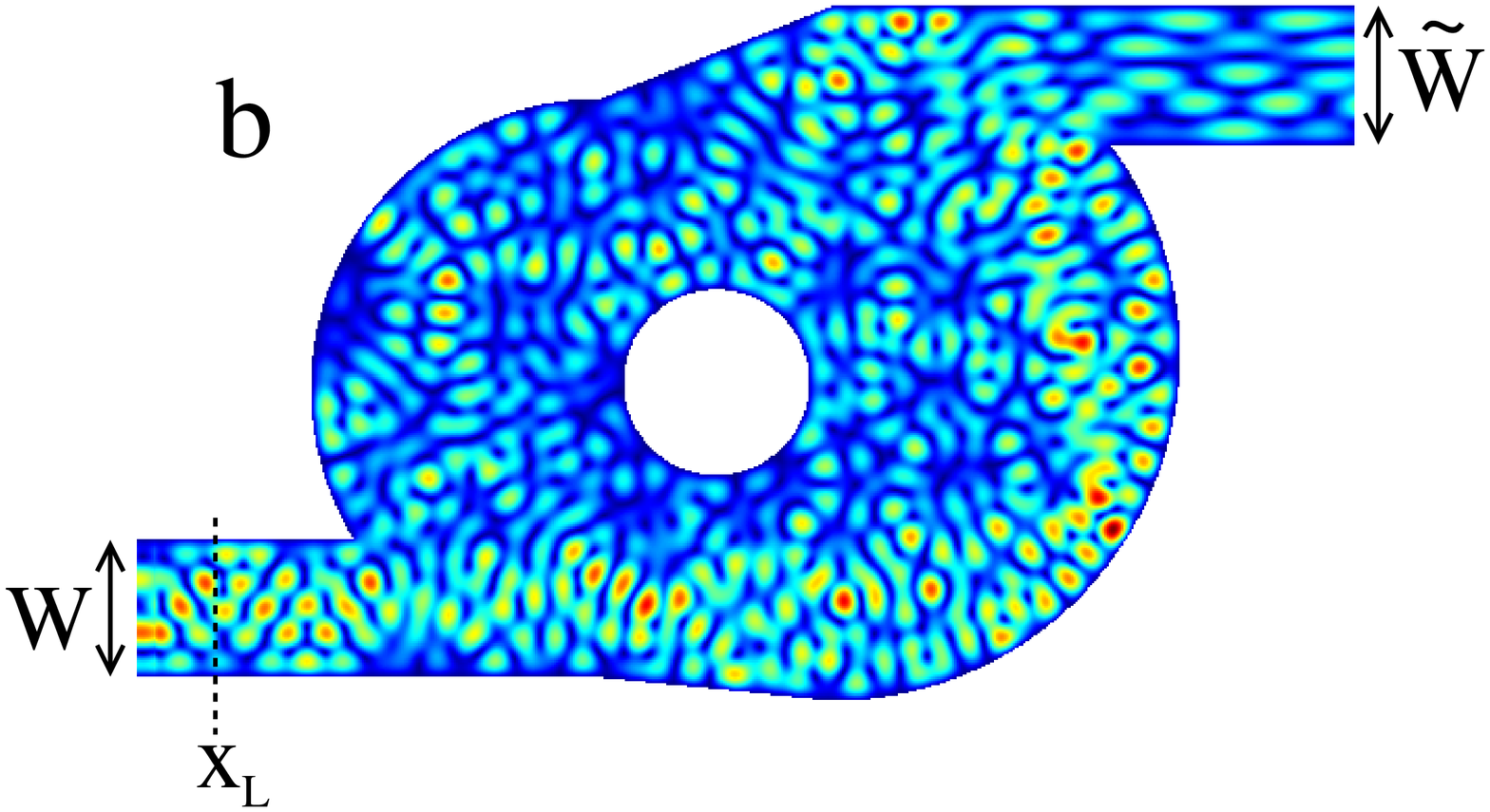}
    \caption{\label{fig:billiard}
      Shapes of the billiards {a and b} under consideration, plotted 
      together with the density of a stationary scattering state.
      {We indicate, in addition, the widths $W$ and $\tilde{W}$ of the
      incident and the transmitted waveguides, respectively, as well as the
      horizontal position $x_L$ at which the incident and {reflected} parts 
      of the scattering wavefunction are decomposed in transverse
      eigenmodes of the waveguide.}
      The semiclassical average that {is} undertaken in order to obtain
      the mean retro-reflection probability involve{s} an average of the 
      retro-reflection probability within a finite window of chemical 
      potentials $\mu$ for different incident channels and different
      locations of the circular and the lower semicircular 
      obstacle in the case of billiards a and b, respectively.
    }
  \end{center}
\end{figure}

The main purpose of this study is to investigate how the scenario of weak
localization is affected by the presence of a weak atom-atom interaction 
within the matter-wave beam.
In lowest order {in the interaction strength}, 
the presence of such an atom-atom interaction is accounted
for by a nonlinear contribution to the effective potential in the 
Schr\"odinger equation describing the motion of the atoms, which is 
proportional to the local density of atoms and which gives rise to the
celebrated Gross-Pitaevskii equation \cite{DalO99RMP}.
The strength of this nonlinear contribution can be controlled by the
{scale} of {the} confinement in the transverse (vertical) spatial
direction.
We shall make, in the following, the simplifying assumption that this 
nonlinearity is present only within the scattering region and vanishes 
within the waveguides.
We furthermore assume that the waveguides are perfectly uniform, and that
the two-dimensional scattering geometry can be described by perfect 
``billiard'' potentials which combine a vanishing potential background 
within the waveguides and the scattering region with
infinitely high hard walls along their boundaries.
{These} assumptions considerably simplify the analytical and numerical 
treatment of the problem, and allow for the identification of well-defined
asymptotic scattering states within the waveguides.
Two such billiard configurations are shown in Fig.~\ref{fig:billiard}.

The dynamics of this matter-wave scattering process is then well modeled
by an inhomogeneous two-dimensional Gross-Pitaevskii equation 
{\cite{ErnPauSch10PRA}}
\begin{equation}
i \hbar \frac{\partial}{\partial t} \Psi(\mathbf{r},t)  =
\left[ - \frac{1}{2m} \left( \frac{\hbar}{i} \nabla - \mathbf{A}(\mathbf{r}) 
\right)^2 + V(\mathbf{r}) + {g\frac{\hbar^2}{2m}}  |\Psi(\mathbf{r},t)|^2 
\right] \Psi(\mathbf{r},t) + S(\mathbf{r},t) \label{eq:GP0}
\end{equation}
with $\mathbf{r} \equiv (x,y)$.
Here $m$ is the mass of the atoms and $V(\mathbf{r})$ represents the 
confinement potential that defines the waveguides and the scattering region.
The effective vector potential $\mathbf{A}(\mathbf{r})$ vanishes within the 
waveguides.
Within the scattering billiard we choose it as 
\begin{equation}
  \mathbf{A}(\mathbf{r}) \equiv \frac{1}{2} B
  \mathbf{e}_z \times (\mathbf{r} - \mathbf{r}_0) 
  = \frac{1}{2} B [(x - x_0) \mathbf{e}_y - (y - y_0) \mathbf{e}_x]
  \label{eq:A}
\end{equation}
where $\mathbf{r}_0\equiv (x_0,y_0)$ represents an arbitrarily chosen 
reference point and $\mathbf{e}_x,\mathbf{e}_y,\mathbf{e}_z$ are the
unit vectors in our spatial coordinate system.
In the presence of an harmonic transverse (vertical) confinement with
oscillation frequency $\omega_\perp\equiv\omega_\perp(\mathbf{r})$, 
the effective two-dimensional interaction strength is given by
$g(\mathbf{r}) = 4 \sqrt{2 \pi} a_s/ a_\perp(\mathbf{r})$ with
$a_\perp(\mathbf{r})\equiv\sqrt{\hbar/[m\omega_\perp(\mathbf{r})]}$
where $a_s$ denotes the $s$-wave scattering length of the atoms.
As stated above, we assume that $g$ is constant within the billiard
and vanishes in the waveguides.

The source amplitude $S(\mathbf{r},t)$ describes the coherent injection
of atoms from the Bose-Einstein condensate within the reservoir trap.
Assuming that only one transverse eigenmode in the waveguide is populated, 
we may write $S$ as
\begin{equation}
  S(\mathbf{r},t) = S_0 \chi_i(y) \delta(x - x_L) 
  {\exp\left(-\frac{i}{\hbar}\mu t\right)}
\end{equation}
where $\chi_i(y)$ denotes the normalized wavefunction associated with the
transverse eigenmode with the excitation index $i$, characterized by the
energy $E_i$, into which the source injects the atoms from the condensate 
(typically one would attempt to achieve coherent injection into the 
transverse ground mode, with $i=1$, in an atom-laser experiment
\cite{GueO06PRL}).
$x_L$ represents an arbitrary longitudinal coordinate within the waveguide 
(which, without loss of generality, is assumed to be oriented along the $x$
axis) 
and $\mu$ is the chemical potential with which the atoms are injected into the
waveguide.
Making the ansatz
\begin{equation}
  \Psi(\mathbf{r},t) \equiv \psi(\mathbf{r},t) 
 {\exp\left(-\frac{i}{\hbar}\mu t\right)}
\end{equation}
we obtain
\begin{equation}
i \hbar \frac{\partial}{\partial t} \psi(\mathbf{r},t)  =
( H - \mu ) \psi(\mathbf{r},t) + {g\frac{\hbar^2}{2m}}
|\psi(\mathbf{r},t)|^2 \psi(\mathbf{r},t)
+ S_0 \chi_i(y) \delta(x - x_L) \label{eq:GPEs}
\end{equation}
with the single-particle Hamiltonian
\begin{equation}
  H = \frac{1}{2m} {\left[ \frac{\hbar}{i} \nabla - \mathbf{A}(\mathbf{r}) 
  \right]^2} + V(\mathbf{r}) \, . \label{eq:H}
\end{equation}

{The time evolution of the scattering wavefunction can be considered to
take place in the presence of an} adiabatically slow increase of the source 
amplitude $S_0$ from zero to a given maximal value. 
In the absence of interaction, this process would necessarily
lead to a stationary scattering state, whose
decomposition into the transverse eigenmodes within the waveguides allows
one to determine the associated channel-resolved reflection and 
transmission amplitudes.
In the special case of a perfectly uniform waveguide without any scattering
potential and in the absence of the vector potential $\mathbf{A}$, 
this stationary state is given by
\begin{equation}
  \psi(x,y) = -i \frac{m S_0}{\hbar p_i^{\mathrm{l}}(\mu)} \chi_i(y) 
  \exp\left[-\frac{i}{\hbar} p_i^{\mathrm{l}}(\mu) \left|x-x_L\right| \right]
 \label{eq:psi0}
\end{equation}
where $p_i^{\mathrm{l}}(\mu) \equiv \sqrt{2 m \mu - E_i}$ 
denotes the longitudinal component of the momentum associated with the 
transverse mode $\chi_i$.
Such a stationary scattering state is, in general, not obtained in the presence
of interaction.
Indeed, a {finite} nonlinearity strength $g$ may{, in combination with
a weak scattering potential,} lead to a permanently
time-dependent, turbulent-like flow across the scattering region
\cite{SkiMay00PRL,SpiZyu00PRL,PauO05PRA,HarO08PRL,ErnPauSch10PRA},
which in dimensionally restricted waveguide geometries should correspond to a 
loss of coherence on a microscopic level of the many-body scattering problem
\cite{ErnPauSch10PRA}.

{In the following, we shall restrict ourselves to rather small 
nonlinearities for which we still obtain, in most cases,
stable quasi-stationary scattering {states} within the billiard under consideration
\cite{rem_stable}. }
{In the subsequent two sections, we shall develop a semiclassical theory 
for the self-consistent scattering state that is obtained as a solution 
of Eq.~(\ref{eq:GPEs}).
Section \ref{sec:semi} focuses on contributions related to coherent 
backscattering, while loop corrections in {next-to} leading order {in the inverse number of energetically accessible channels} are taken into 
account in section \ref{sec:loop}.}

\section{Semiclassical theory of nonlinear coherent backscattering}

\label{sec:semi}

\subsection{{Coherent backscattering in the linear case}}

{The key ingredient of a semiclassical description of this nonlinear 
scattering process is the representation} of the retarded quantum Green 
function
\begin{equation}
  G_0(\mathbf{r},\mathbf{r}',E) \equiv \langle 
  \mathbf{r} | (E - H_0 + i 0)^{-1} | \mathbf{r}' \rangle
  \label{eq:G0}
\end{equation}
in terms of all classical (single-particle) trajectories 
$(\mathbf{p}_\gamma,\mathbf{q}_\gamma)(t)$ within the billiard,
indexed by 
$\gamma$, that propagate from the initial point 
$\mathbf{r}'$ to the final point $\mathbf{r}$ at total energy $E$.
{Here, we deliberately
exclude the vector potential $\mathbf{A}(\mathbf{r})$, i.e.~the underlying
Hamiltonian is given by
\begin{equation}
  H_0 = \frac{\hat{\mathbf{p}}^2}{2m} + V(\mathbf{r})
\end{equation}
where $\hat{\mathbf{p}} \equiv  - i \hbar \nabla$
represents the quantum momentum operator. The}
 semiclassical representation of the Green function can be derived 
from the Fourier transform of the quantum propagator in Feynman's path 
integral representation, which is evaluated in the formal limit 
$\hbar \to 0$ using the method of stationary phase.
It reads {\cite{Gut}}
\begin{equation}
  G_0(\mathbf{r},\mathbf{r}',E) = \sum_\gamma
  A_\gamma(\mathbf{r},\mathbf{r}',E) 
  {\exp\left[\frac{i}{\hbar} S_\gamma(\mathbf{r},\mathbf{r}',E)
    -i \frac{\pi}{2} \mu_\gamma\right]}
  \, . \label{eq:G0sc}
\end{equation}
Here,
\begin{equation}
  S_\gamma(\mathbf{r},\mathbf{r}',E) = \int_{0}^{{T_\gamma}} 
  \mathbf{p}_\gamma(t) \cdot \dot{\mathbf{q}}_\gamma(t) dt \label{eq:S}
\end{equation}
is the classical action integral along the trajectory $\gamma$ ({$T_\gamma$} 
denotes the total propagation time from $\mathbf{r}'$ to $\mathbf{r}$),
$\mu_\gamma$ represents the integer Maslov index that counts the number of 
conjugate points along the trajectory (which, in a billiard, {also
involves} twice the number of bouncings at the walls{, in addition to the 
number of conjugate points inside the billiard}), and
\begin{equation}
  A_\gamma(\mathbf{r},\mathbf{r}',E) = \frac{2\pi}{\sqrt{2 \pi i \hbar}^3}
  \sqrt{ |\det D^2S_\gamma(\mathbf{r},\mathbf{r}',E)|} \label{eq:ampl}
\end{equation}
is an amplitude that smoothly depends on $\mathbf{r}$ and $\mathbf{r}'${,
with}
\begin{equation}
{|\det D^2S_\gamma(\mathbf{r}, \mathbf{r}',E)| = \left| \det
  \frac{\partial(\mathbf{p}',\mathbf{r}',{T})}
       {\partial(\mathbf{r},\mathbf{r}',E)} \right| \, . \label{eq:D2S1}}
\end{equation}
{the Jacobian of the transformation
from the initial phase space variables $(\mathbf{p}',\mathbf{r}')$ and the
propagation time ${T}$ to the final and initial positions 
$(\mathbf{r},\mathbf{r}')$ and the energy $E$.}

The presence of a weak effective magnetic field is now incorporated in a 
perturbative manner {using the eikonal approximation.
As shown in \ref{sec:eikonal}, this} yields the {well-known
modification of the} Green function 
\begin{equation}
  G(\mathbf{r},\mathbf{r}',E) = \sum_\gamma A_\gamma(\mathbf{r},\mathbf{r}',E) 
  \exp\left\{\frac{i}{\hbar}\left[ S_\gamma(\mathbf{r},\mathbf{r}',E)
   - \phi_\gamma(\mathbf{r},\mathbf{r}',E)\right]
   -i\frac{\pi}{2} \mu_\gamma\right\}
  \label{eq:G1sc}
\end{equation}
{with $\phi_\gamma(\mathbf{r},\mathbf{r}',E) = - 
\varphi_\gamma(\mathbf{r},\mathbf{r}',E) - 
\tilde{\varphi_\gamma}(\mathbf{r},\mathbf{r}',E)$ and}
\begin{eqnarray}
  \varphi_\gamma(\mathbf{r},\mathbf{r}',E) & \equiv &
  \frac{1}{m} \int_0^{{T_\gamma}} \mathbf{p}_\gamma(t) \cdot 
  \mathbf{A}[\mathbf{q}_\gamma(t)] dt \, , \label{eq:phipara} \\
  {\varphi^{(\mathrm{d})}_\gamma}(\mathbf{r},\mathbf{r}',E) & \equiv &
  - \frac{1}{2m} \int_0^{{T_\gamma}}\mathbf{A}^2[\mathbf{q}_\gamma(t)] dt
\end{eqnarray}
{where the integration is peformed along the unperturbed trajectory
$\mathbf{q}_\gamma(t)$.}
While the latter (diamagnetic) contribution {$\varphi^{(\mathrm{d})}_\gamma$}
gives only rise to a spatial modulation of the effective potential background 
within the billiard, the former (paramagnetic) contribution $\varphi_\gamma$ 
explicitly breaks the time-reversal symmetry of the system and plays a 
crucial role for the intensity of coherent backscattering.

This expression for the Green function can be directly used in order to 
construct the scattering state $\psi(\mathbf{r})$ that arises as a 
stationary solution of Eq.~(\ref{eq:GPEs}).
We obtain
\begin{equation}
  \psi(\mathbf{r}) = S_0 \int G[ \mathbf{r}, (x_L,y'), \mu ] \chi_i(y') dy'
  \label{eq:scstate}
\end{equation}
where $\chi_i$ represents the energetically lowest transverse eigenmode within
the waveguide.
Assuming billiard-like waveguides with a vanishing potential background and 
infinitely high hard walls along their boundaries, the {{$n$th} 
normalized transverse eigenmode {(}$n>0${)} is given by}
\begin{equation}
  \chi_n(y) = \sqrt{\frac{2}{W}} \sin \left( \frac{p_n y}{\hbar} \right)
  = \frac{1}{2i} \sqrt{\frac{2}{W}} {\left[ \exp\left(\frac{i}{\hbar} p_n y\right) -
  \exp\left(-\frac{i}{\hbar} p_n y\right)\right]\, \mbox{for $0\leq y\leq W$}}
\end{equation}
{and $\chi_n(y)=0$ otherwise.} 
$p_n \equiv n \pi \hbar / W$
is the quantized transverse momentum and $W$ represents the width of the 
waveguide.
We can therefore write
\begin{equation}
  \psi(\mathbf{r}) = \frac{S_0}{i} \sqrt{\frac{\pi \hbar}{W}}
  \left\{ \Anfang{G}\left[\mathbf{r},(x_L,p_i),\mu\right] - 
  \Anfang{G}\left[\mathbf{r},(x_L,-p_i),\mu\right] 
  \right\} \label{scstsc}
\end{equation}
where
\begin{equation}
  \Anfang{G}[\mathbf{r},(x',p_y'),E] \equiv \frac{1}{\sqrt{2\pi\hbar}}
  {\int_0^W} G[\mathbf{r},(x',y'),E] {\exp\left(\frac{i}{\hbar} p_y' y'\right)} dy'
\end{equation}
denotes a partial Fourier transform of $G[\mathbf{r},(x',y'),E]$. 

Inserting the semiclassical expression (\ref{eq:G1sc}) for the Green function 
$G$, this partial Fourier transform can again be evaluated using the 
stationary phase approximation.
The stationary phase condition yields
$(\mathbf{p}_\gamma^{\mathrm{i}})_y[\mathbf{r},(x',y'),E] = p_y'$, i.e.\
$p_y'$ should be the $y$-component of the initial momentum of the trajectory.
The integration over $y'$ yields the prefactor $\sqrt{2\pi i \hbar/\alpha}$
with
\begin{equation}
  \alpha \equiv \frac{\partial^2}{\partial y'^2} S_\gamma[\mathbf{r},(x',y'),E]
  = - \frac{\partial[\mathbf{r},(x',p_y'),E]}{\partial[\mathbf{r},(x',y'),E]} 
  \, .
\end{equation}
Combining it with the prefactor $\sqrt{|\det D^2S_\gamma|}$ according to the 
expression (\ref{eq:D2S1}) and with the other prefactors that are contained
within the amplitude $A_\gamma$, we finally obtain
\begin{equation}
  \Anfang{G}(\mathbf{r},\mathbf{z}',E) = \sum_\gamma
  \Anfang{A}_\gamma(\mathbf{r},\mathbf{z}',E)
  \exp\left\{\frac{i}{\hbar}\left[
    \Anfang{S}_\gamma(\mathbf{r},\mathbf{z}',E) -
    \Anfang{\phi}_\gamma(\mathbf{r},\mathbf{z}',E) 
    \right] {- i\frac{\pi}{2} \Anfang{\mu}_\gamma} \right\}
\label{eq:G1barsc}
\end{equation}
with
\begin{eqnarray}
  \Anfang{A}_\gamma(\mathbf{r},\mathbf{z}',E) & = & 
  {\frac{2\pi\sqrt{i}}{\sqrt{2\pi i \hbar}^3}}
      \sqrt{\left|\det \frac{\partial[\mathbf{p}',(x',y'),{T}]}
        {\partial[\mathbf{r},(x',p_y'),E]}\right|} \, , \\
  \Anfang{\mu}_\gamma & = & \mu_\gamma + \left\{ \begin{array}{c@{\; : \;}l} 
    1 & \frac{\partial^2}{\partial y'^2} 
    S_\gamma(\mathbf{r},\mathbf{r}',E) < 0 \\ 0 & \mbox{otherwise}
  \end{array}\right. \, {,} \\
  {\Anfang{S}_\gamma(\mathbf{r},\mathbf{z}',E)} & {=} &
  {S_\gamma\left\{\mathbf{r},\left[ x', y_\gamma'(\mathbf{r},\mathbf{z}',E) 
      \right], E \right\} + p_y' y_\gamma'(\mathbf{r},\mathbf{z}',E) \, ,}
\end{eqnarray}
and $\Anfang{\phi}_\gamma(\mathbf{r},\mathbf{z}',E)$ defined 
according to {Eq.}~(\ref{eq:Phi}),
where the initial phase-space point of the trajectories $\gamma$ is given
by the combination $\mathbf{z}'\equiv (x',p_y')$
{and $y_\gamma'(\mathbf{r},\mathbf{z}',E)$ denotes the resulting initial 
$y$ coordinate}.

Channel-resolved reflection and transmission amplitudes can now be computed
by projecting $\psi$ onto the transverse eigenmodes of the waveguides.
This involves again a partial Fourier transform of the Green function, this
time in the final coordinate.
In particular, the reflection amplitude into channel $n$ is obtained from
\begin{eqnarray}
  \psi_n & \equiv & {\int_0^W} \chi_n^*(y) \psi(x_L,y) dy \\
  & = & S_0 \frac{\pi \hbar}{W} \left\{ 
  \AnfangEnde{G}[(x_L,p_n),(x_L,p_i),\mu]
  - \AnfangEnde{G}[(x_L,p_n),(x_L,-p_i),\mu] \right. \nonumber \\
  && \left. - \AnfangEnde{G}[(x_L,-p_n),(x_L,p_i),\mu]
  + \AnfangEnde{G}[(x_L,-p_n),(x_L,-p_i),\mu] \right\} \nonumber \\
  & \equiv & S_0 \frac{\pi \hbar}{W} \left[
  \AnfangEnde{G}(\mathbf{z}_n^-,\mathbf{z}_1^+,\mu) - 
  \AnfangEnde{G}(\mathbf{z}_n^+,\mathbf{z}_1^+,\mu) - 
  \AnfangEnde{G}(\mathbf{z}_n^-,\mathbf{z}_1^-,\mu) + 
  \AnfangEnde{G}(\mathbf{z}_n^+,\mathbf{z}_1^-,\mu) \right] 
  \label{eq:psin}
\end{eqnarray}
with 
\begin{eqnarray}
  \Ende{G}[(x,p_y),\mathbf{r}',E] & \equiv &
  \frac{1}{\sqrt{2\pi\hbar}}
  {\int_0^W} G[(x,y),\mathbf{r}',E] {\exp\left(-\frac{i}{\hbar} p_y y\right)} dy
  \label{eq:G1tildesc} \, , \\
  \AnfangEnde{G}[(x,p_y),\mathbf{z}',E] & \equiv &
  \frac{1}{\sqrt{2\pi\hbar}}
  {\int_0^W} \Anfang{G}[(x,y),\mathbf{z}',E] {\exp\left(-\frac{i}{\hbar} p_y y\right)} dy
  \label{eq:G1barbarsc}
\end{eqnarray}
where we define
\begin{equation}
  \mathbf{z}_n^\pm \equiv \left\{ \begin{array}{r@{\;}l} 
    (x_L,\pm p_n) & \mbox{for incoming trajectories (with ${p'_x} > 0$)} \\
    (x_L,\mp p_n) & \mbox{for outgoing trajectories (with $p_x < 0$)}
  \end{array} \right. . \label{eq:zn}
\end{equation}
Similarly as for $\Anfang{G}$, the semiclassical evaluation of this
Fourier transform using Eq.~(\ref{eq:G1barsc}) yields 
{\cite{adist3,adist1,adist2}}
\begin{equation}
  \AnfangEnde{G}(\mathbf{z},\mathbf{z}',E) = \sum_\gamma
  \AnfangEnde{A}_\gamma(\mathbf{z},\mathbf{z}',E)
  \exp\left\{\frac{i}{\hbar}\left[
    \AnfangEnde{S}_\gamma(\mathbf{z},\mathbf{z}',E) -
    \AnfangEnde{\phi}_\gamma(\mathbf{z},\mathbf{z}',E) 
    \right]{-i\frac{\pi}{2}\AnfangEnde{\mu}_\gamma} \right\}
\label{eq:G2barsc}
\end{equation}
with
\begin{eqnarray}
  \AnfangEnde{A}_\gamma(\mathbf{z},\mathbf{z}',E) & = & 
  {\frac{2\pi i}{\sqrt{2\pi i\hbar}^3}}
  \sqrt{\left|\det \frac{\partial[(p_x',p_y'),(x',y'),{T}]}
    {\partial[(x,p_y),(x',p_y'),E]}\right|} \, , \\
  \AnfangEnde{\mu}_\gamma & = & \Anfang{\mu}_\gamma + 
  \left\{ \begin{array}{c@{\; : \;}l} 
    1 & \frac{\partial^2}{\partial y^2} 
    \Anfang{S}_\gamma(\mathbf{r},\mathbf{z}',E) < 0 \\ 0 & \mbox{otherwise}
  \end{array}\right. \, {,} \\
  {\AnfangEnde{S}_\gamma(\mathbf{z},\mathbf{z}',E)} & {=} &
  {\Anfang{S}_\gamma\left\{\left[x,y_\gamma(\mathbf{z},\mathbf{z}',E)\right],
    \mathbf{z}',E\right\} - p_y y_\gamma(\mathbf{z},\mathbf{z}',E) \, ,}
\end{eqnarray}
and
$\AnfangEnde{\phi}_\gamma(\mathbf{z},\mathbf{z}',E)$ defined according to
 {Eq}.~(\ref{eq:Phi}),
where the final phase-space point of the trajectories $\gamma$ is given
by the combination $\mathbf{z}\equiv (x,p_y)$
{(and $y_\gamma$ is the final $y$ coordinate)}.

From {Eq.~(\ref{eq:G2barsc})} it becomes obvious that
subtle interferences between different classical trajectories may give rise
to channel-resolved reflection and transmission probabilities that strongly
fluctuate under variation of the incident chemical potential 
$\mu~{\equiv E}$.
Those fluctuations generally cancel, however, when performing an average
within a finite window of chemical potentials.
Specifically, the calculation of $|\psi_n|^2$ involves sums over pairs of
trajectories $\gamma$ and $\gamma'$, whose contributions contain phase factors
that depend on the difference 
$\AnfangEnde{S}_\gamma - \AnfangEnde{S}_{\gamma'}$ 
of the associated action integrals.
These differences strongly vary with the chemical potential $\mu$ unless the
two trajectories $\gamma$ and $\gamma'$ are somehow correlated.

An obvious correlation arises if the two trajectories happen to be identical,
in which case the phase factor is unity.
In the framework of the \emph{diagonal approximation}, we only take into 
account this specific case, i.e., we approximate the double sum 
$\sum_{\gamma,\gamma'}$ by a single sum $\sum_\gamma$ where 
$\gamma'$ is taken to be identical to $\gamma$.
The energy average $\langle |\psi_n|^2 \rangle$ of $|\psi_n|^2$ is then given
by
\begin{eqnarray}
  \langle |\psi_n|^2 \rangle & \simeq & \langle |\psi_n|^2 
  \rangle_{\mathrm{d}} \label{eq:psin2} \\
  & = & \left|S_0 \frac{\pi \hbar}{W}\right|^2 \left[
  \left\langle\left|\AnfangEnde{G}(\mathbf{z}_n^+,\mathbf{z}_1^+,\mu)
  \right|^2\right\rangle_{\mathrm{d}} +
  \left\langle\left|\AnfangEnde{G}(\mathbf{z}_n^+,\mathbf{z}_1^-,\mu)
  \right|^2 \right\rangle_{\mathrm{d}}
  \nonumber\right. \\ & & + \left.
  \left\langle\left|\AnfangEnde{G}(\mathbf{z}_n^-,\mathbf{z}_1^+,\mu)
  \right|^2\right\rangle_{\mathrm{d}} +
  \left\langle\left|\AnfangEnde{G}(\mathbf{z}_n^-,\mathbf{z}_1^-,\mu)
  \right|^2\right\rangle_{\mathrm{d}} \right] \label{eq:psin2diag}
\end{eqnarray}
with {$\left\langle\left|\AnfangEnde{G}(\mathbf{z},\mathbf{z}',E)
  \right|^2\right\rangle_{\mathrm{d}} = \sum_\gamma \left\langle
  \left|\AnfangEnde{A}_\gamma(\mathbf{z},\mathbf{z}',E)\right|^2 
  \right\rangle$.}

As shown in \ref{sec:sumrules}, this sum is evaluated using 
the generalized Hannay-Ozorio de Almeida sum rule
\cite{HanOzo84JPA,Sie99JPA}.
Defining by $\tau_D$ the ``dwell time'' of the system, i.e.\ the mean 
evolution time that a classical trajectory spends within the billiard 
before escaping to one of the waveguides, and introducing the 
``Heisenberg time'' as $\tau_H \equiv m \Omega / \hbar$ where $\Omega$ 
denotes the area of the billiard, we obtain [see Eq.~(\ref{eq:Gzz})]
\begin{equation}
  \left\langle\left|\AnfangEnde{G}(\mathbf{z},\mathbf{z}',E)
  \right|^2\right\rangle_{\mathrm{d}} =
  \left(\frac{mW}{2\pi\hbar^2}\right)^2 \frac{\tau_D}{\tau_H}
  \frac{1}{\sqrt{2mE - p_y^2}}\frac{1}{\sqrt{2mE - p_y'^2}} \, .
  \label{eq:G12diagsr}
\end{equation}
Inserting this expression into Eq.~(\ref{eq:psin2diag}) and defining
\begin{equation}
  p_n^{\mathrm{l}}(E) \equiv \sqrt{2 m E - p_n^2}
  = \sqrt{2 m E - (n \pi \hbar / W)^2}
\end{equation}
as the longitudinal component of the momentum that is associated with
the transverse mode $\chi_n$ {finally yields}
\begin{equation}
  \langle |\psi_n|^2 \rangle_{\mathrm{d}} = 
  \left| \frac{m S_0}{\hbar} \right|^2 \frac{\tau_D}{\tau_H} 
  \frac{1}{p_n^{\mathrm{l}}(\mu) p_i^{\mathrm{l}}(\mu)} \, .
  \label{eq:psin2srdiag}
\end{equation}
This expression can be used in order to determine the steady current 
$j_n$ of atoms that are reflected into channel $n$, according to
\begin{equation}
  {j_n = \frac{p_n^{\mathrm{l}}(\mu)}{m} \langle |\psi_n|^2 \rangle \, .}
\end{equation}
Dividing it by the incident current which is derived from Eq.~(\ref{eq:psi0})
as 
\begin{equation}
  {j^{\mathrm{i}} = \frac{m |S_0|^2}{\hbar^2 p_i^{\mathrm{l}}(\mu)} \, ,}
\end{equation}
we obtain the reflection probability into channel $n$ as
\begin{equation}
  r_{ni} \equiv j_n / j^{\mathrm{i}} = \tau_D / \tau_H \, .
  \label{eq:r1n}
\end{equation}

The same reasoning can be applied to the outgoing waveguide on the other,
transmitted side of the billiard. 
Again we obtain $t_{{ni}} = \tau_D/\tau_H$ as the probability for transmission
into the transverse channel $n$ of the outgoing waveguide, even if its width
$\tilde{W}$ is different from the width $W$ of the incoming guide.
The total reflection and transmission probabilities $R$ and $T$ are then 
simply related to the numbers of open channels $N_c$ and $\tilde{N}_c$ in
the incoming and outgoing waveguide according to $R = N_c \tau_D/\tau_H$
and $T = \tilde{N}_c \tau_D/\tau_H$, where we evaluate
$N_c = 2 W / \lambda_{\mathrm{dB}}$ and 
$\tilde{N}_c = 2 \tilde{W} / \lambda_{\mathrm{dB}}$ 
in the semiclassical limit, with
$\lambda_{\mathrm{dB}}\equiv 2 \pi \hbar / \sqrt{2 m \mu}$
the de Broglie wavelength of the atoms.
We can furthermore use the general expression {\cite{adist3,RicSie02PRL}}
\begin{equation}
  \tau_D = \frac{\pi \Omega}{(W + \tilde{W})v} \label{eq:tauD}
\end{equation}
for the mean survival time of a classical particle propagating with
velocity $v$ in a chaotic billiard with area $\Omega$ that contains
two openings of width $W$ and $\tilde{W}$, 
{which yields}
\begin{equation}
  {\frac{\tau_D}{\tau_H} = \frac{\lambda_{\mathrm{dB}} / 2}{W+\tilde{W}}
    = \frac{1}{N_c + \tilde{N}_c} \, .} \label{eq:tauDtauH}
\end{equation}
{We then} arrive at the intuitive
results $R = W / ( W + \tilde{W})$ and $T = \tilde{W} / ( W + \tilde{W})$,
i.e.\ the total reflection and transmission probabilities are simply given 
by the relative widths of the corresponding waveguides.

The diagonal approximation therefore yields predictions for reflection
and transmission that are expected for incoherent, classical particles
{in a chaotic cavity}.
It represents {in leading order in the inverse total channel number 
$(N_c+\tilde{N}_c)^{{-1}}$ the contributions} for all channels on the
transmitted side, and for all reflected channels except for the channel
$n={i}$ in which the matter-wave beam is injected into the billiard.
In this incident channel, there is another, equally important possibility
to pair the trajectories $\gamma$ and $\gamma'$ in the double sums that are
involved in the calculation of $|\psi_{{i}}|^2$:
$\gamma'$ can be chosen to be the \emph{time-reversed counterpart} of
$\gamma$, the existence of which is guaranteed by the time-reversal 
symmetry of $H_0$.

Consequently, Eq.~(\ref{eq:psin2}) has to be corrected for the special
case $n={i}$ according to
\begin{equation}
  \langle |\psi_{{i}}|^2 \rangle \simeq \langle |\psi_{{i}}|^2 
  \rangle_{\mathrm{d}} + \langle |\psi_{{i}}|^2 \rangle_{\mathrm{c}}
\end{equation}
where the ``crossed'' or ``Cooperon''{-type} contribution
\begin{eqnarray}
  \langle |\psi_{{i}}|^2 \rangle_{\mathrm{c}} & = &
  \left|S_0 \frac{\pi \hbar}{W}\right|^2 \left[
  \left\langle\left|\AnfangEnde{G}
  (\mathbf{z}_{{i}}^+,\mathbf{z}_{{i}}^+,\mu)
  \right|^2\right\rangle_{\mathrm{c}} +
  \left\langle\left|\AnfangEnde{G}
  (\mathbf{z}_{{i}}^-,\mathbf{z}_{{i}}^-,\mu)
  \right|^2 \right\rangle_{\mathrm{c}}
  \nonumber\right. \\ & & + \left. \left\langle
  \AnfangEnde{G}^*(\mathbf{z}_{{i}}^+,\mathbf{z}_{{i}}^-,\mu) \,
  \AnfangEnde{G}(\mathbf{z}_{{i}}^-,\mathbf{z}_{{i}}^+,\mu)
  \right\rangle_{\mathrm{c}} + \left\langle
  \AnfangEnde{G}^*(\mathbf{z}_{{i}}^-,\mathbf{z}_{{i}}^+,\mu) \,
  \AnfangEnde{G}(\mathbf{z}_{{i}}^+,\mathbf{z}_{{i}}^-,\mu)
  \right\rangle_{\mathrm{c}} \right] \label{eq:psi12cross}
\end{eqnarray}
contains all those combinations of trajectories for which $\gamma'$ is
the time-reversed counterpart of $\gamma$.
Obviously, the action integrals {$\AnfangEnde{S}_\gamma$} and 
Maslov indices {$\AnfangEnde{\mu}_\gamma$} are identical for the
trajectories $\gamma$ and their time-reversed counterparts.
This is not the case, however, for the modification 
{$\AnfangEnde{\phi}_\gamma$} of the action integral that is induced by 
the vector potential, whose paramagnetic part 
{$\AnfangEnde{\varphi}_\gamma$} [Eq.~(\ref{eq:phipara})] 
changes sign when integrating along the trajectory $\gamma$ in the 
opposite direction.
We therefore obtain
\begin{equation}
  \left\langle\AnfangEnde{G}^*(\mathbf{z}',\mathbf{z},E) \,
  \AnfangEnde{G}(\mathbf{z},\mathbf{z}',E)
  \right\rangle_{\mathrm{c}} = \sum_\gamma \left\langle
  \left|\AnfangEnde{A}_\gamma(\mathbf{z},\mathbf{z}',E)\right|^2 
  \exp\left[\frac{2 i}{\hbar} \AnfangEnde{\varphi}_\gamma
    (\mathbf{z},\mathbf{z}',E)\right] \right\rangle \, .
  \label{eq:G12cross}
\end{equation}

To provide some physical insight into the role of this additional phase
factor, we use the representation (\ref{eq:A}) of the vector potential
within the billiard.
Using $\mathbf{p}_\gamma(t) = m \dot{\mathbf{q}}_\gamma(t)$ along
trajectories $\gamma$ generated by $H_0$, the paramagnetic contribution 
to the effective action integral reads then
\begin{equation}
  \AnfangEnde{\varphi}_\gamma(\mathbf{z},\mathbf{z}',E) =
  \frac{B}{2} \, \mathbf{e}_z \cdot \int_0^{{T_\gamma}} \left[
  \mathbf{q}_\gamma(t) - \mathbf{r}_0 \right] \times 
  \dot{\mathbf{q}}_\gamma(t) \, dt \label{eq:phipara1}
\end{equation}
where $\mathbf{r}_0$ is an arbitrarily chosen reference point.
Within the billiard, the trajectories 
$(\mathbf{p}_\gamma,\mathbf{q}_\gamma)(t)$
can be decomposed into segments of straight lines that connect
subsequent reflection points at the billiard boundary.
Denoting those reflection points by 
$\mathbf{q}_\gamma^{(1)}, \ldots, \mathbf{q}_\gamma^{(N-1)}$
and defining $\mathbf{q}_\gamma^{(0)} \equiv \mathbf{r}'$ and
$\mathbf{q}_\gamma^{(N)} \equiv \mathbf{r}$, the initial and final points
of the trajectory, we rewrite Eq.~(\ref{eq:phipara1}) as
\begin{equation}
  \AnfangEnde{\varphi}_\gamma(\mathbf{z},\mathbf{z}',E) =
  B \sum_{j=1}^N a_j {\equiv} B {\mathcal{A}} \label{eq:phipara2}
\end{equation}
where
\begin{equation}
  a_j = \frac{1}{2} \, \mathbf{e}_z \cdot \left[
  (\mathbf{q}_\gamma^{(j-1)} - \mathbf{r}_0) \times
  (\mathbf{q}_\gamma^{(j)} - \mathbf{r}_0) \right]
\end{equation}
is the directed area of the triangle spanned by the reflection points
$\mathbf{q}_\gamma^{(j-1)}$ and $\mathbf{q}_\gamma^{(j)}$ as well as by the
reference point $\mathbf{r}_0$.
Quite obviously, $\AnfangEnde{\varphi}_\gamma$ is independent of
the particular choice of $\mathbf{r}_0$, or of any other gauge transformation
$\mathbf{A} \mapsto \mathbf{A} + \nabla \chi$ that vanishes within the
waveguide, provided the initial and final points $\mathbf{r}'$ and 
$\mathbf{r}$ of the trajectory $\gamma$ are identical or, less restrictively,
lie both within the same, incident waveguide where the vector potential 
vanishes (in which case a straight-line integration 
$\int \mathbf{A} \cdot d\mathbf{q}$ from $\mathbf{r}'$ to $\mathbf{r}$ 
would formally close the trajectory without adding any further contribution
to {$\AnfangEnde{\varphi}_\gamma$}).

The central limit theorem is now applied in order to obtain the
pro{b}ability distribution ${P(T_{\gamma},{\mathcal{A}})}$
{for accumulating the area ${\mathcal{A}}$ after the propagation time 
$T_\gamma$}
\cite{adist3,RicSie02PRL,adist1,adist2}. We have
\begin{equation}
P({T}_{\gamma},{\mathcal{A}})=\frac{1}{\sqrt{\pi
{\eta \Omega^{3/2} v T_{\gamma}}}}
{\exp\left(-\frac{\mathcal{A}^2}{\eta \Omega^{3/2} v T_{\gamma}}
\right)} \label{eq:tadistri}
\end{equation}
where $\Omega$ is the area of the billiard, $v$ is the velocity of 
the particle, and $\eta$ is a dimensionless scaling parameter that
characterizes the geometry of the system and that can be numerically 
computed from the classical dynamics within the billiard
as described in \ref{sec:cladyn}.
This distribution is now used to obtain an average value of the magnetic
phase factor according to
\begin{equation}
  \left\langle\exp\left[\frac{2 i}{\hbar} 
    \AnfangEnde{\varphi}_\gamma(\mathbf{z},\mathbf{z}',E)\right] \right\rangle
  = \int_{-\infty}^{+\infty} P({T}_{\gamma},{\mathcal{A}}) 
  {\exp\left(\frac{2 i}{\hbar} B \mathcal{A} \right)} d{\mathcal{A}}
  = {\exp\left(-\frac{T_\gamma}{\tau_B}\right)}
 \label{eq:phasedecay}
\end{equation}
with
\begin{equation}
  \tau_B \equiv \frac{\hbar^2}{{\eta \Omega^{3/2} v B^2}}
  \label{eq:tB}
\end{equation}
{the characteristic time scale for magnetic dephasing.}

With this information, we can now follow the derivation of the 
Hannay-Ozorio de Almeida sum rule, as explicated in 
\ref{sec:sumrules}, in order to evaluate
the expression (\ref{eq:G12cross}), with the only complication that
each contribution in the sum over trajectories needs to be weighted
by the ``dephasing'' factor $\exp(- T_\gamma /\tau_B)$.
This yields
\begin{equation}
  \left\langle\AnfangEnde{G}^*(\mathbf{z}',\mathbf{z},E) \,
  \AnfangEnde{G}(\mathbf{z},\mathbf{z}',E)
  \right\rangle_{\mathrm{c}} = 
  \left(\frac{mW}{2\pi\hbar^2}\right)^2 
  \left(\frac{\tau_H}{\tau_D} + \frac{\tau_H}{\tau_B} \right)^{-1}
  \frac{1}{\sqrt{2mE - p_y^2}}\frac{1}{\sqrt{2mE - p_y'^2}} \, .
\end{equation}
Hence, we obtain
\begin{equation}
  \langle |\psi_{{i}}|^2 \rangle_{\mathrm{c}} = 
  \left| \frac{m S_0}{\hbar p_i^{\mathrm{l}}(\mu)} \right|^2 
  \left(\frac{\tau_H}{\tau_D} + \frac{\tau_H}{\tau_B} \right)^{-1}
  \label{eq:psin2srcross}
\end{equation}
in very close analogy with Eq.~(\ref{eq:psin2srdiag}), which altogether yields
\begin{equation}
  \langle |\psi_{{i}}|^2 \rangle \simeq \langle |\psi_{{i}}|^2 
  \rangle_{\mathrm{d}}
  + \langle |\psi_{{i}}|^2 \rangle_{\mathrm{c}}
  = \left| \frac{m S_0}{\hbar p_i^{\mathrm{l}}(\mu)} \right|^2
  \left( 1 + \frac{1}{1 + {\tau_D / \tau_B}} \right) \frac{\tau_D}{\tau_H}
  \, .
\end{equation}
This gives rise to {an enhanced} probability for retro-reflection into the 
incident channel $n={i}$, namely
\begin{equation}
   r_{{ii}} = \left( 1 + \frac{1}{1 + {\tau_D / \tau_B}} \right)
   \frac{\tau_D}{\tau_H} = 
   \left( 1 + \frac{1}{1 + {B^2 / B_0^2}} \right) \frac{\tau_D}{\tau_H}
   \label{eq:r11}
\end{equation}
with
\begin{equation}
  B_0 \equiv {\frac{\hbar}{\sqrt{\eta v \tau_D \Omega^{3/2}}} }
  \, , \label{eq:B0}
\end{equation}
as compared to reflection into different channels described by 
Eq.~(\ref{eq:r1n}), which is the characteristic signature of
coherent backscattering.
{Note that, due to conservation of the total flux, increased 
retro-reflection for $n=i$ implies decreased reflection or transmission 
into other channels $n\neq i$. 
This will be subject of Section \ref{sec:loop} below.}

The above prediction (\ref{eq:r11}) is expected to be valid 
{for chaotic cavities} in the semiclassical limit of small $\hbar$ 
(i.e.\ of a small de Broglie wavelength as compared to the size of the
scattering region) and in the limit of small widths of the leads.
Leads of finite widths, as the ones that are considered in the scattering
geometries shown in Fig.~\ref{fig:billiard}, will give rise to non-universal
corrections to Eq.~(\ref{eq:r11}) that are related to short reflected or
transmitted paths.
In particular, the presence of \emph{self-retracing} trajectories, which
are identical to their time-reversed counterparts, affects the probability for
retro-reflection due to coherent backscattering, as those trajectories are
evidently doubly counted in the addition of ladder and crossed contributions.
Hence, the enhancement of this retro-reflection probability with respect to
the incoherent ladder background (\ref{eq:r1n}) will, in practice, be reduced
as compared to Eq.~(\ref{eq:r11}), due to the presence of short and therefore
semiclassically relevant self-retracing trajectories.

\subsection{Diagrammatic representation of nonlinear scattering states}

\label{sec:nldiag}
We now consider the presence of a weak interaction strength $g>0$ in the 
Gross-Pitaevskii equation (\ref{eq:GPEs}).
As a consequence, the scattering process becomes nonlinear and the final
(stationary or time-dependent) scattering state may depend on the 
``history'' of the process, i.e.\ on the initial matter-wave population within
the scattering region as well as on the specific ramping process of the source
amplitude.
{We} shall assume that the scattering region is initially empty (i.e., 
$\psi(\mathbf{r},t) = 0$ for $t \to - \infty$) and that the source amplitude
$S_0$ is adiabatically ramped from zero to a given maximal value $\tilde{S}_0$, 
on a time scale that is much larger than any other relevant time scale of the
scattering system.
This adiabatic ramping is formally expressed as $S_0(t) = \tilde{S}_0 f(t/t_R)$
where $f(\tau)$ is a real dimensionless function that {monotonically} 
increases from $0$ (for $t \to -\infty$) to $1$ (for $t \to \infty$) and 
$t_R \to \infty$ is a very large ramping time scale.
Redefining $\psi(\mathbf{r},t) \equiv f(t/t_R) \tilde{\psi}(\mathbf{r},t)$ and
neglecting terms of the order of $1/t_R$, we obtain from Eq.~(\ref{eq:GPEs})
\begin{equation}
i \hbar \frac{\partial}{\partial t} \tilde{\psi}(\mathbf{r},t)  =
( H - \mu ) \tilde{\psi}(\mathbf{r},t) + \tilde{S}(\mathbf{r},t)
\label{eq:GPEst}
\end{equation}
as effective Gross-Pitaevskii equation for $\tilde{\psi}$, with
\begin{equation}
\tilde{S}(\mathbf{r},t) \equiv S_0 \chi_i(y) \delta(x - x_L) + 
{\tilde{g}(t) \frac{\hbar^2}{2m}}
|\tilde{\psi}(\mathbf{r},t)|^2 \tilde{\psi}(\mathbf{r},t) \label{eq:sourcet}
\end{equation}
and $\tilde{g}(t) \equiv f^2(t/t_R) g$.
For weak enough nonlinearities $g$ and long enough ramping time scales $t_R$,
Eq.~(\ref{eq:GPEst}) can be considered as describing an effectively linear
scattering problem the source term of which is gradually adapted according to
Eq.~(\ref{eq:sourcet}).
We can therefore express the time-dependent scattering wavefunction as
\begin{equation}
  \tilde{\psi}(\mathbf{r},t) = \int d^2 r' G(\mathbf{r},\mathbf{r}',\mu)
  \tilde{S}(\mathbf{r}',t)
\end{equation}
where $G \equiv (\mu - H + i0)^{-1}$ is the Green function of the linear 
scattering problem [see {Eq.~(\ref{eq:G1sc})}].
In the limit of long evolution times $t\to\infty$, we thereby obtain
\begin{equation}
  \psi(\mathbf{r}) = S_0 \int G[ \mathbf{r}, (x_L,y'), \mu ] \chi_i(y') dy'
  + \int d^2 r' G(\mathbf{r},\mathbf{r}',\mu) 
 {g\frac{\hbar^2}{2m} }
  |\psi(\mathbf{r}')|^2 \psi(\mathbf{r}') \label{eq:scstate_g}
\end{equation}
as self-consistent equation for the scattering wavefunction, which generalizes
the expression (\ref{eq:scstate}) obtained for the linear case.

In rather close analogy with the numerical procedure that is employed for
computing a stationary scattering state, we can construct a self-consistent 
solution of Eq.~(\ref{eq:scstate_g}) by starting with the expression 
(\ref{eq:scstate}) for the linear case and by iteratively inserting the
subsequent expressions obtained for $\psi(\mathbf{r})$ on the right-hand side 
of Eq.~(\ref{eq:scstate_g}).
This naturally leads to a power series in the nonlinearity,
\begin{equation}
  \psi(\mathbf{r}) = \psi^{(0)}(\mathbf{r}) + 
  \sum_{n=1}^\infty g^n \delta \psi^{(n)}(\mathbf{r}) \, , \label{eq:scstg}
\end{equation}
where $\psi^{(0)}(\mathbf{r})$ represents the solution of 
Eq.~(\ref{eq:scstate}), i.e.\ the scattering state of the noninteracting 
system.

It is instructive to evaluate the semiclassical representation of the 
first-order correction to the linear scattering wavefunction $\psi^{(0)}$, 
given by
\begin{equation}
  \delta \psi^{(1)}(\mathbf{r}) = \frac{\hbar^2}{2m}
  \int d^2 r' G(\mathbf{r},\mathbf{r}',\mu) 
  |\psi^{(0)}(\mathbf{r}')|^2 \psi^{(0)}(\mathbf{r}') \, . \label{eq:scst1}
\end{equation}
Using the expression (\ref{scstsc}) for the scattering state of the
noninteracting system, we obtain
\begin{eqnarray}
  \delta \psi^{(1)}(\mathbf{r}) & = & \frac{\hbar^2}{2m}
  \frac{S_0}{i} \sqrt{\frac{\pi \hbar}{W}}
  |S_0|^2 \frac{\pi \hbar}{W} \sum_{\nu_1,\nu_2,\nu_3=\pm 1}
  \nu_1\nu_2\nu_3 \nonumber \\
  & & \times \int d^2r' G(\mathbf{r},\mathbf{r}',\mu)
  \Anfang{G}(\mathbf{r}',\mathbf{z}_{{i}}^{\nu_1},\mu)
  \Anfang{G}^*(\mathbf{r}',\mathbf{z}_{{i}}^{\nu_2},\mu)
  \Anfang{G}(\mathbf{r}',\mathbf{z}_{{i}}^{\nu_3},\mu)
\end{eqnarray}
with $\mathbf{z}_{{i}}^{\pm 1} \equiv \mathbf{z}_{{i}}^\pm$ as defined in 
Eq.~(\ref{eq:zn}).
Inserting the semiclassical expansion for the Green function, given by
Eqs.~(\ref{eq:G1sc}) and (\ref{eq:G1barsc}), yields
\begin{eqnarray}
  \delta \psi^{(1)}(\mathbf{r}) & = & \frac{\hbar^2}{2m} \frac{S_0}{i} 
  \sqrt{\frac{\pi \hbar}{W}}
  |S_0|^2 \frac{\pi \hbar}{W} \sum_{\nu_1,\nu_2,\nu_3=\pm 1}
  \nu_1\nu_2\nu_3 \nonumber \\
  & & \times \int d^2r' \sum_{\gamma_0} \sum_{\gamma_1,\gamma_2,\gamma_3}
  A_{\gamma_0}(\mathbf{r},\mathbf{r}',\mu)
  \Anfang{A}_{\gamma_1}(\mathbf{r}',\mathbf{z}_{{i}}^{\nu_1},\mu)
  \Anfang{A}_{\gamma_2}^*(\mathbf{r}',\mathbf{z}_{{i}}^{\nu_2},\mu)
  \Anfang{A}_{\gamma_3}(\mathbf{r}',\mathbf{z}_{{i}}^{\nu_3},\mu) \nonumber \\
  & & \times \exp\left\{\frac{i}{\hbar}\left[      
    S_{\gamma_0}(\mathbf{r},\mathbf{r}',\mu) +
    \Anfang{S}_{\gamma_1}(\mathbf{r}',\mathbf{z}_{{i}}^{\nu_1},\mu) -
    \Anfang{S}_{\gamma_2}(\mathbf{r}',\mathbf{z}_{{i}}^{\nu_2},\mu) +
    \Anfang{S}_{\gamma_3}(\mathbf{r}',\mathbf{z}_{{i}}^{\nu_3},\mu)
    \right]\right\} \nonumber \\
  & & \times \exp\left\{-\frac{i}{\hbar}\left[
    \phi_{\gamma_0}(\mathbf{r},\mathbf{r}',\mu) +
    \Anfang{\phi}_{\gamma_1}(\mathbf{r}',\mathbf{z}_{{i}}^{\nu_1},\mu) -
    \Anfang{\phi}_{\gamma_2}(\mathbf{r}',\mathbf{z}_{{i}}^{\nu_2},\mu) +
    \Anfang{\phi}_{\gamma_3}(\mathbf{r}',\mathbf{z}_{{i}}^{\nu_3},\mu)
    \right]\right\} \nonumber \\
  & & \times \exp\left[ -\frac{i\pi}{2}\left( \mu_{\gamma_0} + 
    \Anfang{\mu}_{\gamma_1} - \Anfang{\mu}_{\gamma_2} + 
    \Anfang{\mu}_{\gamma_3} \right)\right]
  \label{eq:scst1sc}
\end{eqnarray}
where the indices $\gamma_0$ and $\gamma_\ell$ ($\ell=1,2,3$) represent 
trajectories that connect $\mathbf{r}'$ and $\mathbf{r}$ as well as 
$\mathbf{z}_{{i}}^{\nu_\ell}$ and $\mathbf{r}'$, respectively.

Neglecting, as done in Section \ref{sec:semi}, the modification of the
trajectories $\gamma_0$ and $\gamma_{1/2/3}$ due to the presence of the
weak magnetic field, a stationary-phase evaluation of the spatial integral
in Eq.~(\ref{eq:scst1sc}) yields the condition
\begin{equation}
  \mathbf{p}_{\gamma_0}^{\mathrm{i}}(\mathbf{r},\mathbf{r}',\mu) + 
  \mathbf{p}_{\gamma_2}^{\mathrm{f}}(\mathbf{r}',\mathbf{z}_{{i}}^{\nu_2},\mu) =
  \mathbf{p}_{\gamma_1}^{\mathrm{f}}(\mathbf{r}',\mathbf{z}_{{i}}^{\nu_1},\mu) +
  \mathbf{p}_{\gamma_3}^{\mathrm{f}}(\mathbf{r}',\mathbf{z}_{{i}}^{\nu_3},\mu) \, .
\end{equation}
Noting that all involved momenta are evaluated at the same spatial point 
$\mathbf{r}'$, this condition is satisfied if and only if
\begin{equation}
  \mathbf{p}_{\gamma_0}^{\mathrm{i}}(\mathbf{r},\mathbf{r}',\mu)
  = \mathbf{p}_{\gamma_1}^{\mathrm{f}}(\mathbf{r}',\mathbf{z}_{{i}}^{\nu_1},\mu)
  \quad \mbox{and} \quad 
  \mathbf{p}_{\gamma_2}^{\mathrm{f}}(\mathbf{r}',\mathbf{z}_{{i}}^{\nu_2},\mu)
  = \mathbf{p}_{\gamma_3}^{\mathrm{f}}(\mathbf{r}',\mathbf{z}_{{i}}^{\nu_3},\mu)
  \label{eq:l1cond}
\end{equation}
or
\begin{equation}
  \mathbf{p}_{\gamma_0}^{\mathrm{i}}(\mathbf{r},\mathbf{r}',\mu)
  = \mathbf{p}_{\gamma_3}^{\mathrm{f}}(\mathbf{r}',\mathbf{z}_{{i}}^{\nu_3},\mu)
  \quad \mbox{and} \quad 
  \mathbf{p}_{\gamma_2}^{\mathrm{f}}(\mathbf{r}',\mathbf{z}_{{i}}^{\nu_2},\mu)
  = \mathbf{p}_{\gamma_1}^{\mathrm{f}}(\mathbf{r}',\mathbf{z}_{{i}}^{\nu_1},\mu)
  \label{eq:l2cond}
\end{equation}
or
\begin{equation}
  \mathbf{p}_{\gamma_0}^{\mathrm{i}}(\mathbf{r},\mathbf{r}',\mu)
  = - \mathbf{p}_{\gamma_2}^{\mathrm{f}}(\mathbf{r}',\mathbf{z}_{{i}}^{\nu_2},\mu)
  \quad \mbox{and} \quad 
  \mathbf{p}_{\gamma_1}^{\mathrm{f}}(\mathbf{r}',\mathbf{z}_{{i}}^{\nu_1},\mu)
  = - \mathbf{p}_{\gamma_3}^{\mathrm{f}}(\mathbf{r}',\mathbf{z}_{{i}}^{\nu_3},\mu)
  \label{eq:ccond}
\end{equation}
holds true.
The cases (\ref{eq:l1cond}) and (\ref{eq:l2cond}) are essentially equivalent
and imply 
, in case (\ref{eq:l1cond}) [or in case (\ref{eq:l2cond})],
 that the
trajectories $\gamma_2$ and $\gamma_3$ (or $\gamma_2$ and $\gamma_1$) are
identical and that $\gamma_0$ represents the direct continuation of the
trajectory $\gamma_1$ (or $\gamma_3$) from $\mathbf{r}'$ to $\mathbf{r}$.
This latter condition determines the stationary points of $\mathbf{r}'$, 
which have to lie along the trajectories from $\mathbf{z}_{{i}}^{\nu_1}$ 
(or $\mathbf{z}_{{i}}^{\nu_3}$) to $\mathbf{r}$.

Case (\ref{eq:ccond}) is more involved.
It implies, on the one hand, that the time-reversed counterpart of trajectory
$\gamma_3$ {represent} the direct continuation of trajectory 
$\gamma_1$ (using the fact that the scattering system under consideration is, 
in the absence of the magnetic field, invariant with respect to time reversal),
which determines the stationary points of $\mathbf{r}'$ along reflected
trajectories from $\mathbf{z}_{{i}}^{\nu_1}$ to $\mathbf{z}_{{i}}^{\nu_3}$.
On the other hand, $\gamma_0$ represents a part of the time-reversed 
counterpart of trajectory $\gamma_2$, which necessarily implies that
the point of observation $\mathbf{r}$ has to lie along $\gamma_2$.
This latter condition generally represents an additional restriction of 
the set of stationary points in Eq.~(\ref{eq:scst1sc}) (namely that 
$\mathbf{r}'$ lie on the continuation of a trajectory from 
$\mathbf{z}_{{i}}^{\nu_2}$ to $\mathbf{r}$), which substantially reduces the 
weight of contributions resulting from case (\ref{eq:ccond}) as compared 
to those emanating from cases (\ref{eq:l1cond}) and (\ref{eq:l2cond}).
An exception {to} this rule arises if the point of observation $\mathbf{r}$ 
is identical with or lies rather close to $\mathbf{z}_{{i}}^{\nu_2}$, in which
case all contributions resulting from Eqs.~(\ref{eq:l1cond})--(\ref{eq:ccond})
are of comparable order.

In full generality, we can express the first-order correction to the
linear scattering wavefunction in the semiclassical regime as
\begin{equation}
  \delta \psi^{(1)}(\mathbf{r}) = 2 \delta \psi_\ell^{(1)}(\mathbf{r})
  + \delta \psi_c^{(1)}(\mathbf{r}) \label{eq:dpsi1}
\end{equation}
where $\delta \psi_\ell^{(1)}(\mathbf{r})$ and 
$\delta \psi_c^{(1)}(\mathbf{r})$ contain the contributions that
respectively emanate from the cases (\ref{eq:l1cond}), (\ref{eq:l2cond})
as well as from the case (\ref{eq:ccond}).
Considering an observation point $\mathbf{r}$ that lies deep inside the
billiard, we neglect $\psi_c^{(1)}(\mathbf{r})$ for the moment.
The expression for $\delta \psi_\ell^{(1)}(\mathbf{r})$ can be cast in a 
form that is, apart from a source-dependent prefactor, exactly equivalent
to the first-order term 
%PS (\ref{eq:G1born1}) 
in the Born series of a 
perturbed Green function, where the effective perturbation Hamiltonian
$\delta H$ corresponds here to the density 
$|\psi^{(0)}(\mathbf{r})|_{\rm d}^2$ of the noninteracting scattering 
wavefunction as evaluated by the diagonal approximation, i.e.\ to
\begin{equation}
  |\psi^{(0)}(\mathbf{r})|_{\rm d}^2 = |S_0|^2 \frac{\pi \hbar}{W}
  \left[ \sum_\gamma 
    \left| \Anfang{A}_{\gamma}(\mathbf{r},\mathbf{z}_{{i}}^+,\mu) \right|^2 
    + \sum_\gamma 
    \left| \Anfang{A}_{\gamma}(\mathbf{r},\mathbf{z}_{{i}}^-,\mu) \right|^2 
    \right] \, . \label{eq:psi2diag}
\end{equation}
In close analogy with 
{the first-order modification (\ref{eq:dG1sp}) of the semiclassical 
Green function in the presence of a weak perturbation, we then} obtain
\begin{eqnarray}
  \delta \psi_\ell^{(1)}(\mathbf{r}) & = &
  \frac{S_0}{i} \sqrt{\frac{\pi \hbar}{W}} \sum_{\nu = \pm 1} {\nu}
  \sum_{\gamma} \left( - \frac{i}{\hbar} \right) \frac{\hbar^2}{2m}
  \int_0^{{T_\gamma}} |\psi^{(0)}[\mathbf{q}_\gamma(t)]|_{\rm d}^2 \, dt
  \nonumber \\
  && \times \Anfang{A}_{\gamma}(\mathbf{r},\mathbf{z}_{{i}}^{\nu},\mu) 
  \exp\left\{\frac{i}{\hbar}\left[ 
    \Anfang{S}_{\gamma}(\mathbf{r},\mathbf{z}_{{i}}^{\nu},\mu) -
    \Anfang{\phi}_{\gamma}(\mathbf{r},\mathbf{z}_{{i}}^{\nu},\mu) \right] 
  - i {\frac{\pi }{2}\Anfang{\mu}_{\gamma}} \right\} \,.
  \label{eq:dpsi1l}
\end{eqnarray}
$\delta \psi_\ell^{(1)}$ and $\delta \psi_c^{(1)}$ shall, in the following,
 be termed ``ladder'' and ``crossed'' contributions, respectively.

To illustrate this point, it is useful to introduce a diagrammatic 
representation for this nonlinear scattering problem.
Following Ref.~\cite{Wel09AP}, we represent by
\includegraphics[height=2ex]{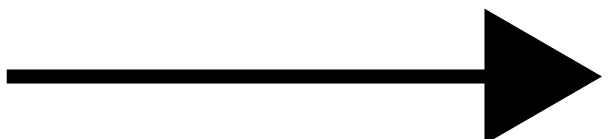}
and
\includegraphics[height=2ex]{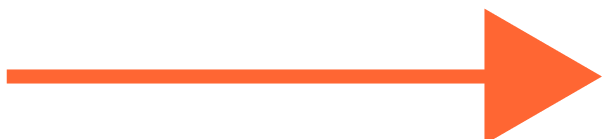}
the Green function $G(\mathbf{r},\mathbf{r}',\mu)$ and its complex conjugate
$G^*(\mathbf{r},\mathbf{r}',\mu)$, respectively.
The (four-legged) vertex 
\includegraphics[height=2ex]{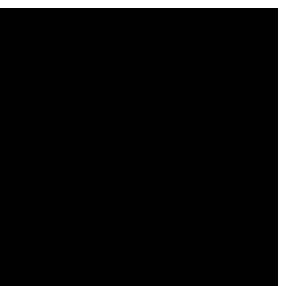}
represents a scattering event of $\psi$ at its own density modulations,
described by the second term of the right-hand side of 
Eq.~(\ref{eq:scstate_g}), and 
\includegraphics[height=2ex]{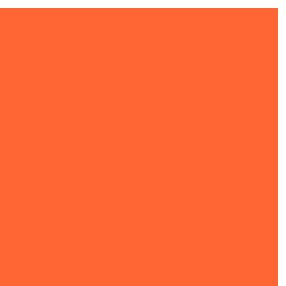}
denotes the corresponding vertex for $\psi^*$, appearing in the complex
conjugate counterpart of Eq.~(\ref{eq:scstate_g}).
The source is depicted by the vertical bar 
\includegraphics[height=2ex]{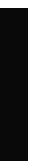},
i.e.~\includegraphics[height=2ex]{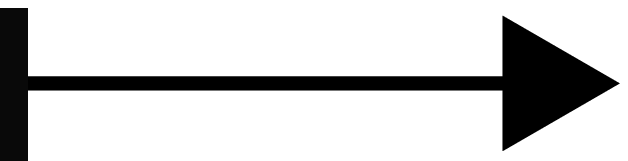}
represents the scattering wavefunction of the noninteracting system, 
given by the convolution of the Green function with the source.
We can then express Eq.~(\ref{eq:scstate_g}) and its complex conjugate as
\begin{eqnarray}
 \centerpic{0.1}{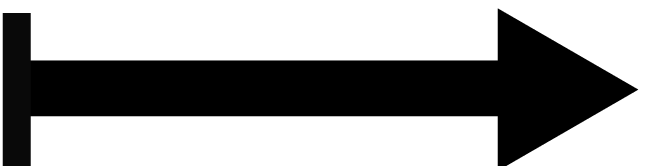}
 &=&
 \centerpic{0.1}{scat_wave_0.eps} +
 \centerpic{0.16}{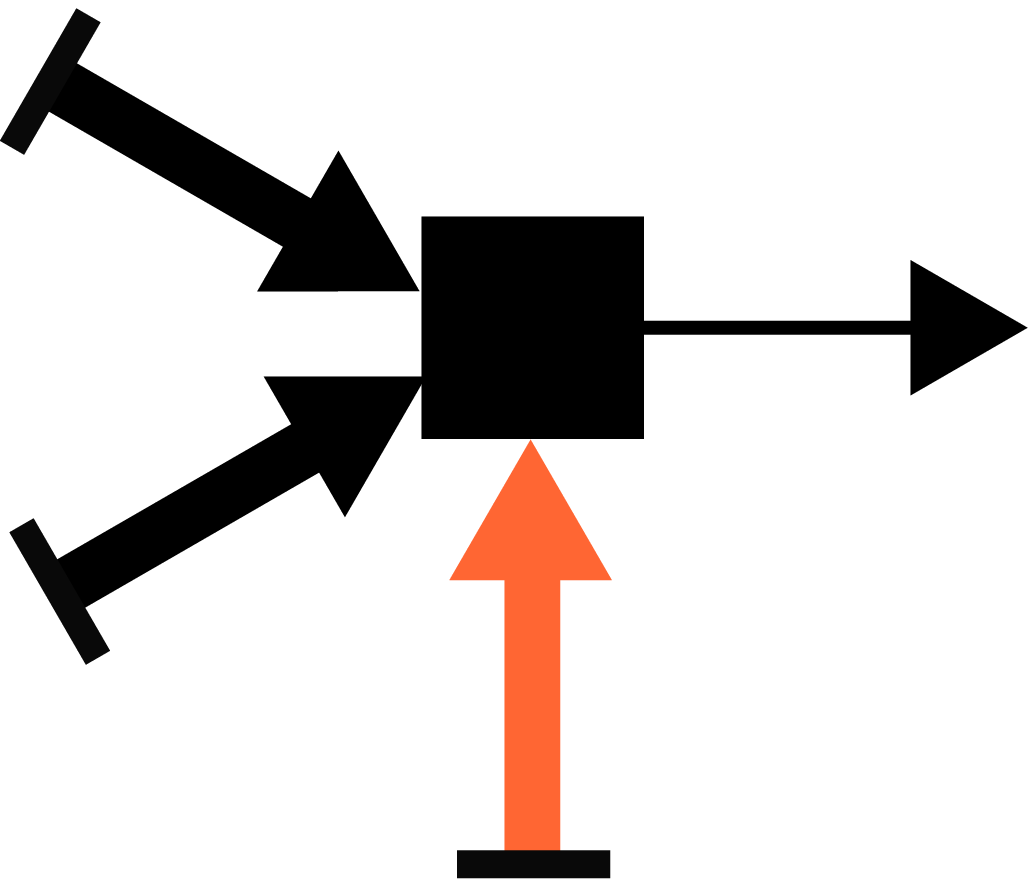}\,,\\
\centerpic{0.1}{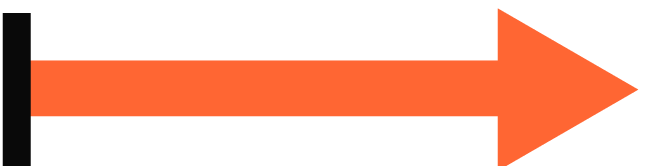}
 &=&
 \centerpic{0.1}{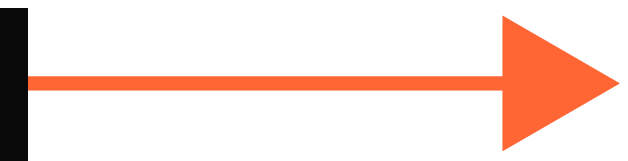} +
 \centerpic{0.16}{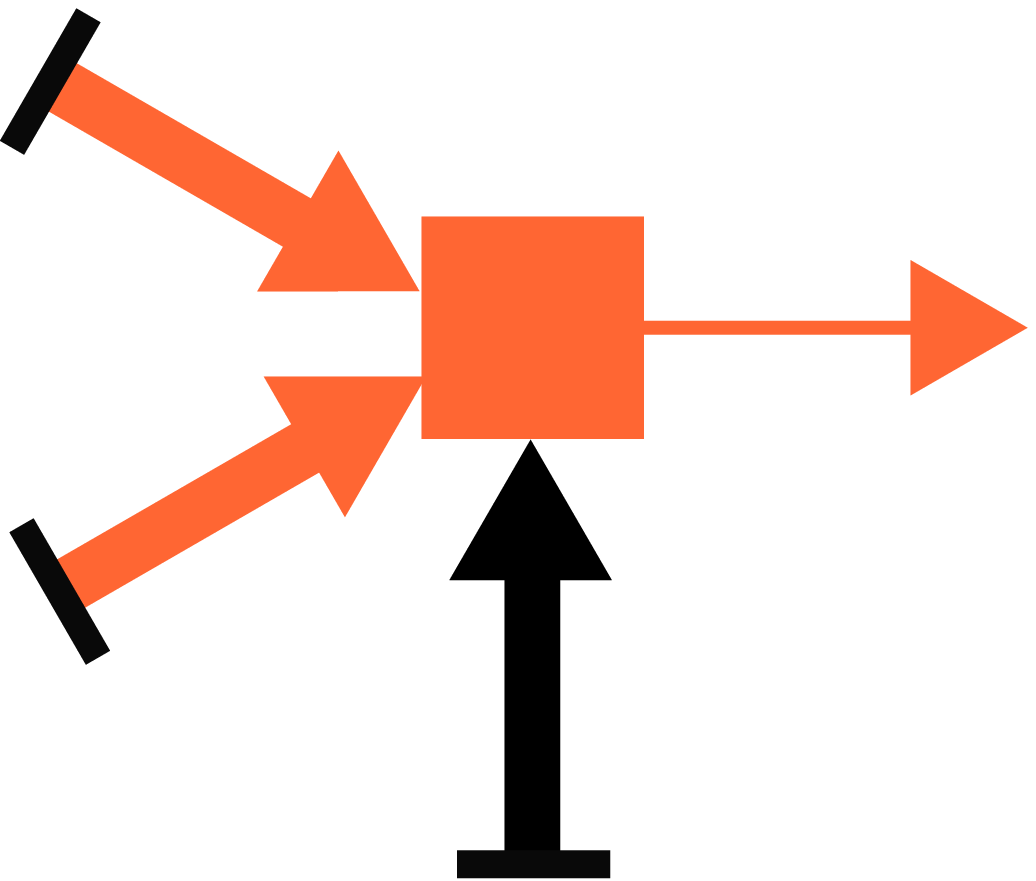}\,,
\end{eqnarray}
where
\includegraphics[height=2ex]{scat_wave.eps}
and
\includegraphics[height=2ex]{scat_wave_conj.eps}
respectively represent the self-consistent stationary scattering wavefunction 
$\psi(\mathbf{r})$ of the nonlinear system and its complex conjugate
$\psi^*(\mathbf{r})$.
Going up to the second order in the power-series expansion (\ref{eq:scstg}),
we obtain the diagrammatic representation
\begin{eqnarray}
\centerpic{0.1}{scat_wave.eps}
&=& \centerpic{0.1}{scat_wave_0.eps}
+\centerpic{0.16}{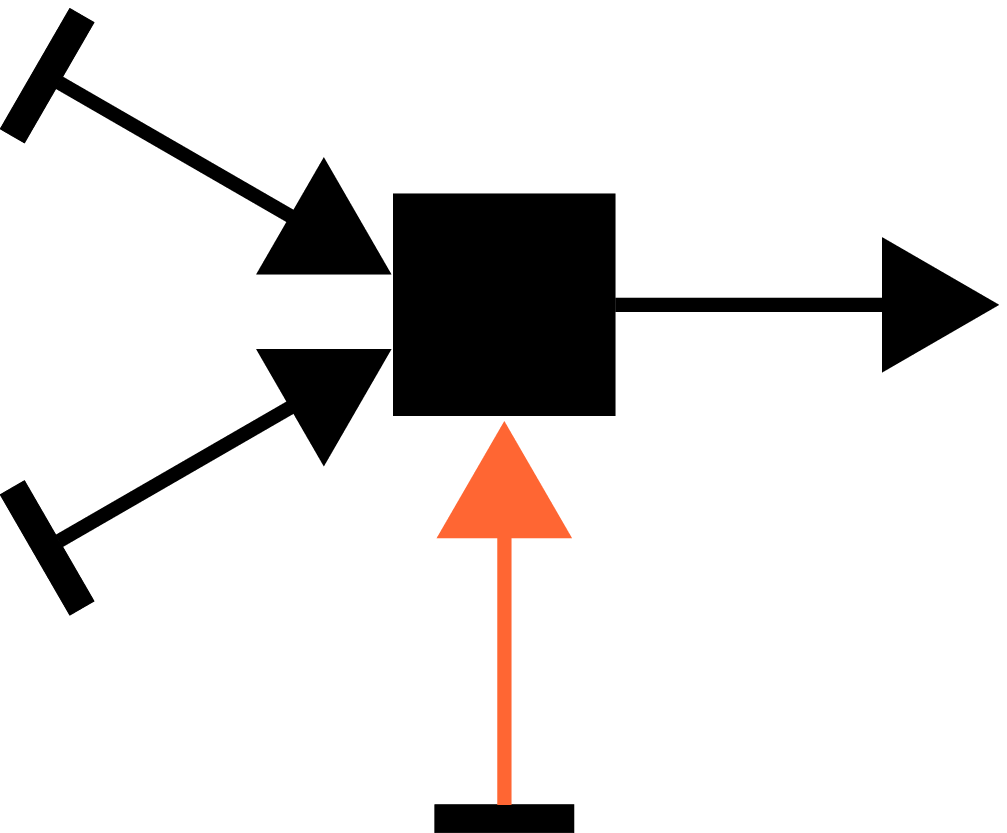}
+\centerpic{0.26}{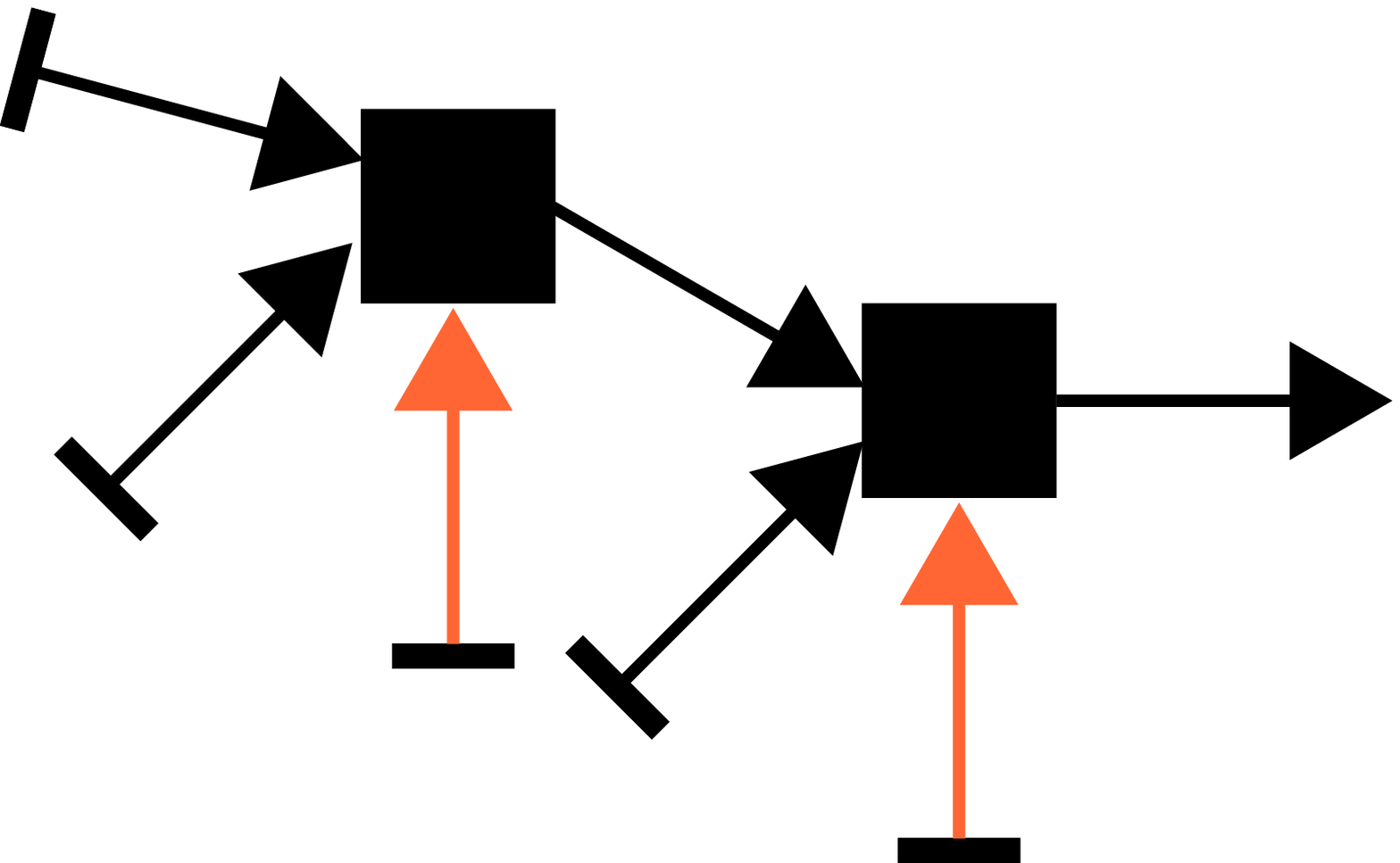}\nonumber\\
&&+\centerpic{0.26}{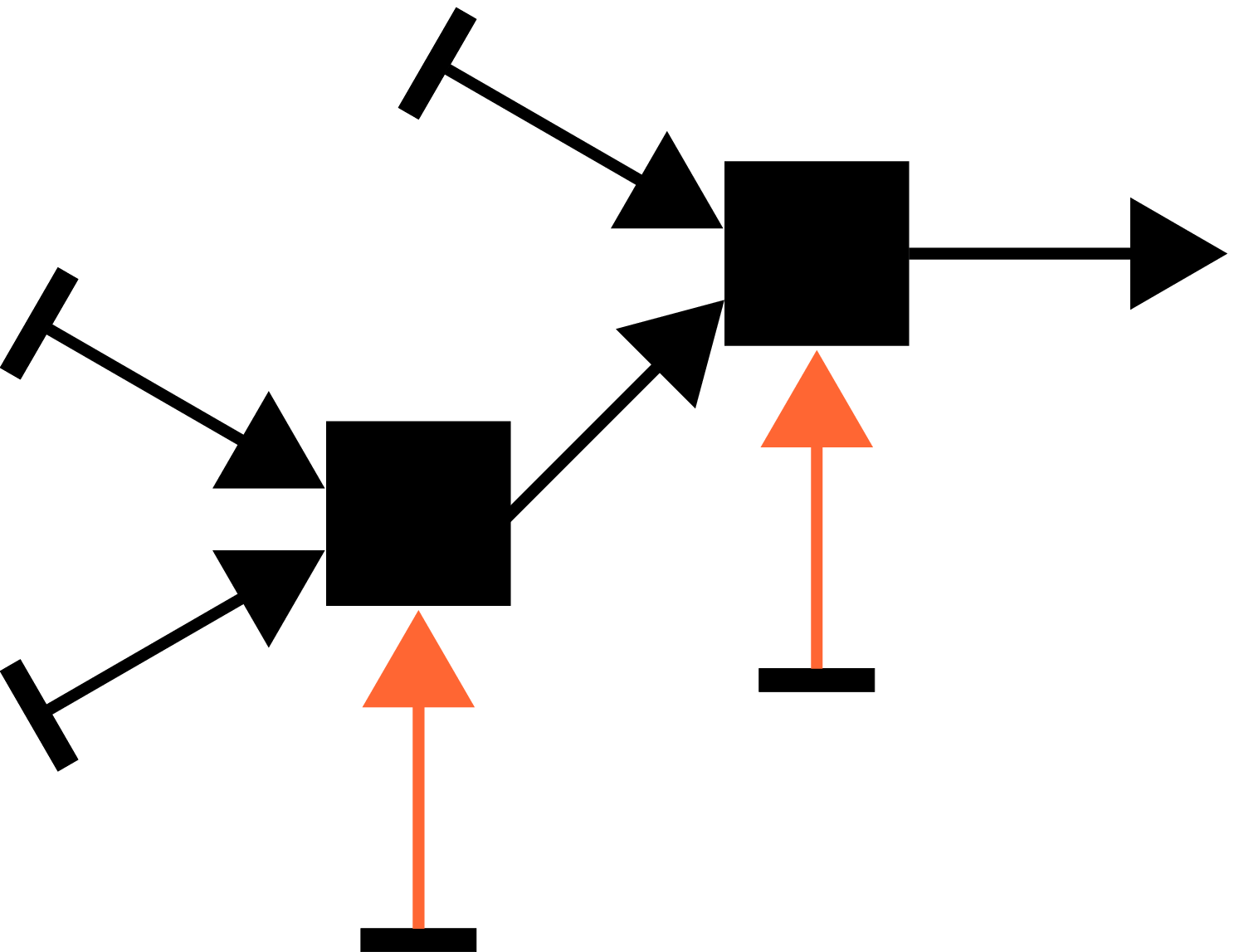}
+\centerpic{0.215}{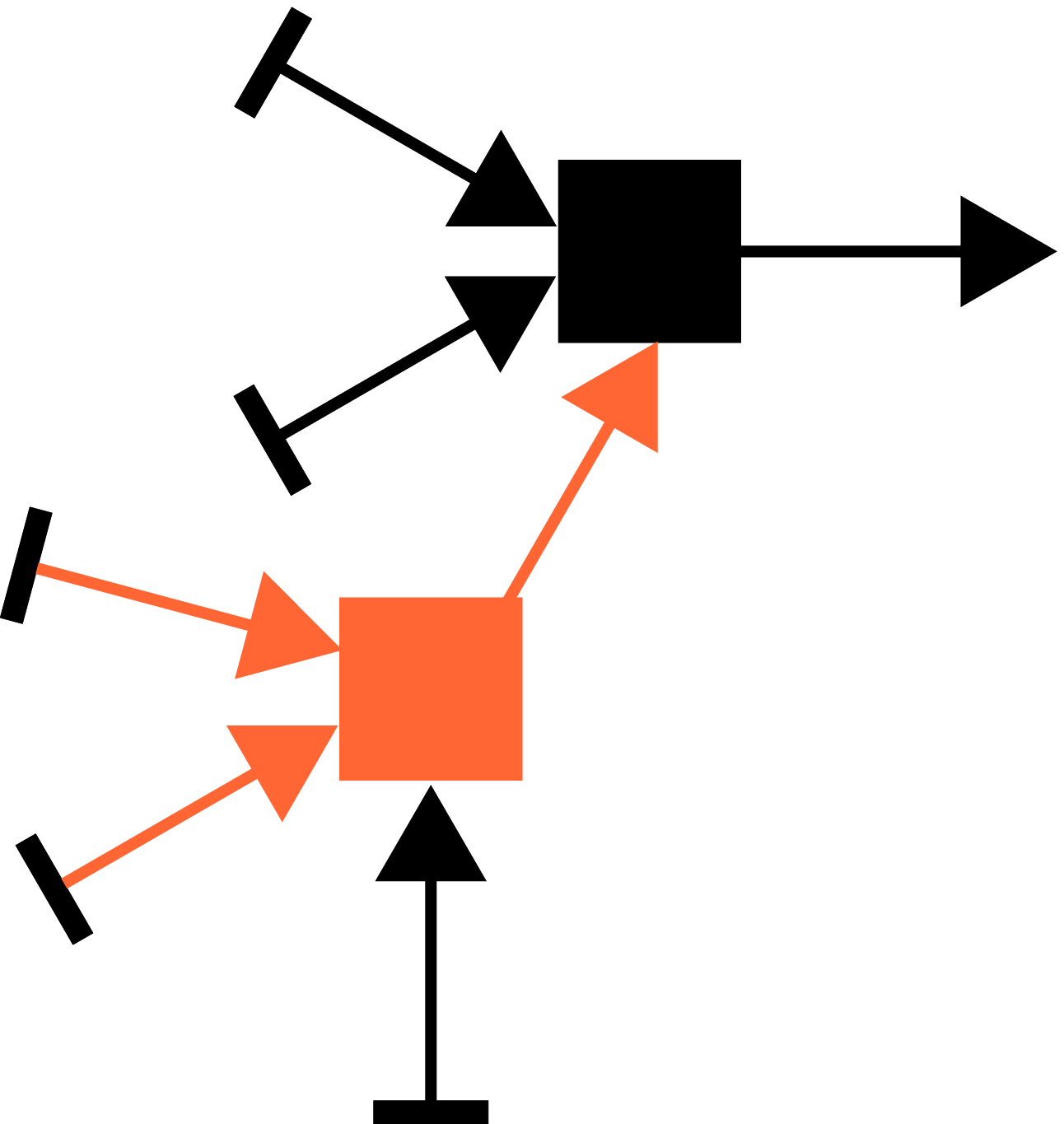}
+\mathcal{O}(g^3)\,.
\end{eqnarray}

The semiclassical evaluation of the first-order term according to 
Eqs.~(\ref{eq:dpsi1}) and (\ref{eq:dpsi1l}), neglecting the contribution
of $\delta \psi_c^{(1)}$, can be expressed as
\begin{equation}
 \centerpic{0.16}{scat_wave_1.eps}\simeq
 2 \; \downpic{0.16}{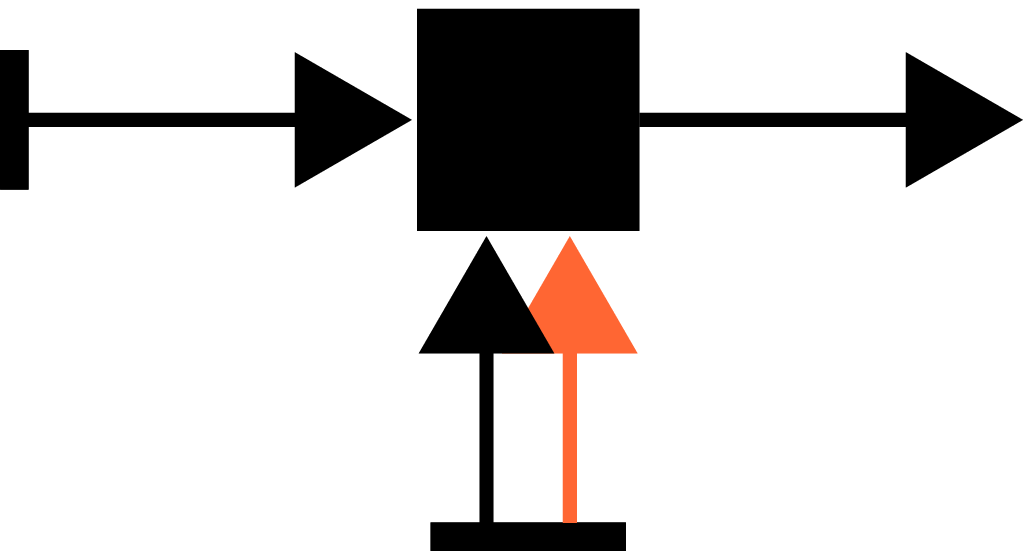} \label{eq:diag_l}
\end{equation}
in diagrammatic terms.
In close analogy with the corresponding ladder diagrams in
disordered systems \cite{WelGre08PRL,WelGre09PRA,Wel09AP}, the parallel arrows
\includegraphics[height=2ex]{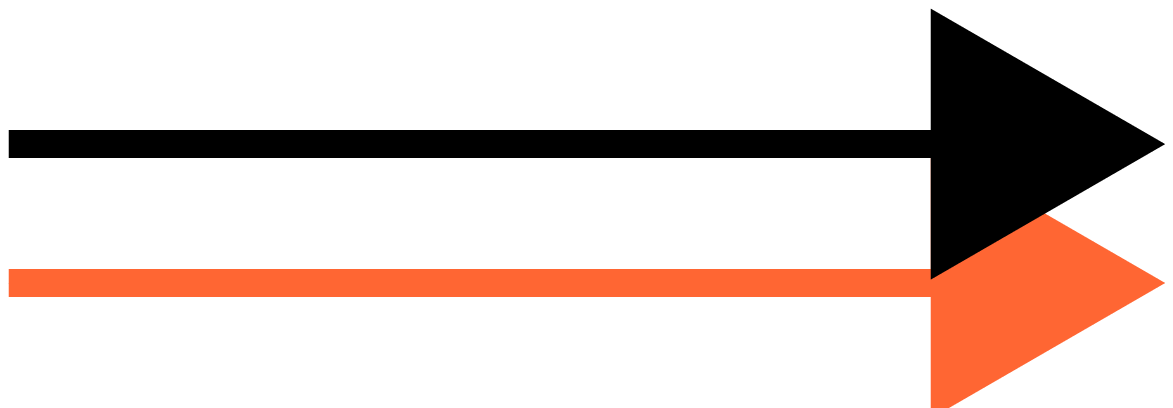}
symbolize the semiclassical evaluation of $G^* G$ in the diagonal
approximation, with $G$ and $G^*$ following the same trajectories that
connect a given initial with a given final point.
The diagram
\includegraphics[height=2ex]{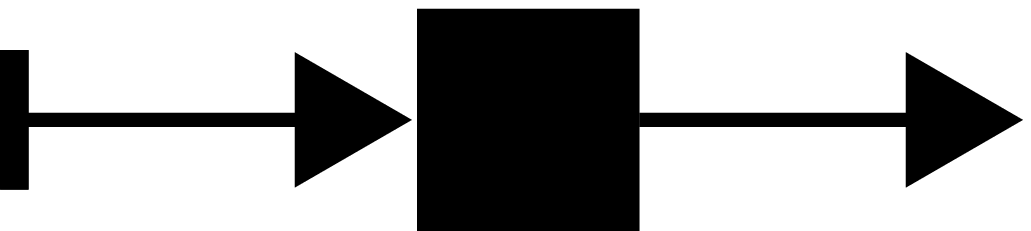}, 
on the other hand, indicates that the nonlinearity event takes place
along a continuous trajectory that connects the source with a given 
final point at the end of the arrow.
{As already discussed above, the factor 2 in Eqs.~(\ref{eq:dpsi1}) 
and (\ref{eq:diag_l}) originates from the two equivalent conditions 
(\ref{eq:l1cond}) and (\ref{eq:l2cond}). 
In other words, the red arrow on the left-hand side of Eq.~(\ref{eq:diag_l}) 
can be paired with either one of the two incoming black arrows.}

\subsection{Ladder contributions}

\label{sec:ladder}

It is suggestive to pursue the analogy with the Born series of a linear
Green function and to introduce a modified Green function $G_\ell$
(the $\ell$ stands for ``ladder contributions''), symbolized by
\includegraphics[height=2ex]{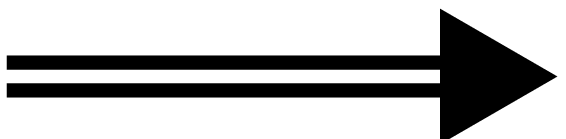},
in which the contribution of the density-induced perturbation is summed up
to all orders in the nonlinearity $g$.
The Dyson equation that this Green function satisfies is represented as
\begin{eqnarray}
\centerpic{0.1}{ladder_GF.eps}
&=&
\centerpic{0.1}{lin_GF.eps}
+2\;\downpic{0.16}{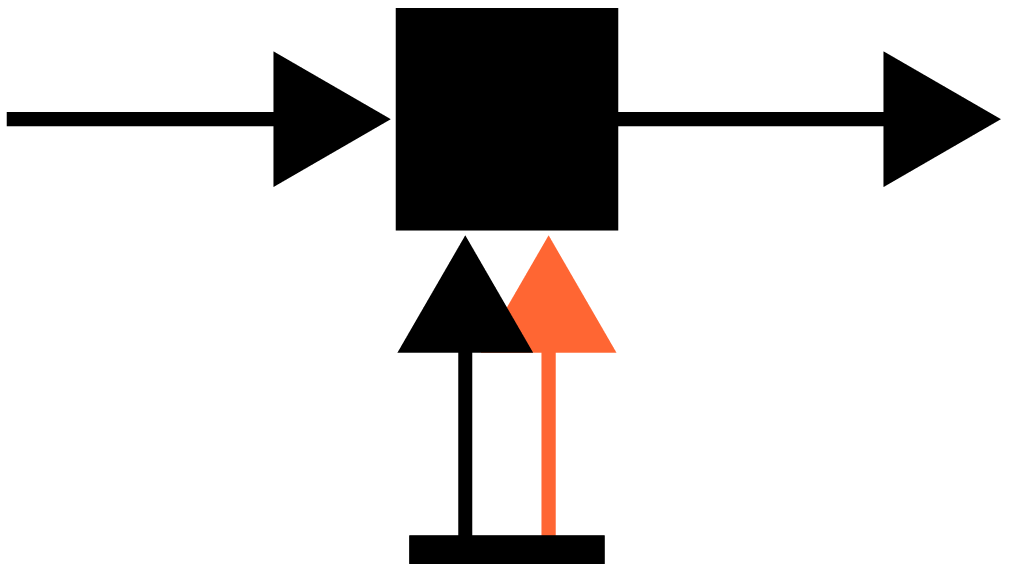}
+4\;\downpic{0.26}{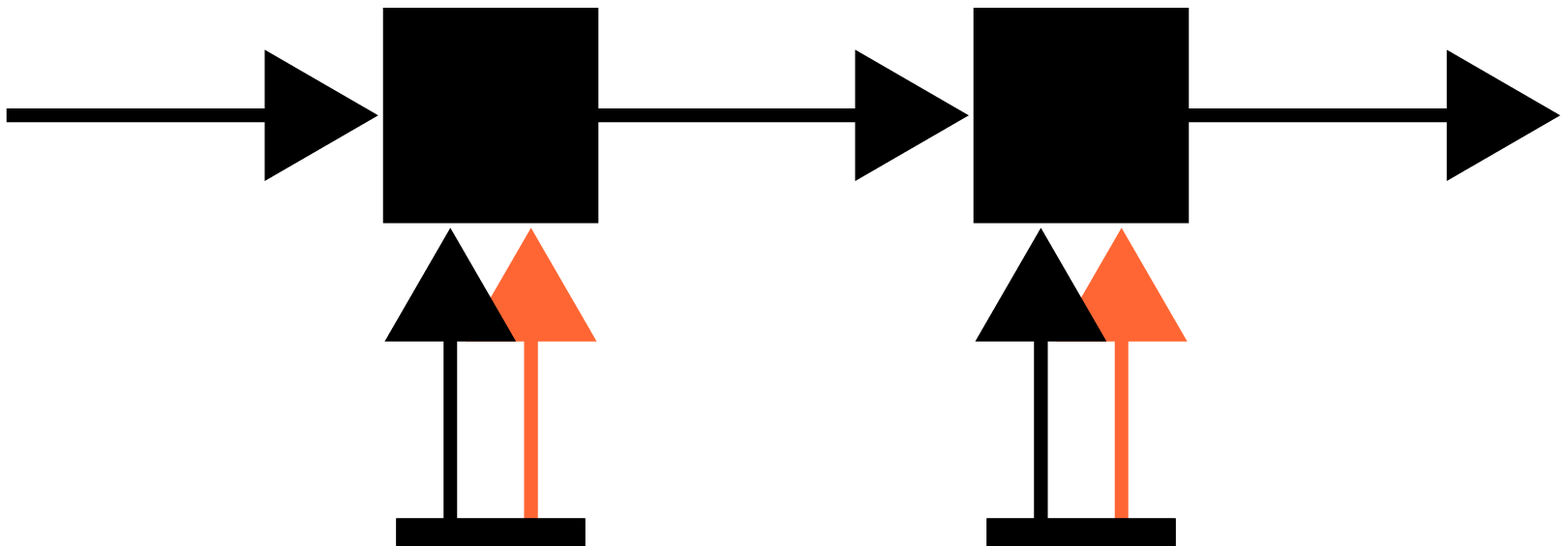}
+\ldots\nonumber\\
&=&
\centerpic{0.1}{lin_GF.eps}
\label{eq:dysonnl}
+2\;\downpic{0.16}{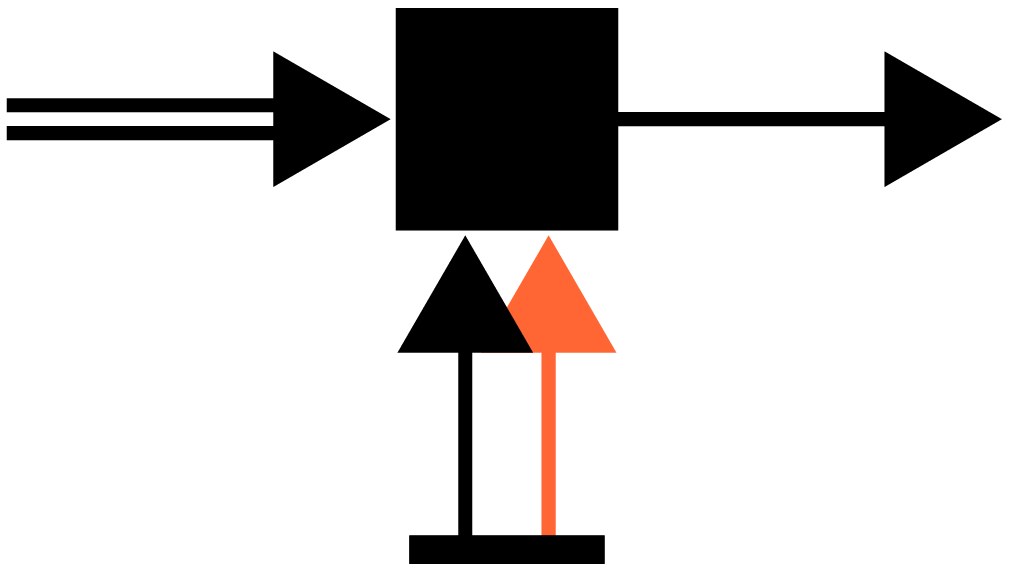}\,.
\end{eqnarray}
Applying the stationary phase approximation, the explicit expression for this
modified Green function reads, in analogy with Eq.~(\ref{eq:G1scApp}),
\begin{equation}
  G_\ell(\mathbf{r},\mathbf{r}',\mu) = \sum_\gamma 
  A_\gamma(\mathbf{r},\mathbf{r}',\mu) 
  \exp\left\{\frac{i}{\hbar} \left[ S_\gamma(\mathbf{r},\mathbf{r}',\mu)
    - \phi_\gamma(\mathbf{r},\mathbf{r}',\mu)
    - \chi_\gamma(\mathbf{r},\mathbf{r}',\mu) \right]{-i\frac{\pi}{2}\mu_\gamma}\right\}
  \label{eq:G2}
\end{equation}
with
$\chi_\gamma(\mathbf{r},\mathbf{r}',\mu) \equiv 2 g {(\hbar^2/2m)} 
\int_0^{{T_\gamma}} |\psi^{(0)}[\mathbf{q}_\gamma(t)]|_{\rm d}^2 \, dt$.
On this level, the nonlinearity therefore induces an effective modification 
of the action integral along the trajectory $\gamma$, in close analogy with
the {change in action for the dynamics in the presence of a weak static
disorder potential \cite{RicUllJal96PRB}}.
This modification, however, does not at all affect the calculation of mean
densities within the billiard using the diagonal approximation:
evaluating the wavefunction $\psi(\mathbf{r})$ according to 
Eq.~(\ref{scstsc}) with $\Anfang{G}$ being replaced by $\Anfang{G}_\ell$,
we would essentially obtain 
$|\psi(\mathbf{r})|_{\rm d}^2 = |\psi^{(0)}(\mathbf{r})|_{\rm d}^2$,
the latter being given by Eq.~(\ref{eq:psi2diag}) {where the phases 
$\chi_\gamma$ appearing in Eq.~(\ref{eq:G2}) drop out}.

The same reasoning applies if we replace $\psi^{(0)}$ by $\psi$ in the
definition of the nonlinearity-induced modification of the effective action
associated with the trajectory $\gamma$, i.e., we (re-)de\-fine
\begin{equation}
  \chi_\gamma(\mathbf{r},\mathbf{r}',\mu) \equiv g \frac{\hbar^2}{m}
  \int_0^{{T_\gamma}} |\psi[\mathbf{q}_\gamma(t)]|_{\rm d}^2 \, dt \label{eq:chi}
\end{equation}
and use this expression in the definition of $G_\ell$ according to 
Eq.~(\ref{eq:G2}).
This amounts to replacing the diagrammatic representation (\ref{eq:dysonnl})
by
\begin{equation}
 \centerpic{0.1}{ladder_GF.eps}
=\centerpic{0.1}{lin_GF.eps}
+2\;\downpic{0.16}{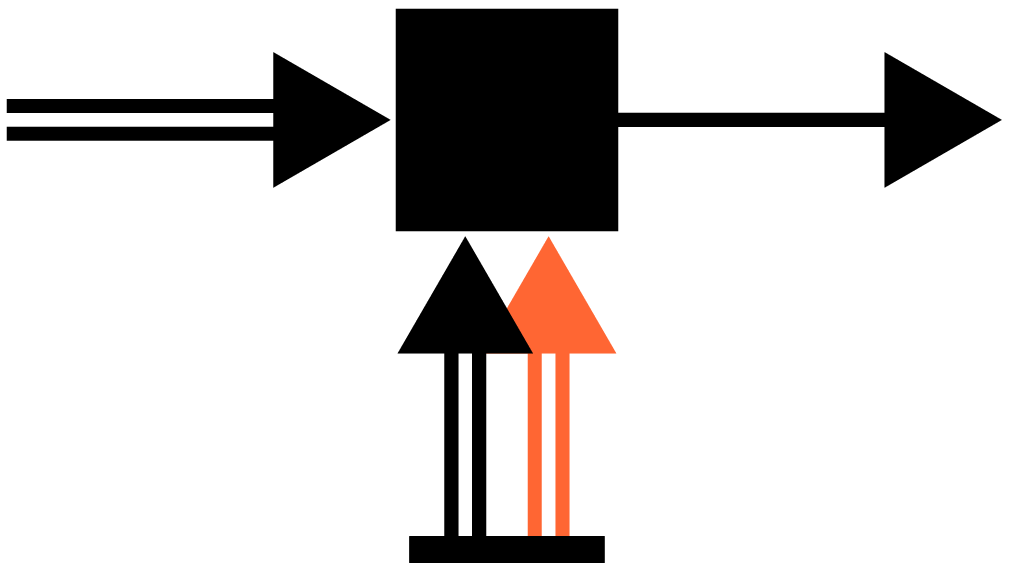}
\label{eq:modifiedG}
\end{equation}
which, when being expanded in powers of $g$ and evaluated using the
stationary phase approximation, involves all possible 
ladder-type (parallel) pairings of $G$ and $G^*$, i.e.,
\begin{eqnarray}
\centerpic{0.1}{ladder_GF.eps}
&=&
\centerpic{0.1}{lin_GF.eps}
+2\;\downpic{0.16}{ladder_GF_1.eps}
+4\;\downpic{0.26}{ladder_GF_2a.eps}
\nonumber\\
&&+4\;\downpic{0.17}{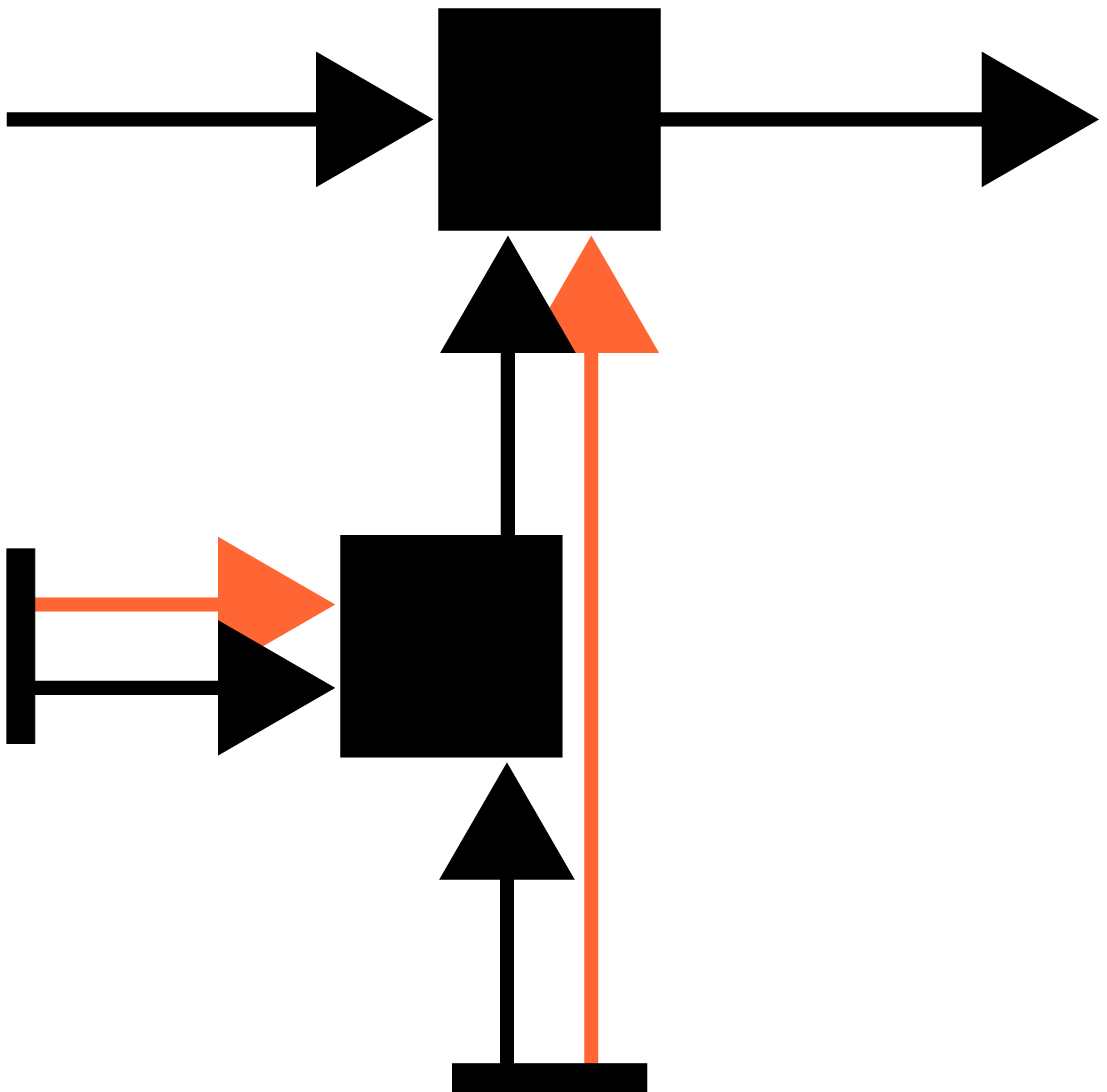}
+4\;\downpic{0.17}{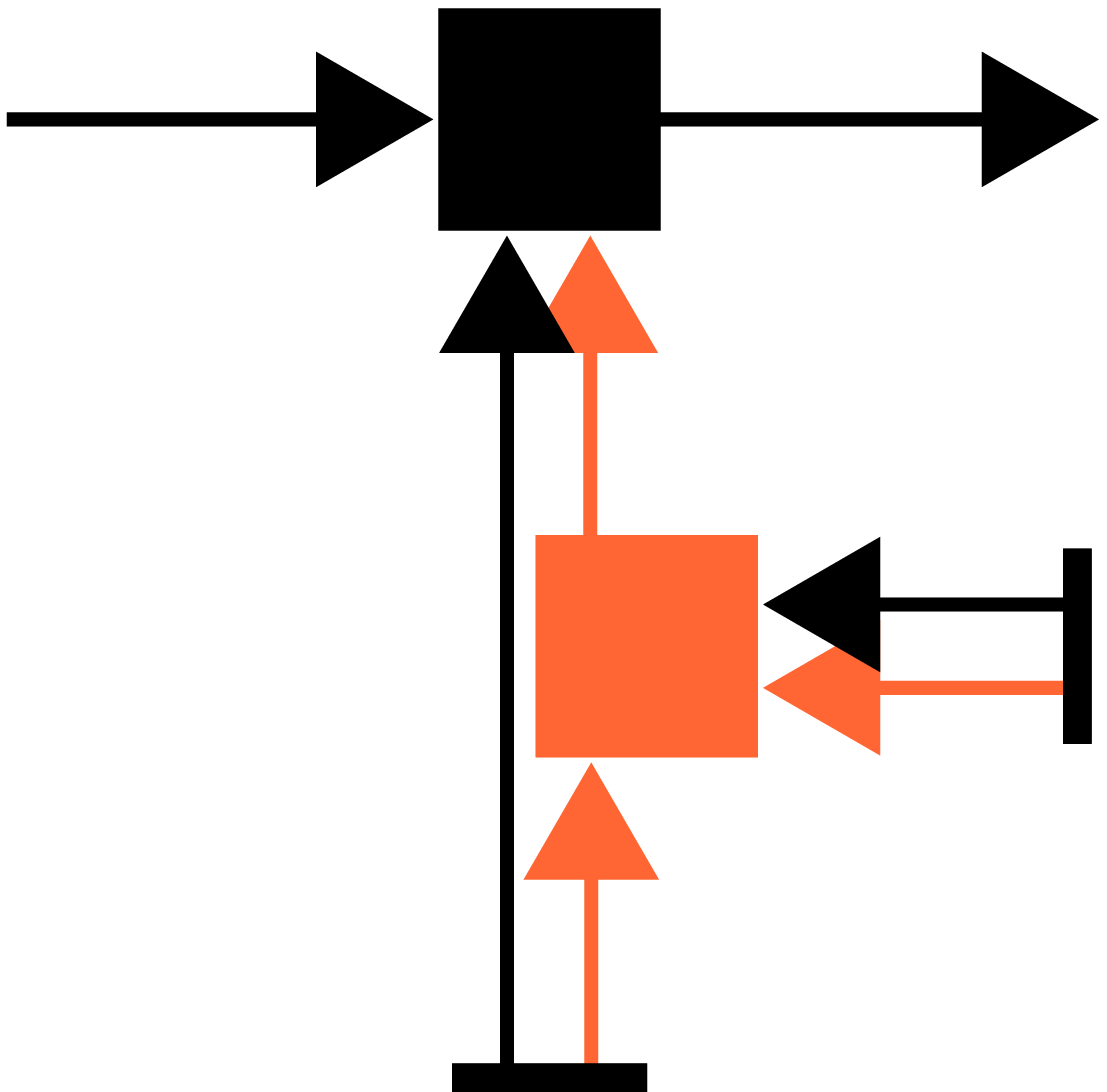}
+ \mathcal{O}(g^3) \label{eq:Gl}
\end{eqnarray}
up to second order in $g$.
The mean density within the billiard as evaluated using the diagonal 
approximation is then given by
\begin{equation}
  |\psi(\mathbf{r})|_{\rm d}^2 = |S_0|^2 \frac{\pi \hbar}{W}
  \left[ \sum_\gamma 
    \left| \Anfang{A}_{\gamma}(\mathbf{r},\mathbf{z}_{{i}}^+,\mu) \right|^2 
    + \sum_\gamma 
    \left| \Anfang{A}_{\gamma}(\mathbf{r},\mathbf{z}_{{i}}^-,\mu) \right|^2 
    \right] \label{eq:psi2diagnl}
\end{equation}
as in the case of the linear scattering problem [see Eq.~(\ref{eq:psi2diag})]
\cite{rem_complexg}.

It is worthwhile to calculate the energy average of the density within 
the billiard using the Hannay-Ozorio de Almeida sum rule \cite{Sie99JPA}.
As shown in \ref{sec:sumrules}, we have [see Eq.~(\ref{eq:Grz})]
\begin{eqnarray}
  \sum_\gamma \left\langle
  \left|\Anfang{A}_\gamma(\mathbf{r},\mathbf{z}',\mu)\right|^2\right\rangle
  & = & \frac{m^2 W}{2 \pi \hbar^4} \frac{\tau_D}{\tau_H} 
  \frac{1}{\sqrt{2 m \mu - p_y'^2}}\,.
  \label{eq:coordmixsr}
\end{eqnarray}

This eventually yields
\begin{equation}
  \left\langle|\psi(\mathbf{r})|^2\right\rangle_{\rm d} = 
  \left| \frac{m S_0}{\hbar} \right|^2 \frac{\tau_D}{\tau_H} 
  \frac{1}{\hbar p_i^{\mathrm{l}}(\mu)} =
  \frac{m j^{\mathrm{i}}}{\hbar} \frac{\tau_D}{\tau_H} =
  \frac{\tau_D}{\Omega} j^{\mathrm{i}} \label{eq:psi2dnl}
\end{equation}
when being expressed in terms of the incident current
$j^{\mathrm{i}} = m |S_0|^2/[\hbar^2 p_i^{\mathrm{l}}(\mu)]${.
The mean density is therefore obtained from an equidistribution of the
population in the case of a stationary flow, which is given by the ratio
of the feeding rate $j^{\mathrm{i}}$ and the decay rate $\tau_D^{-1}$.}

\subsection{Crossed contributions}
\label{sec:crossed}

{As seen above, the nonlinear ladder contributions vanish on average. 
However, we have so far} neglected the influence of terms arising from the 
association of trajectories according to the remaining (and less intuitive)
case (\ref{eq:ccond}).
As was argued above, the contributions of such terms to the local density is 
generally suppressed with respect to the ladder-type contributions arising
from the cases (\ref{eq:l1cond}) and (\ref{eq:l2cond}), due to the fact that
case (\ref{eq:ccond}) requires not only the time-reversed counterpart of 
trajectory $\gamma_3$ to represent the  direct continuation of trajectory 
$\gamma_1$, but also that the point of observation $\mathbf{r}$ {lie} on 
the trajectory $\gamma_2$ connecting the source with the interaction point
$\mathbf{r}'$.
In the case of retro-reflection into the incident channel, however, where
$\mathbf{r}$ lies directly at the location of the source, this latter 
condition is {satisfied} by default, and we should therefore expect a finite
contribution from this ``crossed'' association of trajectories to the
probability of coherent backscattering.

It is instructive to first compute the influence of such crossed terms in
linear order in the nonlinearity.
We evaluate for this purpose the remaining term 
$\delta \psi_c^{(1)}(\mathbf{r})$ in Eq.~(\ref{eq:dpsi1}) that is associated 
with the case (\ref{eq:ccond}).
The requirement that the time-reversed counterpart of $\gamma_3$ {represent}
the direct continuation of $\gamma_1$ allows one to apply the stationary phase
approximation in order to evaluate the spatial integral in 
Eq.~(\ref{eq:scst1sc}).
In close analogy with {Eq.~(\ref{eq:dpsi1l})}, we then
obtain a single sum over all trajectories $\gamma$ that connect the initial 
phase-space point $\mathbf{z}_{{i}}^{\nu_1}$ with the final point 
$\mathbf{z}_{{i}}^{\nu_3}$ ($\nu_1,\nu_3=\pm 1$) both being associated with the 
incident channel $\chi_i(y)$.
An important extension as compared to the  structure of 
Eq.~(\ref{eq:dpsi1l}) is provided by the paramagnetic 
contribution (\ref{eq:phipara}) to the effective action integral, which
changes its sign under the time-reversal of the trajectory $\gamma_3$.

Calculating the overlap of $\delta \psi_c^{(1)}(\mathbf{r})$ with the incident
channel, we obtain the associated first-order correction to the backscattering
amplitude as
\begin{eqnarray}
  \delta \psi_{{i}}^{(c)} & \equiv & {\int_0^W} \chi_{{i}}^*(y) 
  \delta \psi_c^{(1)}(x_L,y) dy \\
  & = & S_0 \frac{\pi \hbar}{W} \sum_{\nu_1,\nu_3 = \pm 1} 
  ( - \nu_1\nu_3 ) \sum_\gamma
  \AnfangEnde{A}_\gamma(\mathbf{z}_{{i}}^{\nu_3},\mathbf{z}_{{i}}^{\nu_1},\mu) 
  \nonumber \\ & &
  \times \exp\left\{\frac{i}{\hbar}\left[
    \AnfangEnde{S}_\gamma(\mathbf{z}_{{i}}^{\nu_3},\mathbf{z}_{{i}}^{\nu_1},\mu) -
    \AnfangEnde{\phi}_\gamma(\mathbf{z}_{{i}}^{\nu_3},\mathbf{z}_{{i}}^{\nu_1},\mu) 
    \right] - {i \frac{\pi }{2}\AnfangEnde{\mu}_{\gamma}} \right\}
  \nonumber \\ & &
  \times \left( - \frac{i}{\hbar} \right) \frac{\hbar^2}{2m} \int_0^{{T_\gamma}}
  C^{(0)}[\mathbf{q}_\gamma(t)] \exp\left\{-\frac{2i}{\hbar} 
  \Ende{\varphi}_\gamma[\mathbf{z}_{{i}}^{\nu_3},\mathbf{q}_\gamma(t), \mu] 
  \right\}{dt} \label{eq:dpsi1c}
\end{eqnarray}
where we define 
\begin{eqnarray}
  C^{(0)}(\mathbf{r}) & = & |S_0|^2 \frac{\pi \hbar}{W}
  \left\{ \sum_{\gamma_2} 
    \left| \Anfang{A}_{\gamma_2}(\mathbf{r},\mathbf{z}_{{i}}^+,\mu) \right|^2
    \exp\left[-\frac{2i}{\hbar} 
    \Anfang{\varphi}_{\gamma_2}(\mathbf{r},\mathbf{z}_{{i}}^+, \mu) \right]
    \right. \nonumber \\ &&  \left.  + \sum_{\gamma_2} 
    \left| \Anfang{A}_{\gamma_2}(\mathbf{r},\mathbf{z}_{{i}}^-,\mu) \right|^2 
    \exp\left[-\frac{2i}{\hbar} 
    \Anfang{\varphi}_{\gamma_2}(\mathbf{r},\mathbf{z}_{{i}}^-, \mu) \right]
    \right\} \label{eq:Crosseddensity}
\end{eqnarray}
as ``crossed density'' within the billiard.
The latter quantity can be interpreted as the semiclassical evaluation of
\begin{equation}
  \mathcal{C}^{(0)}(\mathbf{r}) = |S_0|^2 \frac{\pi \hbar}{W}
  \left[ \Anfang{G}^*(\mathbf{r},\mathbf{z}_{{i}}^+, \mu)
    \Ende{G}(\mathbf{z}_{{i}}^+,\mathbf{r}, \mu)
    + \Anfang{G}^*(\mathbf{r},\mathbf{z}_{{i}}^-, \mu)
    \Ende{G}(\mathbf{z}_{{i}}^-,\mathbf{r}, \mu) \right]
\end{equation}
within the diagonal approximation.
In contrast to the actual density within the billiard, given in leading order
by the expression (\ref{eq:psi2diagnl}), $C^{(0)}(\mathbf{r})$ is, in general, 
not invariant under gauge transformations 
$\mathbf{A} \mapsto \mathbf{A} + \nabla \chi$ of the effective vector 
potential $\mathbf{A}(\mathbf{r})$, due to the presence of the phase factors
containing the paramagnetic contribution to the effective action integral.
The combination of those phase factors with the corresponding one arising in 
Eq.~(\ref{eq:dpsi1c}), however, gives rise to an overall expression that is
invariant under gauge transformations.

To verify this, we introduce for each point $\mathbf{r}$ within the billiard
a straight-line trajectory, denoted by the index $\omega$, that connects this
point to a fixed reference point $\mathbf{r}_L$ within the incident lead, 
given, e.g., by $\mathbf{r}_L\equiv(x_L,W/2)$.
This straight-line trajectory can be defined as
\begin{eqnarray}
  \mathbf{q}_\omega(t) & \equiv & \mathbf{r} + \frac{t}{{T_\omega}} 
  (\mathbf{r}_L - \mathbf{r}) \, , \\
  \mathbf{p}_\omega(t) & \equiv & 
  \frac{\hbar}{{T_\omega}} (\mathbf{r}_L - \mathbf{r}) \, ,
\end{eqnarray}
with ${T_\omega} \equiv m |\mathbf{r}_L - \mathbf{r}| / \sqrt{2 m \mu}$
\cite{rem_omega}.
We now define for each ``incident'' trajectory $\gamma$, i.e.\ which 
connects a phase-space point $\mathbf{z}$ within the incident lead to a
spatial point $\mathbf{r}$ within the billiard, its ``completion'' as 
$\bar{\gamma} \equiv \omega \circ \gamma$.
In physical terms, $\bar{\gamma}$ traces the motion of a particle that 
follows $\gamma$ and is then scattered back to the incident lead due to the 
presence of a local perturbation within the billiard (a point scatterer)
at position $\mathbf{r}$.
We obviously have the relation
\begin{equation}
  \Anfang{\varphi}_{\bar{\gamma}}(\mathbf{r}_L,\mathbf{r},\mathbf{z},\mu)
  = \varphi_{\omega}(\mathbf{r}_L,\mathbf{r},\mu) +
  \Anfang{\varphi}_{\gamma}(\mathbf{r},\mathbf{z},\mu)
\end{equation}
for the paramagnetic action integral along the trajectory $\bar{\gamma}$.
As integrations of $\mathbf{p}(t) \cdot \mathbf{A}[\mathbf{q}(t)]$ along
paths that are entirely contained within the incident lead obviously vanish 
due to the local absence of the vector potential, we can state that 
$\Anfang{\varphi}_{\bar{\gamma}}(\mathbf{r}_L,\mathbf{r},\mathbf{z},\mu)$
is invariant under gauge transformations.
Analogously, a trajectory $\gamma'$ that leads from a spatial point 
$\mathbf{r}$ within the billiard to a phase-space point $\mathbf{z}$ 
within the incident lead is ``completed'' as 
$\bar{\gamma}' \equiv \gamma'  \circ \omega$ with the associated paramagnetic
action integral
\begin{equation}
  \Ende{\varphi}_{\bar{\gamma}'}(\mathbf{z},\mathbf{r},\mathbf{r}_L,\mu)
  = \Ende{\varphi}_{\gamma'}(\mathbf{z},\mathbf{r},\mu) 
  + \varphi_{\omega}(\mathbf{r},\mathbf{r}_L,\mu) \, .
\end{equation}

The crossed density (\ref{eq:Crosseddensity}) can therefore be re-expressed 
in terms of such completed trajectories $\bar{\gamma_2}$ through
\begin{equation}
  C^{(0)}(\mathbf{r}) = c^{(0)}(\mathbf{r}) \exp\left[\frac{2i}{\hbar}
    \varphi_\omega(\mathbf{r}_L,\mathbf{r},\mu) \right]
  \label{eq:Crossedfactorize}
\end{equation}
where its gauge-invariant part is introduced as 
\begin{eqnarray}
  c^{(0)}(\mathbf{r}) & \equiv & |S_0|^2 \frac{\pi \hbar}{W}
  \left\{ \sum_{\gamma_2} 
    \left| 
     \Anfang{A}_{{\gamma}_2}
     (\mathbf{r}_L,\mathbf{r},\mathbf{z}_{{i}}^+,\mu) \right|^2
    \exp\left[-\frac{2i}{\hbar} 
    \Anfang{\varphi}_{\bar{\gamma}_2}
    (\mathbf{r}_L,\mathbf{r},\mathbf{z}_{{i}}^+, \mu) \right]
    \right. \nonumber \\ &&  \left.  + \sum_{\gamma_2} 
    \left| \Anfang{A}_{{\gamma}_2}
    (\mathbf{r}_L,\mathbf{r},\mathbf{z}_{{i}}^-,\mu) \right|^2 
    \exp\left[-\frac{2i}{\hbar} 
    \Anfang{\varphi}_{\bar{\gamma}_2}
    (\mathbf{r}_L,\mathbf{r},\mathbf{z}_{{i}}^-, \mu) \right]
    \right\} \, . \label{eq:crosseddensity}
\end{eqnarray}
As in the case of ``ordinary'' backscattering trajectories, the energy average
of the paramagnetic phase factor of $\bar{\gamma}$ yields, in analogy with
Eq.~(\ref{eq:phasedecay}),
\begin{equation}
  \left\langle\exp\left[-\frac{2 i}{\hbar} \Anfang{\varphi}_{\bar{\gamma}}
    (\mathbf{r}_L,\mathbf{r},\mathbf{z},\mu)\right] \right\rangle
  = {\exp\left(-\frac{T_\gamma}{\tau_B}\right)}
  \label{eq:phaseaverage}
\end{equation}
with $\tau_B$ the characteristic time scale associated with the magnetic field,
defined by Eq.~(\ref{eq:tB}).
We neglect in this expression the contribution of $T_\omega$ to the total
propagation time of $\bar{\gamma}$ {(which is, in fact, canceled 
in the nonlinear diagrams contributing to the backscattering probability to be
discussed below, as the latter involve, by construction, flux integrals along
closed paths)} and assume  
$T_{\bar{\gamma}} \simeq T_\gamma$.
In perfect analogy with the derivation of the energy-averaged density within
the billiard, we then obtain [see Eqs.~(\ref{eq:G2cl}) and (\ref{eq:Grz})]
\begin{equation}
  \left\langle c^{(0)}(\mathbf{r}) \right\rangle =
  \left| \frac{m S_0}{\hbar} \right|^2 
  \left( \frac{\tau_H}{\tau_D} + \frac{\tau_H}{\tau_B} \right)^{-1}
  \frac{1}{\hbar p_i^{\mathrm{l}}(\mu)} =
  \frac{\tau_D}{\tau_H} \frac{1}{1 + \tau_D / \tau_B} \, \frac{m}{\hbar} 
  j^{\mathrm{i}} \equiv \left\langle c^{(0)} \right\rangle
  \label{eq:avcrosseddensity}
\end{equation}
for the energy average of the gauge-invariant part of the crossed density.

This expression can be used in order to evaluate the first-order correction
to the crossed contribution $\langle |\psi_{{i}}|^2 \rangle_{\mathrm{c}}^{(g)}$ 
of the nonlinear backscattering probability according to
\begin{equation}
  \langle |\psi_{{i}}|^2 \rangle_{\mathrm{c}}^{(g)} =
  \langle |\psi_{{i}}|^2 \rangle_{\mathrm{c}}^{(0)} + g \left[
  \left\langle \psi_{{i}}^* \delta \psi_{{i}}^{(c)} \right\rangle +
  \left\langle \left(\delta \psi_{{i}}^{(c)}\right)^* \psi_{{i}} \right\rangle \right] 
  + \mathcal{O}(g^2) \label{eq:psi12cross1}
\end{equation}
where $\langle |\psi_{{i}}|^2 \rangle_{\mathrm{c}}^{(0)}$ represents the
linear crossed contribution as defined in Eq.~(\ref{eq:psi12cross}).
Within the diagonal approximation, we obtain
\begin{eqnarray}
  \left\langle \psi_{{i}}^* \delta \psi_{{i}}^{(c)} \right\rangle & = &
  - \frac{i}{\hbar} \frac{\hbar^2}{2m} 
  \left|S_0 \frac{\pi \hbar}{W}\right|^2
  \sum_{\nu_1,\nu_3 = \pm 1} \sum_\gamma \left\langle \left| 
  \AnfangEnde{A}_\gamma(\mathbf{z}_{{i}}^{\nu_3},\mathbf{z}_{{i}}^{\nu_1},\mu) 
  \right|^2 \right\rangle \int_0^{{T_\gamma}} dt 
  \left\langle c^{(0)}[\mathbf{q}_\gamma(t)] \right\rangle
  \nonumber \\ && \times
  \left\langle \exp\left\{-\frac{2i}{\hbar} \Ende{\varphi}_{\bar{\gamma}}
  [\mathbf{z}_{{i}}^{\nu_3},\mathbf{q}_\gamma(t),\mathbf{r}_L,\mu] \right\} +
  \exp\left\{\frac{2i}{\hbar} \Anfang{\varphi}_{\bar{\gamma}}
  [\mathbf{r}_L,\mathbf{q}_\gamma(t),\mathbf{z}_{{i}}^{\nu_1},\mu]
  \right\} \right\rangle \label{eq:dpsi12cross}
\end{eqnarray}
where, in the second line of Eq.~(\ref{eq:dpsi12cross}), we account for the
fact that each trajectory $\gamma$ can be paired with itself as well as
with its time-reversed counterpart, the latter giving rise to a different
paramagnetic phase factor.
For both possibilities of the pairing, the remaining piece of the trajectory
$\gamma$, respectively connecting $\mathbf{q}_\gamma(t)$ with 
$\mathbf{z}_{{i}}^{\nu_3}$ as well as $\mathbf{z}_{{i}}^{\nu_1}$ with 
$\mathbf{q}_\gamma(t)$, can be ``completed'' by combining it with the
straight-line trajectory $\omega$ from $\mathbf{q}_\gamma(t)$ to $r_L$
that is introduced through the factorization (\ref{eq:Crossedfactorize}).

We can now perform the energy average of the paramagnetic phase factors
according to Eq.~(\ref{eq:phaseaverage}), taking into account that the
effective propagation times of the pieces of trajectories under consideration
equal ${T_\gamma}-t$ as well as $t$, respectively, for the two phase factors
appearing in the second line of Eq.~(\ref{eq:dpsi12cross}) (the additional
contribution of the straight-line trajectory $\omega$ to the total propagation
time is neglected).
{For both phase factors, this gives} rise to integrals that are 
straightforwardly evaluated as 
\begin{equation}
  \int_0^{{T_\gamma}} {\exp\left(-\frac{{t}}{\tau_B}\right)} dt = \tau_B 
  \left[ 1 - {\exp\left(-\frac{T_\gamma}{\tau_B}\right)} \right] \, . \label{eq:phaseaverageint}
\end{equation}
We therefore obtain
\begin{equation}
  \left\langle \psi_{{i}}^* \delta \psi_{{i}}^{(c)} \right\rangle =
  - \frac{i}{\hbar}  \frac{\hbar^2}{2m} 
  \left|S_0 \frac{\pi \hbar}{W}\right|^2 \left\langle c^{(0)} \right\rangle
  \sum_{\nu_1,\nu_3 = \pm 1} 2\tau_B \sum_\gamma \left\langle \left| 
  \AnfangEnde{A}_\gamma(\mathbf{z}_{{i}}^{\nu_3},\mathbf{z}_{{i}}^{\nu_1},\mu) \right|^2
  \right\rangle \left[ 1 - {\exp\left(-\frac{T_\gamma}{\tau_B}\right)} \right]
\end{equation}
which, after applying the Hannay-Ozorio de Almeida sum rule 
{[see Eq.~(\ref{eq:Gzz})]}, is evaluated as
\begin{equation}
  \left\langle \psi_{{i}}^* \delta \psi_{{i}}^{(c)} \right\rangle =
  - i \left| 
  \frac{m S_0}{\hbar p_i^{\mathrm{l}}(\mu)} \right|^4
  \left(\frac{\tau_H}{\tau_D} + \frac{\tau_H}{\tau_B} \right)^{-2} 
  \frac{p_i^{\mathrm{l}}(\mu) \tau_D}{m}
  \label{eq:back_refl_o1} %label added by JM
\end{equation}
using the expression (\ref{eq:avcrosseddensity}) for the average of the
crossed density $\left\langle c^{(0)} \right\rangle$.
As this expression is purely imaginary, the modification of the 
backscattering probability due to the presence of the nonlinearity
vanishes in first order in $g$, as seen from Eq.~(\ref{eq:psi12cross1}).

Going beyond the first order in $g$, we can express the full nonlinear 
coherent backscattering probability, as evaluated using the semiclassical 
stationary phase approximation, in diagrammatic terms according to
\begin{eqnarray}
 \centerpic{0.08}{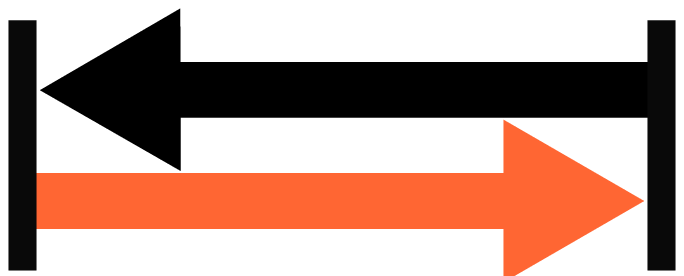}
&=&\centerpic{0.08}{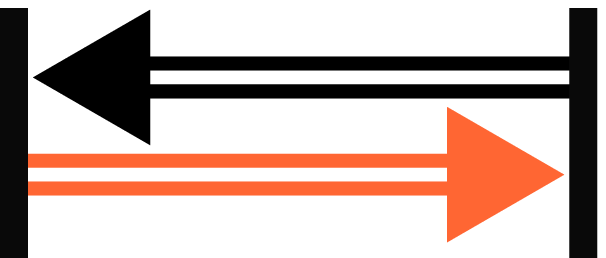}
 +\downpic{0.16}{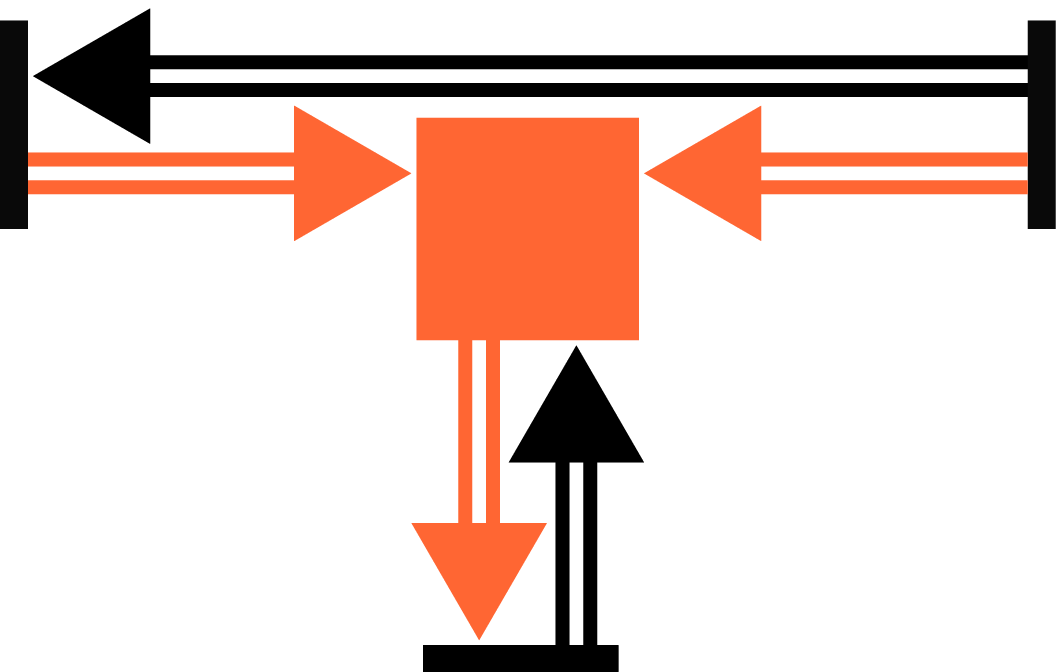}
 +\uppic{0.16}{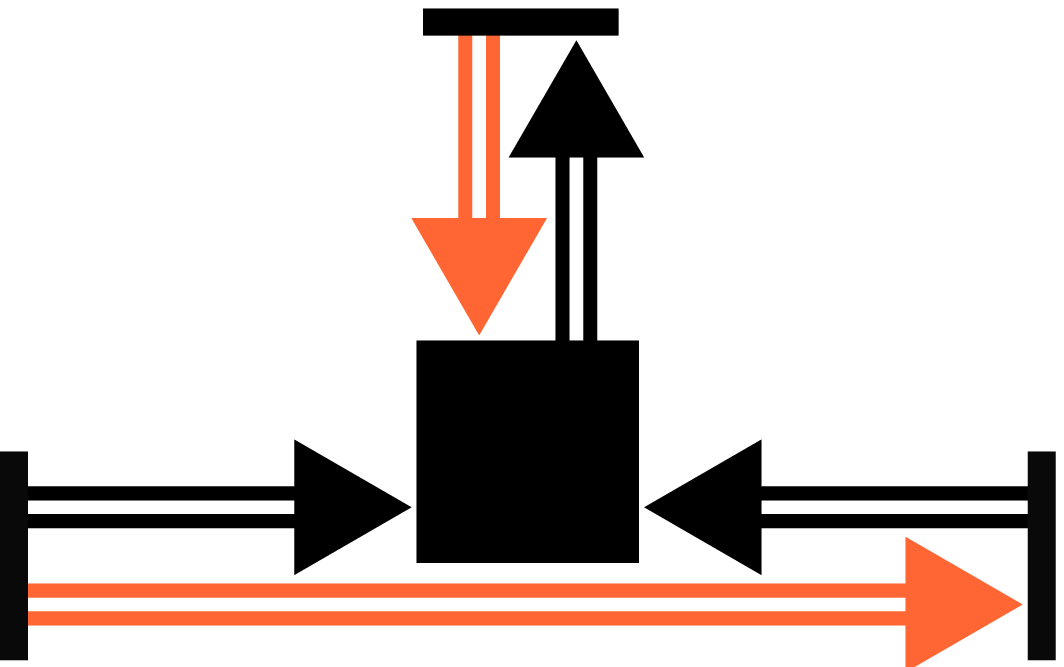}
\nonumber\\[-4ex]
&&+\centerpic{0.16}{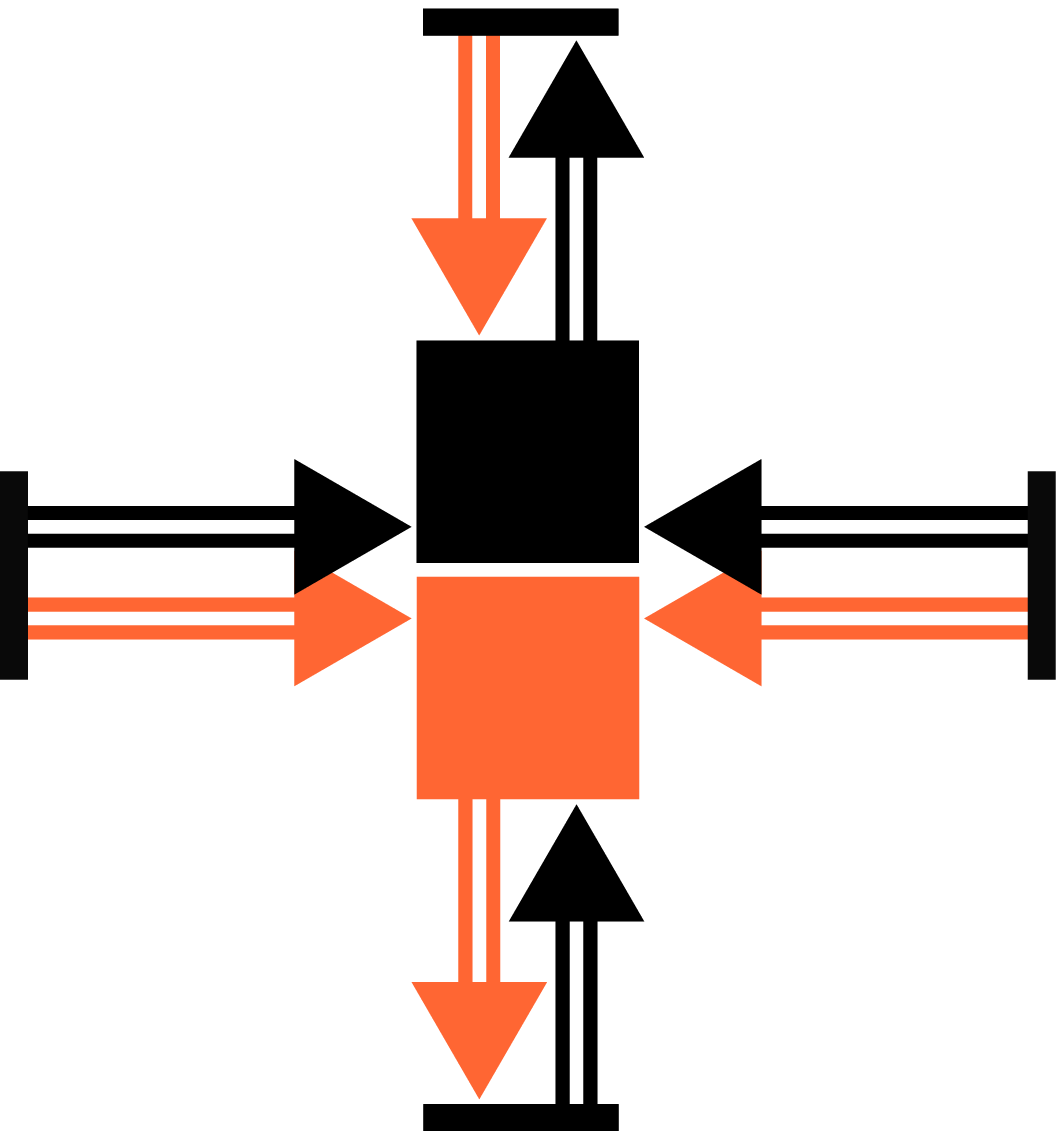}
+2\;\uppic{0.16}{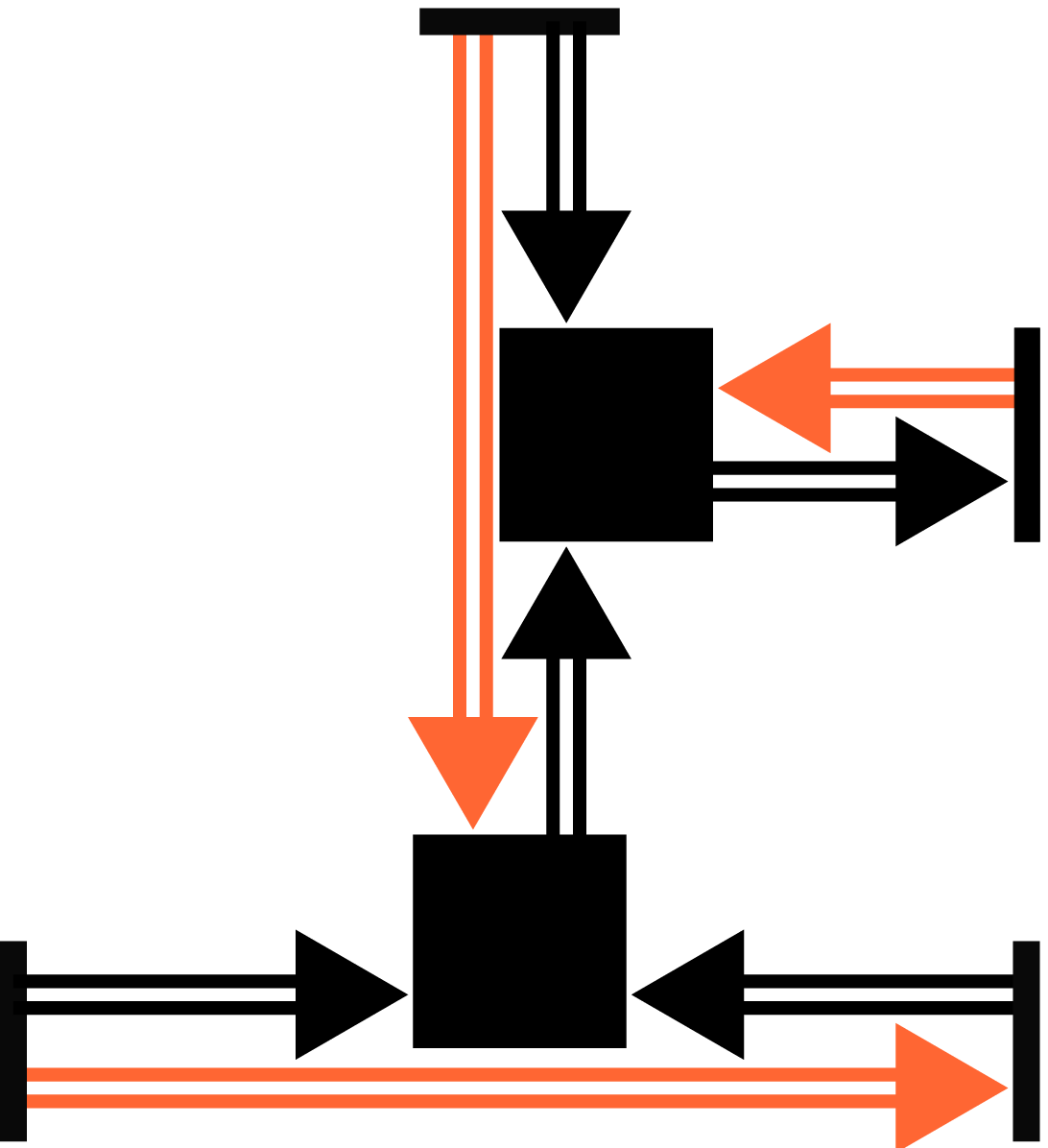}
+2\;\downpic{0.16}{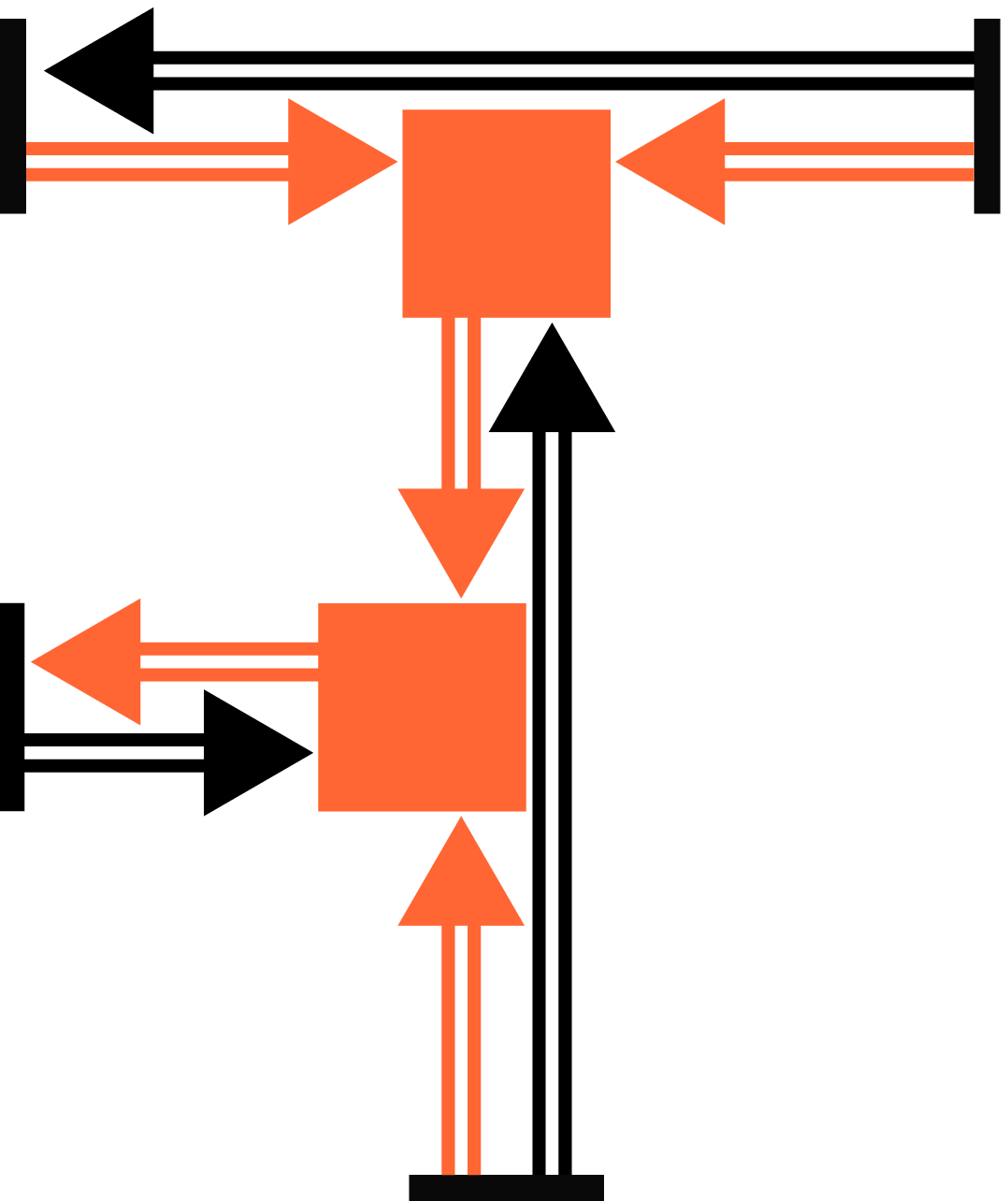}
+\mathcal{O}(g^3)\,. \label{eq:crosseddiags}
\end{eqnarray}
Here,
\includegraphics[height=2ex]{ladder_GF.eps}
represents, according to Eq.~(\ref{eq:modifiedG}), the modified Green 
function $G_\ell$ due to the inclusion of ladder contributions.
All types of ladder diagrams that were discussed in the previous subsection
\ref{sec:ladder} are therefore implicitly included in this representation.
As in Eq.~(\ref{eq:Gl}), the prefactors $2$ symbolize the fact 
that two different possibilities of pairings have to be counted
for certain diagrams.

In analogy with the derivation undertaken in Ref.~\cite{Wel09AP},
this series of diagrams can be exactly summed yielding
\begin{eqnarray}
 \centerpic{0.08}{crossed_complete.eps}
 =\centerpic{0.08}{crossed_0.eps}
 +\downpic{0.16}{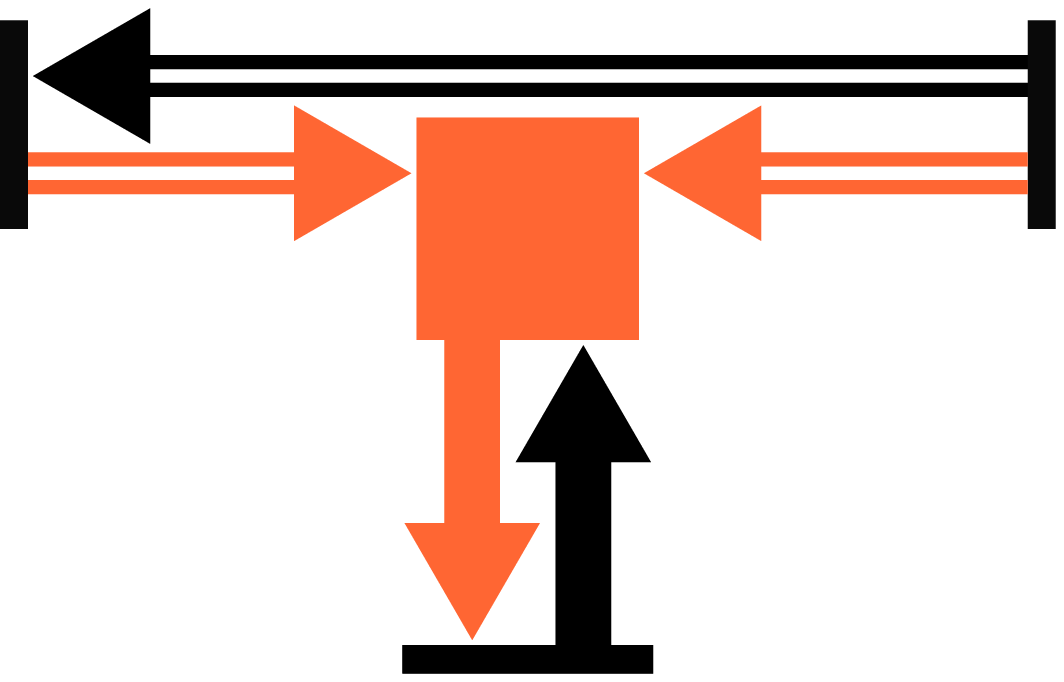}
 +\uppic{0.16}{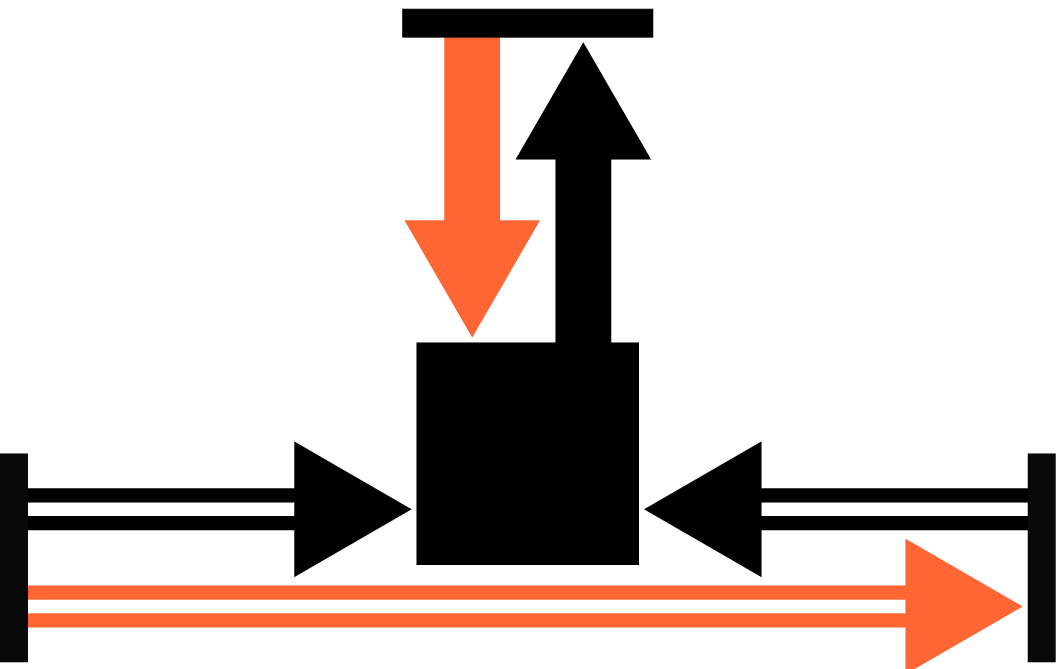}
 +\centerpic{0.16}{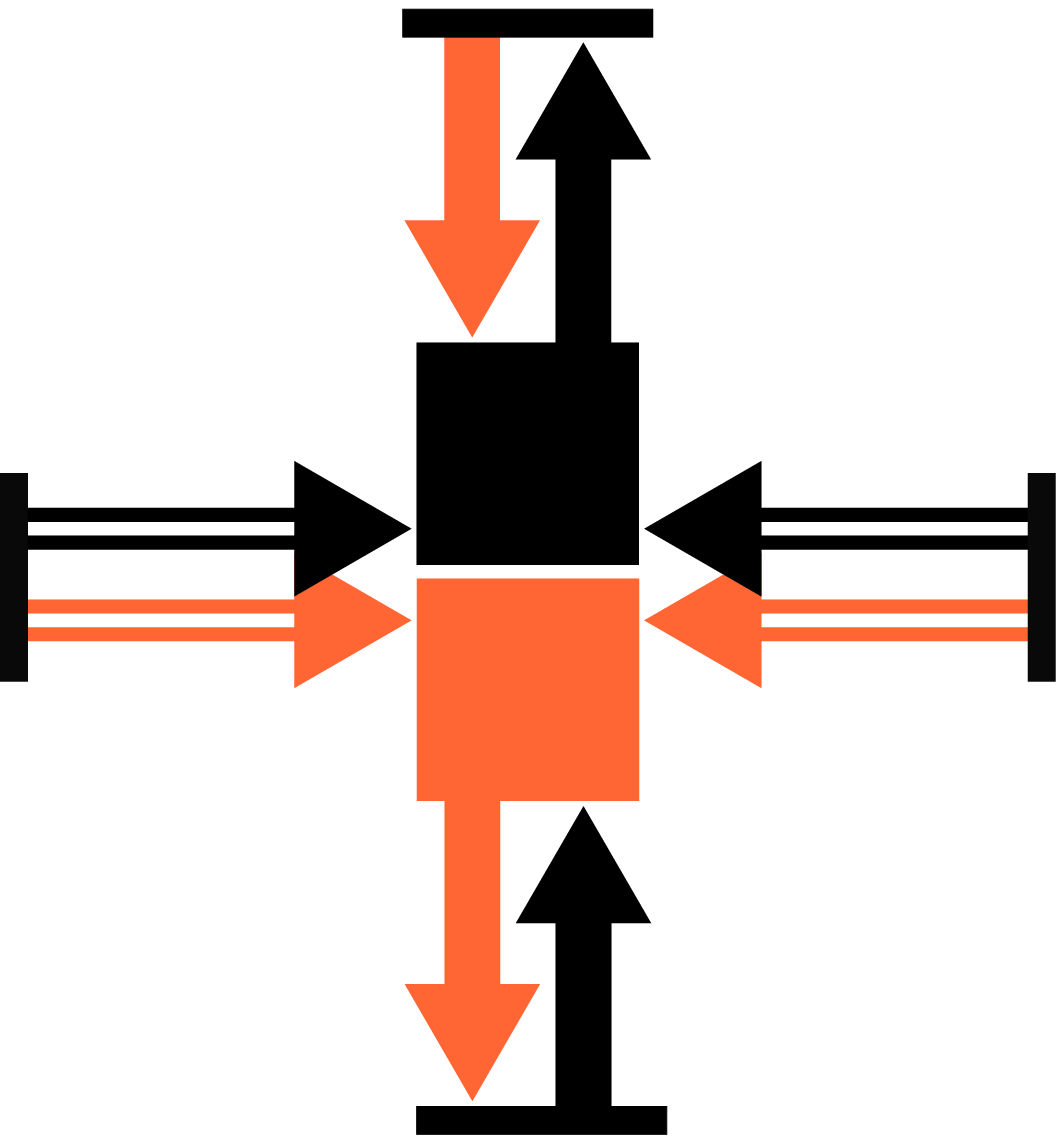}
\label{eq:fullbackscattering}
\end{eqnarray}
where we define the nonlinear crossed density
\includegraphics[height=2ex]{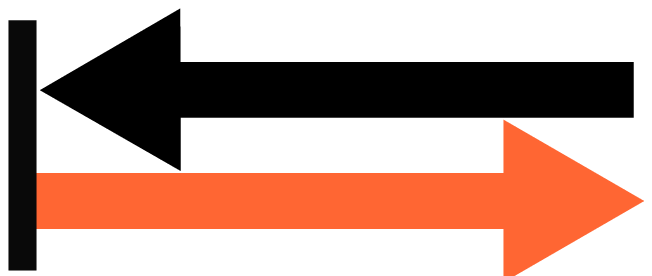}
and its complex conjugate
\includegraphics[height=2ex]{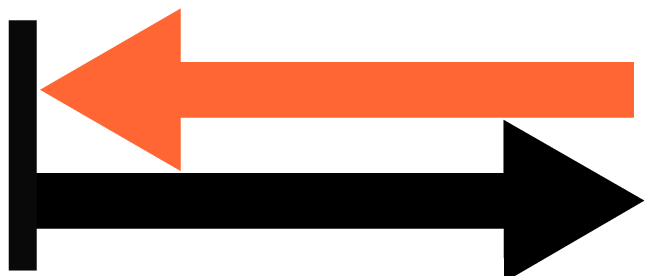}
in a self-consistent manner through
\begin{eqnarray}
\centerpic{0.1}{crossed_density.eps}
&=&\centerpic{0.1}{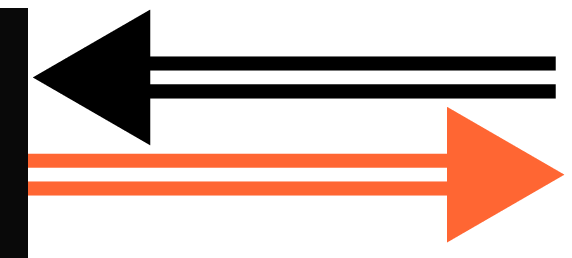}
+2\;\uppic{0.16}{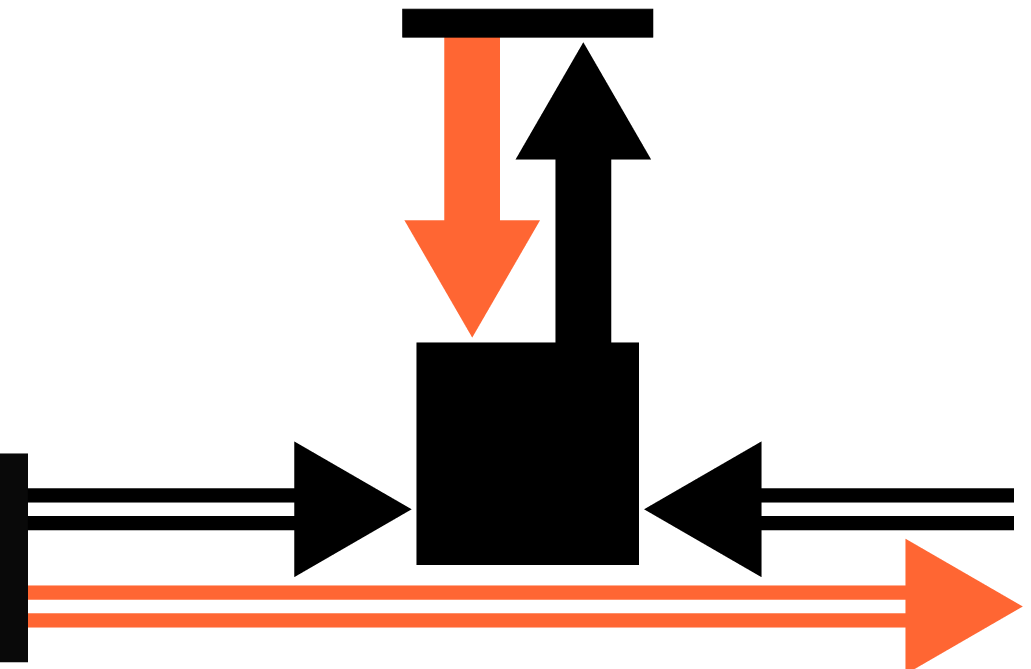}\,,
 \label{eq:nonlincrosseddensity1}\\
\centerpic{0.1}{crossed_density_conj.eps}
&=&\centerpic{0.1}{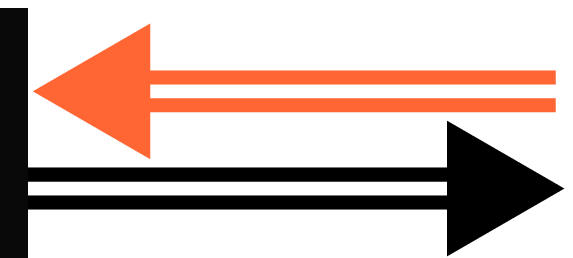}
+2\;\uppic{0.16}{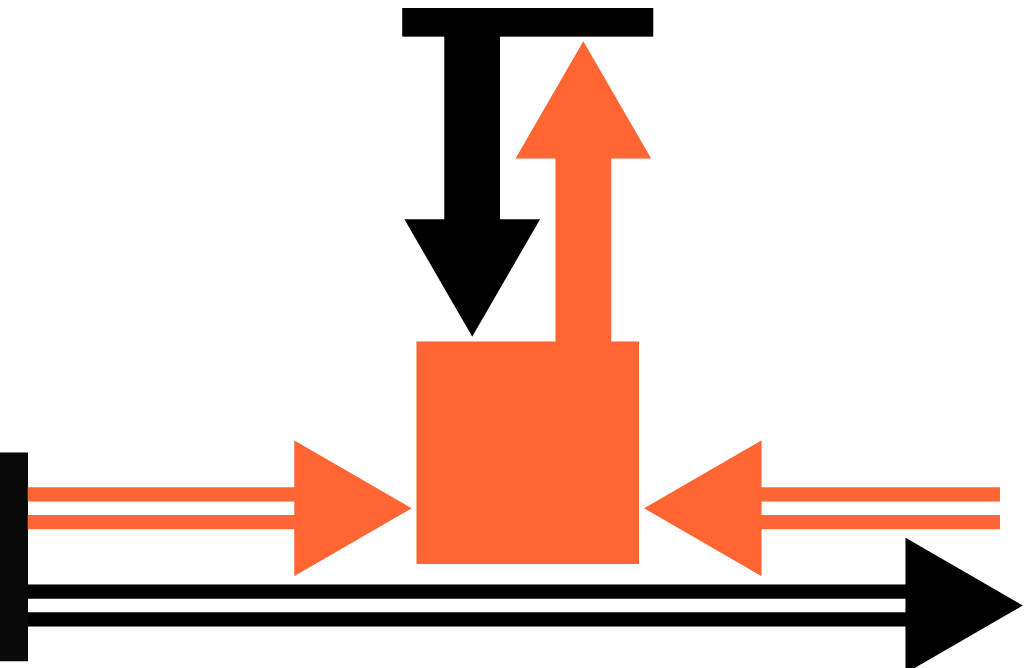}\,,
 \label{eq:nonlincrosseddensity2}
\end{eqnarray}
This nonlinear crossed density can be expressed through a transport equation
of the form
\begin{eqnarray}
  C^{(g)}(\mathbf{r}) & = & C^{(0)}(\mathbf{r}) +
  2 g \frac{\hbar^2}{2m} |S_0|^2 \frac{\pi \hbar}{W} 
  \sum_{\nu_1,\nu_2=\pm 1} \nu_1\nu_2 
  \nonumber \\ && \times
  \int d^2r' C^{(g)}(\mathbf{r}') 
  G_\ell(\mathbf{r}',\mathbf{r},\mu)
  \Anfang{G}_\ell(\mathbf{r}',\mathbf{z}_{{i}}^{\nu_1},\mu)
  \Anfang{G}_\ell^*(\mathbf{r},\mathbf{z}_{{i}}^{\nu_2},\mu)
  \label{eq:Cg}
\end{eqnarray}
which involves the modified Green function (\ref{eq:G2}) that takes into 
account the average shift of the effective potential within the billiard
due to the presence of the nonlinearity.
Being invariant under time-reversal, the nonlinearity-induced contribution
$\chi_\gamma$ (\ref{eq:chi}) to the action integral does {not} play any role for 
the determination of the nonlinear crossed density.
Indeed, applying the stationary phase and diagonal approximations in 
Eq.~(\ref{eq:Cg}), we obtain
\begin{eqnarray}
  C^{(g)}(\mathbf{r}) & = & C^{(0)}(\mathbf{r}) 
  + g |S_0|^2 \frac{\pi \hbar^3}{m W} \sum_{\nu=\pm 1}
  \left| \Anfang{A}_{\gamma}(\mathbf{r},\mathbf{z}_{{i}}^\nu,\mu) \right|^2
  \nonumber \\ && \times
  \left( - \frac{i}{\hbar} \right) \int_0^{{T_\gamma}}
  C^{(g)}[\mathbf{q}_\gamma(t)] \exp\left\{-\frac{2i}{\hbar}
  \varphi_\gamma[\mathbf{r},\mathbf{q}_\gamma(t), \mu] \right\} dt
\end{eqnarray}
which does not involve any reference to $\chi_\gamma$.

Quite obviously, the nonlinear crossed density $C^{(g)}(\mathbf{r})$ is
not invariant under gauge transformations of the effective vector potential.
In perfect analogy with $C^{(0)}(\mathbf{r})$, however, we can describe the
explicit gauge dependence of $C^{(g)}(\mathbf{r})$ in terms of a phase factor
that contains the paramagnetic contribution of a straight-line trajectory
$\omega$ from $\mathbf{r}$ to the reference point $\mathbf{r}_L$ within the 
incident lead \cite{rem_omega}.
In analogy with Eq.~(\ref{eq:Crossedfactorize}), we therefore propose 
\begin{equation}
  C^{(g)}(\mathbf{r}) \equiv c^{(g)}(\mathbf{r}) \exp\left[\frac{2i}{\hbar}
    \varphi_\omega(\mathbf{r}_L,\mathbf{r},\mu) \right]
  \label{eq:nlCrossedfactorize}
\end{equation}
as definition for the gauge-invariant part $c^{(g)}(\mathbf{r})$ of the
nonlinear crossed density, which in turn satisfies the gauge-invariant 
transport equation
\begin{eqnarray}
  c^{(g)}(\mathbf{r}) & = & c^{(0)}(\mathbf{r}) 
  + g |S_0|^2 \frac{\pi \hbar^3}{m W} \sum_{\nu=\pm 1}
  \left| \Anfang{A}_{\gamma}(\mathbf{r},\mathbf{z}_{{i}}^\nu,\mu) \right|^2
  \nonumber \\ && \times
  \left( - \frac{i}{\hbar} \right) \int_0^{{T_\gamma}} 
  c^{(g)}[\mathbf{q}_\gamma(t)] \exp\left\{-\frac{2i}{\hbar}
  \varphi_{\bar{\gamma}}[\mathbf{r},\mathbf{q}_\gamma(t),\mathbf{r}_L,\mu] 
  \right\} dt \ .
\end{eqnarray}
We can now compute the energy average of $c^{(g)}(\mathbf{r})$ by assuming
that it is, as the one for $c^{(0)}(\mathbf{r})$ 
[see Eq.~(\ref{eq:avcrosseddensity})], independent of the position 
$\mathbf{r}$ within the billiard, which is to be verified 
\textit{a posteriori}.
Using Eqs.~(\ref{eq:phaseaverage}), (\ref{eq:avcrosseddensity}),
(\ref{eq:phaseaverageint}), {(\ref{eq:G2cl}), and (\ref{eq:Grz})}, 
we obtain 
\begin{eqnarray}
  \left\langle c^{(g)} \right\rangle & = & \left\langle c^{(0)} \right\rangle
  - i g |S_0|^2 \frac{\pi \hbar^2}{m W}
  \left\langle c^{(g)} \right\rangle \sum_{\nu = \pm 1} \sum_\gamma 
  \left\langle \left| \Anfang{A}_{\gamma}(\mathbf{r},\mathbf{z}_{{i}}^\nu,\mu)
  \right|^2 \right\rangle\left[1- {\exp\left(-\frac{T_\gamma}{\tau_B}\right)}\right] 
  \nonumber \\ & = & 
  \left\langle c^{(0)} \right\rangle - i g \frac{\hbar \tau_D}{m}
  \left| \frac{m S_0}{\hbar} \right|^2 
  \left( \frac{\tau_H}{\tau_D} + \frac{\tau_H}{\tau_B} \right)^{-1}
  \frac{1}{\hbar p_i^{\mathrm{l}}(\mu)} \left\langle c^{(g)} \right\rangle
  \nonumber \\ & = &
  \left\langle c^{(0)} \right\rangle - i g \frac{\hbar \tau_D}{m} 
  \left\langle c^{(0)} \right\rangle \left\langle c^{(g)} \right\rangle
\end{eqnarray}
which is straightforwardly solved as
\begin{equation}
  \left\langle c^{(g)} \right\rangle = \frac{\left\langle c^{(0)} \right\rangle}
  {1 + i g \tau_D \frac{\hbar}{m} \left\langle c^{(0)} \right\rangle} \, .
  \label{eq:cg}
\end{equation}

We are now in a position to evaluate the full nonlinear coherent 
backscattering probability according to the diagrammatic representation
(\ref{eq:fullbackscattering}).
Denoting the linear crossed contribution to the backscattering probability by
\begin{equation}
  c_{{ii}}^{(0)} \equiv \left| \frac{\hbar p_i^{\mathrm{l}}(\mu)}{m S_0} \right|^2
  \langle |\psi_{{i}}|^2 \rangle_{\mathrm{c}} = 
  \left(\frac{\tau_H}{\tau_D} + \frac{\tau_H}{\tau_B} \right)^{-1}
\end{equation}
we have, as a generalization of Eq.~(\ref{eq:r11}),
\begin{equation}
  r_{{ii}} = \frac{\tau_D}{\tau_H} + c_{{ii}}^{(g)}
\end{equation}
with
\begin{equation}
  c_{{ii}}^{(g)} = c_{{ii}}^{(0)} +
  \left(\frac{\pi \hbar^2 p_i^{\mathrm{l}}(\mu)}{m W}\right)^2 
  \left\langle g \delta c_{{ii}}^{(1)} + g \left(\delta c_{{ii}}^{(1)}\right)^*
  + g^2 \delta c_{{ii}}^{(2)} \right\rangle
\end{equation}
where we introduce
\begin{equation}
  \delta c_{{ii}}^{(1)} = \frac{\hbar^2}{2m}
  \sum_{\nu_1,\nu_2,\nu_3,\nu_4=\pm 1} \nu_1\nu_2\nu_3\nu_4 
  \int d^2r C^{(g)}(\mathbf{r})
  \Anfang{G}_\ell(\mathbf{r},\mathbf{z}_{{i}}^{\nu_1},\mu)
  \Anfang{G}_\ell(\mathbf{r},\mathbf{z}_{{i}}^{\nu_2},\mu)
  \AnfangEnde{G}_\ell^*(\mathbf{z}_{{i}}^{\nu_3},\mathbf{z}_{{i}}^{\nu_4},\mu)
\end{equation}
and
\begin{eqnarray}
  \delta c_{{ii}}^{(2)} & = & \left(\frac{\hbar^2}{2m}\right)^2
  \sum_{\nu_1,\nu_2,\nu_3,\nu_4=\pm 1} \nu_1\nu_2\nu_3\nu_4 
  \int d^2r C^{(g)}(\mathbf{r}) 
  \Anfang{G}_\ell(\mathbf{r},\mathbf{z}_{{i}}^{\nu_1},\mu)
  \Anfang{G}_\ell(\mathbf{r},\mathbf{z}_{{i}}^{\nu_2},\mu)
  \nonumber \\ && \times
  \int d^2r' \left[ C^{(g)}(\mathbf{r}')\right]^*
  \Anfang{G}_\ell^*(\mathbf{r}',\mathbf{z}_{{i}}^{\nu_3},\mu)
  \Anfang{G}_\ell^*(\mathbf{r}',\mathbf{z}_{{i}}^{\nu_4},\mu)
\end{eqnarray}
as contributions that result from the nonlinear diagrams in 
Eq.~(\ref{eq:fullbackscattering}).
Again, stationary phase and diagonal approximations are employed in order
to evaluate these contributions, and we also use 
Eqs.~(\ref{eq:nlCrossedfactorize}) and (\ref{eq:cg}) in order to express the
nonlinear crossed density $C^{(g)}(\mathbf{r})$.
This yields for the energy average 
\begin{eqnarray}
  \left\langle \delta c_{{ii}}^{(1)} \right\rangle & = & 
  - i \frac{\hbar}{2m} \left\langle c^{(g)} \right\rangle
  \sum_{\nu_1,\nu_2} \sum_\gamma \left\langle \left|
  \AnfangEnde{A}_\gamma(\mathbf{z}_{{i}}^{\nu_1},\mathbf{z}_{{i}}^{\nu_2},\mu) \right|^2
  \right\rangle
  \\ && \times \int_0^{{T_\gamma}} dt
  \left\langle \exp\left\{-\frac{2i}{\hbar} \Ende{\varphi}_{\bar{\gamma}}
  [\mathbf{z}_{{i}}^{\nu_1},\mathbf{q}_\gamma(t),\mathbf{r}_L,\mu] \right\} +
  \exp\left\{\frac{2i}{\hbar} \Anfang{\varphi}_{\bar{\gamma}}
  [\mathbf{r}_L,\mathbf{q}_\gamma(t),\mathbf{z}_{{i}}^{\nu_2},\mu]
  \right\} \right\rangle \nonumber \\
  & = & - i \frac{\hbar \tau_D}{m} \left\langle c^{(g)} \right\rangle
  \left( \frac{m W}{\pi \hbar^2 p_i^{\mathrm{l}}(\mu)} \right)^2
  \left( \frac{\tau_H}{\tau_D} + \frac{\tau_H}{\tau_B} \right)^{-1}
\end{eqnarray}
(the real part of which is nonzero due to the fact that 
$\left\langle c^{(g)} \right\rangle$ is complex) and
\begin{eqnarray}
  \left\langle \delta c_{{ii}}^{(2)} \right\rangle & = & 
  \frac{\hbar^2}{2m^2} \left|\left\langle c^{(g)} \right\rangle\right|^2
  \sum_{\nu_1,\nu_2} \sum_\gamma \left\langle \left|
  \AnfangEnde{A}_\gamma(\mathbf{z}_{{i}}^{\nu_1},\mathbf{z}_{{i}}^{\nu_2},\mu) \right|^2
  \right\rangle
  \\ && \times 
  \int_0^{{T_\gamma}} dt \int_0^{{T_\gamma}} dt'
  \left\langle \exp\left\{-\frac{2i}{\hbar} \Anfang{\varphi}_{\bar{\gamma}}
  [\mathbf{r}_L,\mathbf{q}_\gamma(t),\mathbf{z}_{{i}}^{\nu_2},\mu] +
  \frac{2i}{\hbar} \Anfang{\varphi}_{\bar{\gamma}}
  [\mathbf{r}_L,\mathbf{q}_\gamma(t'),\mathbf{z}_{{i}}^{\nu_2},\mu]
  \right\} \right\rangle \nonumber \\
  & = & \frac{\hbar^2 \tau_D^2}{m^2} \left|\left\langle c^{(g)} 
  \right\rangle\right|^2
  \left( \frac{m W}{\pi \hbar^2 p_i^{\mathrm{l}}(\mu)} \right)^2
  \left( \frac{\tau_H}{\tau_D} + \frac{\tau_H}{\tau_B} \right)^{-1}  
\end{eqnarray}
where we evaluate
\begin{eqnarray}
  \lefteqn{
    \int_0^{{T_\gamma}} dt \int_0^{{T_\gamma}} dt'
    \left\langle \exp\left\{-\frac{2i}{\hbar} \Anfang{\varphi}_{\bar{\gamma}}
    [\mathbf{r}_L,\mathbf{q}_\gamma(t),\mathbf{z}_{{i}}^{\nu_2},\mu] +
    \frac{2i}{\hbar} \Anfang{\varphi}_{\bar{\gamma}}
    [\mathbf{r}_L,\mathbf{q}_\gamma(t'),\mathbf{z}_{{i}}^{\nu_2},\mu]
    \right\} \right\rangle
  } \nonumber \\
  & = & \int_0^{{T_\gamma}} dt \int_0^t dt'
    \left\langle \exp\left\{-\frac{2i}{\hbar} \varphi_{\gamma}
    [\mathbf{q}_\gamma(t),\mathbf{q}_\gamma(t'),\mu] \right\} \right\rangle
    \quad + \quad \mathrm{c.c.} \nonumber \\
    & = & 2 \tau_B^2 \left[ \frac{{T_\gamma}}{\tau_B} + 
      \exp\left(-\frac{{T_\gamma}}{\tau_B}\right) - 1 \right] \, .
    \label{eq:cii_2_flux}
\end{eqnarray}
Altogether, we then obtain
\begin{equation}
  c_{{ii}}^{(g)} = c_{{ii}}^{(0)}
  \left| 1 - i g \tau_D \frac{\hbar}{m} \left\langle c^{(g)} 
  \right\rangle \right|^2 = c_{{ii}}^{(0)} \left| 1 - 
  \frac{i g \tau_D \frac{\hbar}{m} \left\langle c^{(0)} \right\rangle}
  {1 + i g \tau_D \frac{\hbar}{m} \left\langle c^{(0)} \right\rangle} 
  \right|^2 = \frac{c_{{ii}}^{(0)}}{1 + 
  \left( g \tau_D \frac{\hbar}{m} \left\langle c^{(0)} \right\rangle \right)^2}
\end{equation}
which together with Eq.~(\ref{eq:avcrosseddensity}) yields
\begin{equation}
  c_{{ii}}^{(g)} = \frac{c_{{ii}}^{(0)}}{1 + 
  \left( g j^{\mathrm{i}} \tau_D c_{{ii}}^{(0)} \right)^2}
  = \frac{\displaystyle \frac{\tau_D}{\tau_H} \frac{1}{1 + \tau_D / \tau_B}
  {\frac{m}{\hbar} j^{\mathrm{i}}}}
  {\displaystyle 1 + \left( g j^{\mathrm{i}} \tau_D \frac{\tau_D}{\tau_H}
  \frac{1}{1 + \tau_D / \tau_B} \right)^2 }
  \label{eq:c11gendresult}
\end{equation}
where $j^{\mathrm{i}}$ is the incident current. 
The probability for retro-reflection into the incident channel $n=i$
is then obtained as
\begin{equation}
  r_{ii} = \frac{\tau_D}{\tau_H} + 
  \frac{\tau_D / \tau_H}{\displaystyle 1 + \frac{\tau_D}{\tau_B} +
    \frac{\left( g j^{\mathrm{i}} \tau_D^2 / \tau_H \right)^2}
    {1 + \tau_D / \tau_B}} \, .
\end{equation}
{This is the main result of Section \ref{sec:semi}.
It essentially states that the presence of the nonlinearity $g$ constitutes
another dephasing mechanism in addition to the magnetic field.}

\subsection{Alternative approach in terms of nonlinearity blocks}
\label{sec:alt_approach}

\begin{figure}[t]
  \begin{center}
    \includegraphics[width=0.5\textwidth]{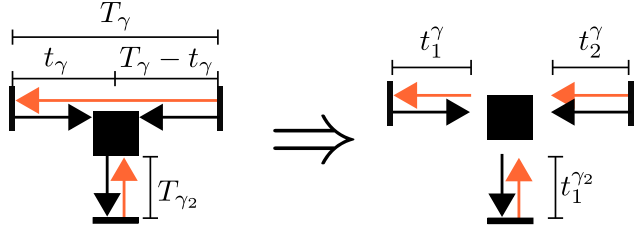}    
  \end{center}
  \caption{\label{fig:separation}
    Nonlinear diagram corresponding to the calculation of
    $\langle\psi_i^{*}(\delta\psi_i^{(c)})\rangle$.
    The left-hand side shows the diagram together with the durations
    that represent the integration variables appearing in 
    Eq.~(\ref{eq:psi_dpsi_alt}).
    The right-hand side illustrates the variable transformation
    leading to Eq.~(\ref{eq:psi_dpsi_neu}).
    In effect, the diagram can be cut into individual ``links'' 
    the contributions of which can be determined by separate integrals.
  }
\end{figure}

Inspired from Refs.~\cite{WelGre08PRL,WelGre09PRA,Wel09AP}, we outline, in
this subsection, an alternative approach to determine the nonlinearity-induced
modifications to the retro-reflection probability, which will become useful in
the subsequent section on loop contributions.
We reconsider for that purpose the calculation of
$\langle\psi_i^{*}(\delta\psi_i^{(c)})\rangle$
{on the basis of Eq.~(\ref{eq:dpsi12cross}), which was done using 
the expressions (\ref{eq:phasedecay}) and (\ref{eq:phaseaverage}) for the 
average} magnetic phase factors and the sum rules 
(\ref{eq:Grz}) and (\ref{eq:Gzz}).
If we deliberately keep the occurring integrations over 
trajectory durations as they appear in the sum rules 
[see Eq.~(\ref{eq:sc_amplitude_transform})], we obtain as an intermediate 
result
\begin{eqnarray}
 \left\langle \psi_i^{*}\delta\psi_i^{(c)}\right\rangle
  &=&-i\left|\frac{m S_0}{\hbar p_i^\mathrm{l}(\mu)}\right|^4
   {\frac{p_i^\mathrm{l}(\mu)}{2m\tau_H^2}}
   \int_0^\infty dT_{\gamma}\int_0^{T_\gamma}dt_\gamma \int_0^\infty dT_{\gamma_2} 
   \exp\left(-\frac{T_{\gamma_2}}{\tau_D}-\frac{T_{\gamma_2}}{\tau_B}\right)
   \nonumber\\ &&\times 
   \left[\exp\left(-\frac{T_\gamma}{\tau_D}-\frac{t_\gamma}{\tau_B}\right)
    +\exp\left(-\frac{T_\gamma}{\tau_D}-\frac{T_\gamma-t_\gamma}{\tau_B}
    \right)\right] \, {,} \label{eq:psi_dpsi_alt}
\end{eqnarray}
which transforms into
\begin{eqnarray}
\left\langle \psi_i^{*}\delta\psi_i^{(c)}\right\rangle
  &=&-i\left|\frac{m S_0}{\hbar p_i^\mathrm{l}(\mu)}\right|^4
   {\frac{p_i^\mathrm{l}(\mu)}{2m\tau_H^2}}
   \int_0^\infty dt_1^{\gamma}\int_0^\infty dt_2^\gamma \int_0^\infty 
   dt_1^{\gamma_2}  \exp\left(-\frac{t_1^{\gamma_2}}{\tau_D}-
   \frac{t_1^{\gamma_2}}{\tau_B}\right)\nonumber\\
   &&\times \left[\exp\left(-\frac{t_1^\gamma+t_2^\gamma}{\tau_D}
     -\frac{t_1^\gamma}{\tau_B}\right)
    +\exp\left(-\frac{t_1^\gamma+t_2^\gamma}{\tau_D}-
    \frac{t_2^\gamma}{\tau_B}\right)\right] \label{eq:psi_dpsi_neu}
\end{eqnarray}
{after applying the variable transformation
$t_1^\gamma \equiv t_\gamma$, $t_2^\gamma \equiv T_\gamma-t_\gamma$,
$t_1^{\gamma_2} \equiv T_{\gamma_2}$ that is motivated from 
Ref.~\cite{MueHeuBraHaa07NJP}.}
{Figure \ref{fig:separation} illustrates this variable transformation
in the corresponding nonlinear diagram.
In effect, the transformation} cuts the diagram into 
{separate pieces of trajectories which we shall}, as done in 
{Ref.}~\cite{MueHeuBraHaa07NJP}, refer to as ``links''
{and which are connected with each other at the ``nonlinearity block''
\includegraphics[height=2ex]{vertex.eps}.
Each link gives rise to a separate integration} yielding either $\tau_D$ 
for {ladder-type links}
\includegraphics[height=2ex]{parallelarrow.eps} 
 or $\tau_D/(1+\tau_D/\tau_B)$ for {crossed-type links}
\includegraphics[height=2ex]{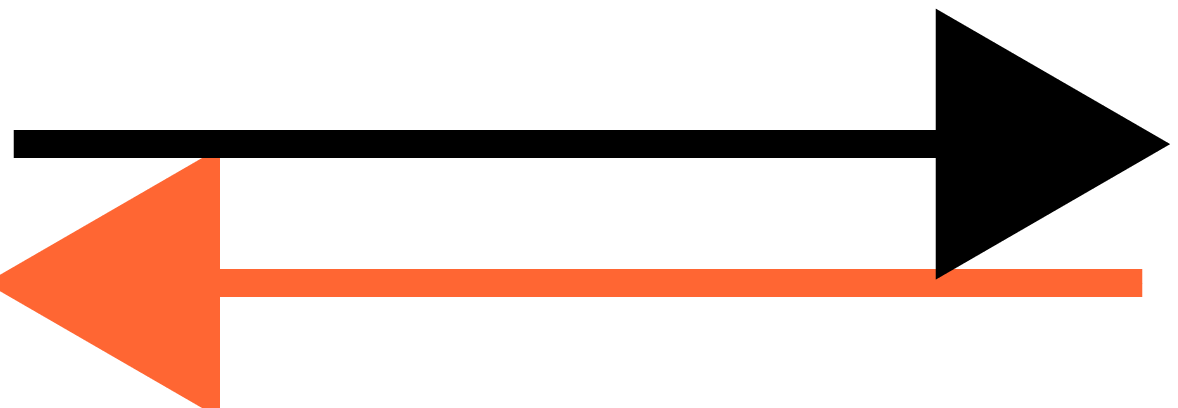}.
{Applying this rule to the diagram under consideration, we obtain
again the expression (\ref{eq:back_refl_o1}) for
$\langle\psi_i^{*}(\delta\psi_i^{(c)})\rangle$.}

\begin{figure}[t]
 \begin{center}
  \includegraphics[width=0.6\textwidth]{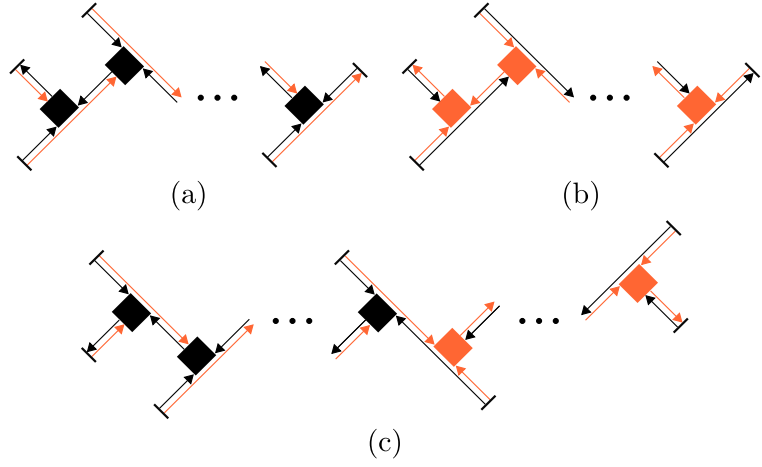}
 \end{center}
 \caption{\label{fig:kth_refl} 
   {Relevant diagrams for the calculation of the nonlinear contribution 
     to the backscattering probability{.
     Diagrams (a), (b), and (c) respectively correspond to the second, 
     the third, and the fourth term on the right-hand side of 
     Eq.~(\ref{eq:fullbackscattering}) as well as to the second, the third,
     and the fourth line of Eq.~(\ref{eq:c11alt}).}}}
\end{figure}

{This reasoning can be generalized to more complicated diagrams involving
more than one nonlinearity block.
{Under consideration of the sum rules (\ref{eq:Grz}) -- (\ref{eq:Gzz})} 
and of the combinatorial prefactors $2$ arising in front of each nonlinearity 
block [see Eq.~\ref{eq:Gl}], we can state the following rules:
\begin{enumerate}[(1)]
\item each source \includegraphics[height=2ex]{bar.eps}
  contributes a factor $\hbar/p_i^\mathrm{l}(\mu)$;
\item each arrow emanating from a source 
  \includegraphics[height=2ex]{scat_wave_0.eps}
  contributes a factor $S_0$
  (and each conjugate arrow
  \includegraphics[height=2ex]{scat_wave_0_conj.eps}
  a factor $S_0^*$);
\item each trajectory, scattering from lead to lead or ending at a nonlinearity
  event within the billiard, contributes a factor $m^2/(\hbar^4 \tau_H)$;
\item each nonlinearity event 
  \includegraphics[height=2ex]{vertex.eps}
  in the scattering wavefunction
  contributes a factor $-i g \hbar / m$
  (and each nonlinearity event
  \includegraphics[height=2ex]{complex_vertex.eps}
  in the conjugate wavefunction contributes a factor $i g \hbar / m$);
\item each ladder-type link
  \includegraphics[height=2ex]{parallelarrow.eps} 
  contributes a factor $\tau_D$;
\item each crossed-type link
  \includegraphics[height=2ex]{antiparallelarrow.eps}
  contributes a factor $(1/\tau_D + 1/\tau_B)^{-1}$.
\setcounter{mycounter}{\value{enumi}}
\end{enumerate}
Using these rules, we can re-calculate the crossed contribution $c_{ii}^{(g)}$
to the retro-reflection intensity.
In contrast to Section \ref{sec:crossed}, we do not explicitly need to 
introduce the nonlinear crossed density $C^{(g)}(\mathbf{r})$ as done in
Eq.~(\ref{eq:Cg}).
Instead, we directly sum over all possible combinations of crossed diagrams
as they are depicted in Fig.~\ref{fig:kth_refl}.
Together with the contribution of the diagram 
\includegraphics[height=2ex]{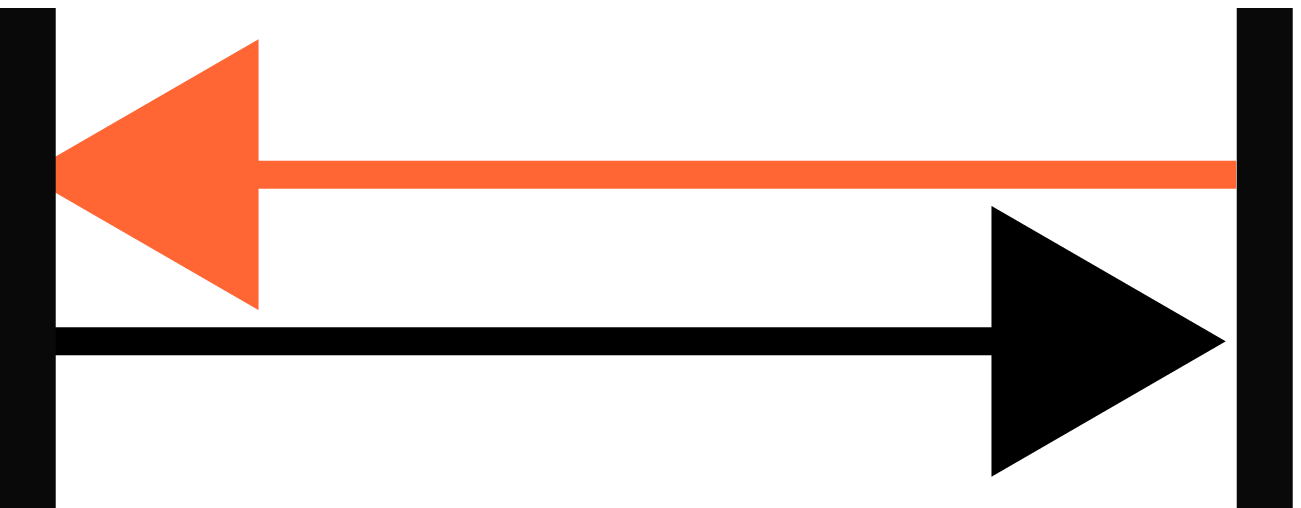}, this yields
\begin{eqnarray}
  \langle |\psi_{{i}}|^2 \rangle_{\mathrm{c}} & = &
  |S_0|^2 \left(\frac{\hbar}{p_i^{\mathrm{l}}(\mu)}\right)^2
  \frac{m^2}{\hbar^4{\tau_H}}
  \left(\frac{{1}}{\tau_D} + \frac{{1}}{\tau_B} \right)^{-1}
  \nonumber \\ && 
  + \sum_{k=1}^{\infty} \left( - i \frac{\hbar}{m} g
  {\tau_D}\right)^k
  \left[ |S_0|^2 \frac{m^2}{\hbar^4{\tau_H}}
    \left(\frac{{1}}{\tau_D} + \frac{{1}}{\tau_B} \right)^{-1} 
    \right]^{k+1} \left(\frac{\hbar}{p_i^{\mathrm{l}}(\mu)}\right)^{k+2}
  \nonumber \\ && 
  + \sum_{k=1}^{\infty} \left( i \frac{\hbar}{m} g {\tau_D}\right)^k
  \left[ |S_0|^2 \frac{m^2}{\hbar^4{\tau_H}}
    \left(\frac{{1}}{\tau_D} + \frac{{1}}{\tau_B} \right)^{-1} 
    \right]^{k+1} \left(\frac{\hbar}{p_i^{\mathrm{l}}(\mu)}\right)^{k+2}
  \nonumber \\ && 
  + \sum_{k,k'=1}^{\infty} i^{k-k'}
  \left( \frac{\hbar}{m} g {\tau_D}\right)^{k+k'}
  \left[ |S_0|^2 \frac{m^2}{\hbar^4{\tau_H}}
    \left(\frac{{1}}{\tau_D} + \frac{{1}}{\tau_B} \right)^{-1} 
    \right]^{k+k'+1} \left(\frac{\hbar}{p_i^{\mathrm{l}}(\mu)}\right)^{k+k'+2}
  \nonumber \\ & = & 
  \frac{\displaystyle
    \left| \frac{m S_0}{\hbar p_i^{\mathrm{l}}(\mu)} \right|^2
    \left(\frac{\tau_H}{\tau_D} + \frac{\tau_H}{\tau_B} \right)^{-1}
  }{\displaystyle
    1+ \left[ g \tau_D
    \frac{m |S_0|^2}{\hbar^2 p_i^{\mathrm{l}}(\mu)}
    \left(\frac{\tau_H}{\tau_D} + \frac{\tau_H}{\tau_B} \right)^{-1}
    \right]^{2}
  } \label{eq:c11alt}
\end{eqnarray}
which is perfectly equivalent to Eq.~(\ref{eq:c11gendresult}).}

\begin{figure}[tb]
  \begin{center}
    \includegraphics[width=0.6\linewidth]{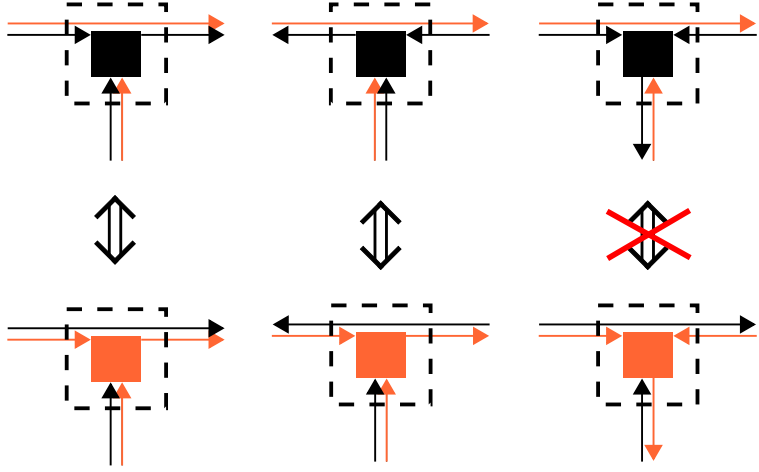}
  \end{center}
  \caption{\label{fig:basicblocks} 
    Basic building blocks with which all possible diagrams can be 
    constructed.
    As shown in the left and central columns, blocks with ladder-type input
    can be incorporated in both the $\psi$ and the $\psi^*$ branch for the
    same orientation and pairing of the trajectories outside the block
    region (depicted by the dashed square).
    This is not the case for blocks with crossed-type input shown in the
    right column, due to the mismatch of trajectories entering and 
    leaving the block.
    {Consequently, the contributions resulting from those blocks do not
      cancel each other, in contrast to the blocks with ladder-type input,
      but give rise to a finite modification of the reflection and
      transmission probabilities in the presence of the nonlinearity.}}
\end{figure}

{The approach on the basis of nonlinearity blocks also provides an
alternative understanding why blocks with ladder-type input [i.e.\
where a ladder pairing is employed along the trajectory that ends at 
the nonlinearity event, corresponding to the cases (\ref{eq:l1cond}) 
and (\ref{eq:l2cond}) in Section \ref{sec:nldiag}, and displayed in the
left and central columns of Fig.~\ref{fig:basicblocks}], do not affect the
mean values of densities and currents of the propagating condensate.
We remark for this purpose that the individual factor provided by each
nonlinearity block is purely \emph{imaginary} (as stated above by rule 4),
with a negative imaginary part for blocks 
\includegraphics[height=2ex]{vertex.eps}
that are incorporated within $\psi$ and with a positive imaginary part for 
blocks \includegraphics[height=2ex]{complex_vertex.eps}
within $\psi^*$.
Two diagrams that are almost identical except for the incorporation of
one single nonlinearity block, which is placed within $\psi$ in one of the
diagrams and within $\psi^*$ in the other diagram, will therefore cancel
each other in summations over all possible diagrams, as they contribute 
with equal amplitudes and opposite signs.
As illustrated in Fig.~\ref{fig:basicblocks}, this is the case for each 
diagram containing a block with ladder-type input, which has a counterpart 
in which this block is incorporated in the opposite manner.
Such diagrams do therefore not need to be considered for the calculation of
mean densities or currents of the propagating condensate.
Blocks with crossed-type input, on the other hand, can, in general, not be 
paired with canceling counterparts, which is shown in the right column of 
Fig.~\ref{fig:basicblocks}.}

\begin{figure}[t]
  \begin{center}
    \includegraphics[width=0.2\linewidth]{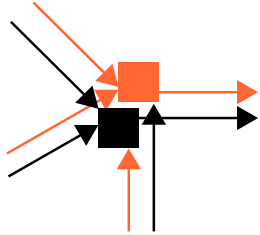}
  \end{center}
  \caption{\label{fig:strongnl}
    {
      Example for a nonlinear diagram that is not accounted for in the present
      diagrammatic theory.
      This diagram requires the presence of two nonlinearity blocks at the
      same spatial location and is therefore not expected to be of relevance
      for weak nonlinearity strengths.
    }
  }
\end{figure}

{Let us finally point out that the validity of the present diagrammatic theory 
is still limited to weak nonlinearity strengths, despite the above summations 
to infinite order in $g$.
This is illustrated in Fig.~\ref{fig:strongnl} which shows an example for a
diagram of second order in $g$ that is not accounted for in our diagrammatic
theory.
{It represents diffraction of the matter wave by short-ranged spatial
fluctuations of the nonlinear term $g\frac{\hbar^2}{2m} |\Psi({\bf r},t)|^2$
in the Gross-Pitaevskii equation (\ref{eq:GP0}). 
As it requires the presence of two nonlinearity events within a
region of the order of one wavelength,} its contribution is strongly
suppressed in the semiclassical regime
as compared to other diagrams of second order in $g$ in which the
nonlinearity blocks are spatially uncorrelated.
We do expect, however, that diagrams of the type shown in 
Fig.~\ref{fig:strongnl} will become relevant for large nonlinearity strengths,
possibly in the regime in which the scattering process destabilizes and
develops turbulent-like flow.}

\section{Loop corrections}
\label{sec:loop}

In the previous section{, we developed a semiclassical description
of weak localization in the presence of a weak atom-atom interaction
restricting ourselves to the diagonal approximation.}
{This theory will fail} to describe the occurring phenomena quantitatively, 
{as it violates} current conservation both in the {absence and in the
presence of the} nonlinearity. 
The reason for this {failure} lies in the use of the diagonal approximation, 
i.e.~we only used identical or time-reversed trajectories when our methods 
demanded correlated trajectory pairs. 
However, as originally shown in {Refs.}~\cite{SieRic01PST,Sie02JPA} 
for the spectral form factor and in {Ref.}~\cite{RicSie02PRL} for the 
Landauer conductance in the transport of electrons through two-dimensional 
uniformly hyperbolic systems with time-reversal symmetry, there is{,
besides the above mentioned one,} a second type of correlated trajectory 
pairs giving significant contributions to the {reflection and} transmission 
probabilities, the {so-called {``loops'' or} ``Sieber-Richter pairs''}. 
These are pairs of trajectories {with nearly identical initial and final
conditions;
as illustrated in Fig.~\ref{fig:loop}, one of the two trajectories undergoes}
a self-crossing with a small crossing angle $\epsilon$ whereas the other one 
avoids {that} crossing.
More generally speaking, as originally {worked out} in 
Refs.~\cite{TurRic03JPA,Spe03JPA} for the spectral form factor,
these trajectories {exhibit} an encounter in phase space with 
{their} time-reversed {counterparts,} which allows {for the existence
of} a partner trajectory {that switches} from the original trajectory to 
the time-reversed counterpart.
We shall adopt this more general phase space picture to derive
the corrections to weak localization in the linear and in the nonlinear case.

\subsection{Loop corrections in the linear case}

\label{sec:loop_linear}

\begin{figure}[tb]
  \begin{center}
    \includegraphics[width=0.8\linewidth]{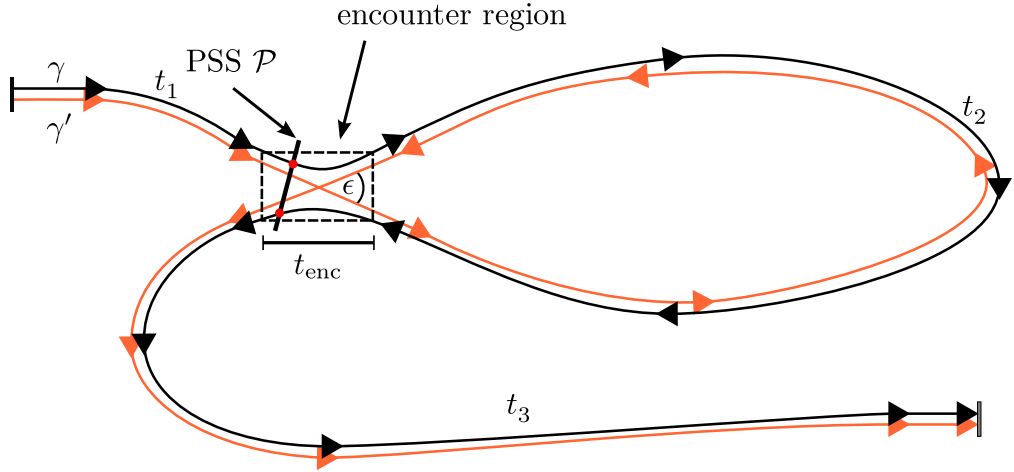}
  \end{center}
  \caption{\label{fig:loop}
  Sketch of a trajectory $\gamma$ which experiences a self encounter and its
corresponding partner trajectory $\gamma'$. In configuration space, which is depicted here,
one trajectory crosses itself with a small crossing angle $\epsilon$ whereas the other one avoids the
crossing. Also shown is the position of a possible Poincar\'e surface of section
$\mathcal{P}$ used to determine the action difference. We  note here
that this sketch is widely overestimating the crossing angle $\epsilon$ and
underestimating the trajectory lengths before and after the encounter region.}
\end{figure}

Our calculation of the contributions to weak
localization in the case $g=0$ mainly follows
{Refs.}~\cite{MueHeuBraHaaAlt05PRE,MueHeuBraHaa07NJP}.
We {shall} restrict ourselves here and 
{in the following} to the case of {at most} one 
{Sieber-Richter pair} per diagram, as {the presence of} more 
{such pairs} would only result in higher{-}order contributions in 
$\tau_D/\tau_H=1/{(N_c+\tilde{N}_c)}$.
{For the sake of definiteness, we shall consider the problem
of determining the transmission probability 
$t_{\tilde{n}i} \equiv j_{\tilde{n}} / j^{\mathrm{i}}$ that is
associated with the scattering process from the incident channel $i$ in the
left lead to the final channel ${\tilde{n}}$ in the right lead.
Our purpose is therefore} to calculate
\begin{equation}
  \left\langle|\psi_{{\tilde{n}}}^{(0)}|^2\right\rangle =
  \left| S_0\frac{\pi\hbar}{\sqrt{W\tilde{W}}} \right|^2 
  \sum_{\nu_1,\nu'_1=\pm1}\sum_{\nu_2,\nu'_2=\pm1}\nu_1\nu'_1\nu_2\nu'_2 
  \left\langle
\AnfangEnde{G}\left(\mathbf{z}_{\tilde{n}}^{\nu_1},\mathbf{z}_{i}^{\nu_1'},
\mu\right)\AnfangEnde{G}^{*}\left(\mathbf{z}_{\tilde{n}}^{\nu_2},\mathbf{z}_{i}^{\nu_2'},
\mu\right)\right\rangle \, .
  \label{eq:psi_squared}  
\end{equation}

As this quantity involves a product of two Green functions{,} we are 
{concerned} with sums over pairs of classical trajectories 
$\gamma$, $\gamma'$ here.
In the context of the diagonal approximation{,} we already {evaluated
in Section \ref{sec:semi}} the most {dominant contribution to this
transmission probability, for which} $\gamma'$ is identical 
{to $\gamma$}.

The next group of systematic{ally} correlated trajectories 
{consists of pairs $\gamma$, $\gamma'$ that exhibit, 
as sketched in Fig.~\ref{fig:loop}, a self{-}encounter in phase space
\cite{SieRic01PST,RicSie02PRL,MueHeuBraHaa07NJP}.} 
Their action difference {can be} determined {by defining a}
Poincar\'e surface of section $\mathcal{P}$ {within the encounter region,}
which is {oriented} perpendicular to $\gamma$ {on the first passage of
this trajectory through it, i.e., which is pierced by the first stretch of
$\gamma$ at its origin.
Linearizing the classical dynamics in the vicinity of this trajectory, we can 
define two basis vectors $\mathrm{e}_s$ and $\mathrm{e}_u$ within the 
two-dimensional surface of section $\mathcal{P}$ that are respectively 
oriented along the stable and unstable manifold of $\gamma$.
The} action difference {between $\gamma$ and $\gamma'$} is then 
{evaluated} as {\cite{TurRic03JPA, Spe03JPA}}
\begin{equation}
 \Delta S_{\gamma,\gamma'} {\equiv} 
 \AnfangEnde{S}_\gamma-\AnfangEnde{S}_{\gamma'}= su
\end{equation}
{where $s$ and $u$ denote the coordinates with respect to the basis vectors
$\mathrm{e}_s$ and $\mathrm{e}_u$, respectively, at which the trajectory 
$\gamma$ pierces through $\mathcal{P}$ for the second time.
Obviously, $\Delta S_{\gamma,\gamma'}$ can be sufficiently small, i.e.\ of the
order of $\hbar$, if, as depicted in Fig.~\ref{fig:loop}, one of the two 
trajectories exhibits a self-crossing in configuration space with a very small
crossing angle $\epsilon$ \cite{SieRic01PST,RicSie02PRL}.
The partner trajectory, whose existence and uniqueness is granted by the 
chaoticity of the classical dynamics, will then avoid that self-crossing and
follow the loop in between the two piercings through $\mathcal{P}$ in the
opposite direction.}

{In order to evaluate the contributions of such Sieber-Richter pairs to
Eq.~(\ref{eq:psi_squared}), we need to determine the probability
of a trajectory $\gamma$ to exhibit a near-encounter in phase space.
Due to ergodicity, the probability density for the trajectory 
$\gamma$ to pierce again through the Poincar\'e surface of section in the 
opposite direction at given coordinates $s$ and $u$ and after a given 
propagation time $\tilde{t}_2$ after the first piercing} is given by the 
Liouville measure ${\delta[\mu-H_0(\mathbf{p},\mathbf{q})]}/\Sigma(\mu)$
{with $(\mathbf{p},\mathbf{q})$ the coordinates of the second piercing 
in the full phase space and} $\Sigma(\mu) {\equiv} \int d^2q' \int d^2 p'
\delta[\mu-H_0(\mathbf{p}',\mathbf{q}')]$ {the phase-space} volume of the 
energy shell.
If we want to calculate the probability density for a trajectory 
$\gamma$ with a given total propagation time $T$ to have a partner 
trajectory $\gamma'$ with a given action difference 
$\Delta S_{\gamma,\gamma'} \equiv \Delta S$,
we are tempted to integrate this Liouville measure over all 
``intermediate'' propagation times $\tilde{t}_2$ between the first and
the second intersection through the Poincar\'e surface of section 
$\mathcal{P}$, over all ``initial'' propagation times $\tilde{t}_1$ 
from the incident channel to the first intersection through $\mathcal{P}$, 
over all ``final'' propagation times $\tilde{t}_3$ from the second 
intersection through $\mathcal{P}$ to the outgoing channel, 
as well as over all possible phase-space coordinates $s$, $u$ that $\gamma$
exhibits within $\mathcal{P}$ at its second piercing, with the requirements 
that $su = \Delta S$ and $\tilde{t}_1 + \tilde{t}_2 + \tilde{t}_3 = T$.
This na\"{i}ve integration would, however, lead to \emph{multiple countings}
of such trajectory pairs.
Indeed, the placement of the Poincar\'e surface of section $\mathcal{P}$ is
not unique, but can be shifted along the first stretch of the trajectory
$\gamma$.
This generally will lead to different coordinates $s,u$ of the second
stretch of $\gamma$ when passing through $\mathcal{P}$, but the product
$su$ of these coordinates will not change, provided the second piercing point
of $\gamma$ is also in a sufficiently close neighborhood of the origin of 
$\mathcal{P}$ such that the linearization of the classical dynamics around
$\gamma$ is still valid (see also the calculations in
\ref{sec:loopappendix}).

{The contribution of an individual Sieber-Richter pair with an action
difference $\Delta S$ would, when performing the above-mentioned integration, 
therefore effectively be overweighted by a factor 
$t_\mathrm{enc}\equiv t_\mathrm{enc}(\Delta S)$ that corresponds to the 
typical ``duration'' of the encounter, i.e., the typical propagation time
within which one of the trajectories ``sees'' the other one within a
distance that is within the linearization region of its transverse dynamics.
Defining by $\lambda$ the Lyapunov exponent of the ergodic system
\cite{hyperbolicity}, and introducing $c$ as the maximal distance along 
the stable and unstable manifolds $\mathrm{e}_s$, $\mathrm{e}_u$ for the 
linearization of the transverse dynamics within the Poincar\'e surface of 
section $\mathcal{P}$ to be valid (i.e., we require that $-c < s,u < c$; 
the precise value of $c$, which is related to the Ehrenfest time of the
system as pointed out in \ref{sec:encint}, will not be of
relevance in the end), we can define
\cite{MueHeuBraHaaAlt05PRE,MueHeuBraHaa07NJP}
\begin{equation}
  t_\mathrm{enc}(su) = \frac{1}{\lambda} \ln\left(\frac{c^2}{|su|}\right) \, .
\end{equation}
This duration $t_\mathrm{enc}(su)$ reflects the fact that some minimal time 
is needed for the two nearby trajectory stretches to part from each other, 
in order to form the loop on one end and to exit toward different leads 
on the other end of the encounter region.}

{In view of these considerations, we define 
(see also Ref.~\cite{MueHeuBraHaa07NJP})
\begin{equation}
 w(\Delta S,t_2;T)= \int\limits_0^\infty\int\limits_0^\infty dt_1 dt_3 
 \int\limits_{-c}^{c}\int\limits_{-c}^{c} ds  du 
 \frac{\delta(su-\Delta S)}{\Sigma(\mu)t_\mathrm{enc}(su)}
   \delta[t_1 + t_2 + t_3 + 2 t_\mathrm{enc}(su) - T] 
 \label{eq:partnerdensity1}
\end{equation}
as the probability density for a trajectory $\gamma$ with the total 
propagation time $T$ to have a partner trajectory $\gamma'$ with an
action difference $\Delta S$ and a loop duration $t_2$.
This loop duration $t_2$ as well as the initial and final propagation times 
$t_1$ and $t_3$ that appear in the integrations in 
Eq.~(\ref{eq:partnerdensity1}) are, as illustrated in Fig.~\ref{fig:loop}, 
defined not with respect to the particular placement of the Poincar\'e surface 
of section, but with respect to the location of the encounter region along 
the trajectory.
Using}
$\AnfangEnde{A}_\gamma{\simeq}\AnfangEnde{A}_{\gamma'}$ {and}
$\AnfangEnde{\mu}_\gamma{\simeq}\AnfangEnde{\mu}_{\gamma'}$ for the 
{trajectory pair $\gamma,\gamma'$,} the loop contributions {to} $\langle
|\psi_{{\tilde{n}}}^{(0)}|^2\rangle$ are calculated as
\begin{eqnarray}
 \left\langle \left|\psi_{{\tilde{n}}}^{(0)}\right|^2 \right\rangle_\textrm{loop}
 &=&{\frac{\left|S_0\right|^2(\pi\hbar)^2}{W\tilde{W}}\sum_{\nu,\nu'=\pm1}
 \int_{-\infty}^{\infty} d(\Delta S) \int_0^\infty dt_2 \sum_{\gamma}
 \left\langle\left|\AnfangEnde{A}_\gamma\left(\mathbf{z}_{\tilde{n}}^{\nu},\mathbf{z}_{i}^{\nu'},\mu
 \right)\right|^2\right\rangle  } \nonumber\\
 &&{\times w(\Delta S,t_2;T_\gamma) 
   \exp\left(\frac{i}{\hbar} \Delta S\right)
   \exp\left(-\frac{t_2}{\tau_B}\right)}\, ,
\end{eqnarray}
where the dephasing factor $\exp(-t_2/\tau_B)$ originates from
$\varphi_\gamma\left[\mathbf{r}_\gamma (t_1+t_\mathrm{enc}+t_2),
\mathbf{r}_\gamma(t_1+t_\mathrm{enc}), \mu\right]$,
the flux integral along the loop
according to Eq.~(\ref{eq:phasedecay}). 
We neglect here the contribution of the flux inside the encounter 
region, which will be discussed in the next subsection.

When applying the sum rule (\ref{eq:Gzz}), we have to
use a modified survival probability 
$\exp\left[-(T-t_\mathrm{enc})/\tau_d\right]$
in Eq.~(\ref{eq:ergod}). 
{Indeed,} if the first stretch of the encounter lies {within the
billiard,} the second one {does so as well}, thus {the trajectory
does not risk to escape during its second passing through the encounter 
region.
The} relevant time for the survival probability has {therefore} to be 
reduced by the {duration $t_\mathrm{enc}$} of this second stretch.
{We then obtain}
\begin{eqnarray}
 \left\langle \left|\psi_{{\tilde{n}}}^{(0)}\right|^2 \right\rangle_\textrm{loop}
 &=&{ \left|\frac{m S_0}{\hbar}\right|^2 
   \frac{1}{\tau_Hp_{\tilde{n}}^\mathrm{l}(\mu)p_i^\mathrm{l}(\mu)}}
 \int_0^\infty dt_1 \int_0^\infty dt_2 \int_0^\infty dt_3 
 \exp\left(-\frac{t_1+t_2+t_3}{\tau_D}-\frac{t_2}{\tau_B}\right)
 \nonumber\\
 &&\times\int_{-c}^c ds \int_{-c}^c du 
 \frac{1}{\Sigma(\mu)t_\mathrm{enc}({su})} \exp\left(\frac{i}{\hbar}su\right)
 \exp\left(-\frac{t_\mathrm{enc}({su})}{\tau_D}
\right)\nonumber\\
 &=&-\frac{m}{p_{\tilde{n}}^\mathrm{l}(\mu)} j^\mathrm{i}
 \left(\frac{\tau_D}{\tau_H}\right)^2 \frac{1}{1+{\tau_D/\tau_B}} \,,
 \label{eq:g0loop}
\end{eqnarray} 
as shown in \ref{sec:encint}.
In a similar way as for nonlinearity blocks (see the discussion in Section 
\ref{sec:alt_approach}),
the encounter region cuts the diagram into three "links" 
contributing either $\tau_D$ or $\tau_D/(1+\tau_D/\tau_B).$

{The same derivation can be applied in order to calculate the loop
contributions to the reflection probability into channel $n$, leading to
exactly the same result as in Eq.~(\ref{eq:g0loop}).
We therefore obtain}
\begin{equation}
 \delta \left(t_{\tilde{n}i}^{(0)}\right)^\textrm{loop}
 {= \delta \left(r_{ni}^{(0)}\right)^\textrm{loop}}
 =-\left(\frac{\tau_D}{\tau_H}\right)^2\frac{1}{1+{\tau_D/\tau_B}} 
 \label{eq:loopg0}
\end{equation}
{as loop contributions to the reflection and
transmission probabilities $r_{ni}^{(0)}$ and $t_{\tilde{n}i}^{(0)}$
in the absence of interaction.
These corrections do} indeed restore current conservation in leading
semiclassical order.
Combining all ladder, crossed, and loop contributions that are evaluated 
in Eqs.~(\ref{eq:r1n}), (\ref{eq:r11}), and (\ref{eq:loopg0}), respectively, 
we obtain with Eq.~(\ref{eq:tauDtauH})
\begin{equation}
  r_{ii}^{{(0)}} = \frac{\tau_D}{\tau_H} \left( 1 + 
  \frac{\tau_H-\tau_D}{\tau_H} \frac{1}{1 + \tau_D / \tau_B} \right) 
  = \frac{1}{N_c+\tilde{N}_c} 
  \left( 1 + \frac{N_c+\tilde{N}_c - 1}{N_c+\tilde{N}_c}  \; 
  \frac{1}{1 + {B^2/B_0^2}} \right) \label{eq:r11wl}
\end{equation}
for the probability of retro-reflection into the incident channel $n=i$,
as well as
\begin{equation}
  r_{ni}^{{(0)}} = t_{\tilde{n}i}^{{(0)}} = \frac{\tau_D}{\tau_H} \left( 1 -
  \frac{\tau_D}{\tau_H} \frac{1}{1 + \tau_D / \tau_B} \right) =
  \frac{1}{N_c+\tilde{N}_c} 
  \left( 1 - \frac{1}{N_c+\tilde{N}_c}  \; \frac{1}{1 + {B^2/B_0^2}}
  \right)
  \label{eq:rn1wl}
\end{equation}
for the probabilit{ies of} reflection into a different channel 
$n \neq i$ {and of} transmission into channel $\tilde{n}$.
This yields the total reflection and transmission probabilities
\begin{eqnarray}
  {R}^{{(0)}} & {=} & {\frac{N_c}{N_c+\tilde{N}_c} + 
  \frac{\tilde{N}_c}{(N_c+\tilde{N}_c)^2} \; \frac{1}{1 + {B^2/B_0^2}}
  \, ,} \\
  {T}^{{(0)}} & {=} & {\frac{\tilde{N}_c}{N_c+\tilde{N}_c} -
  \frac{\tilde{N}_c}{(N_c+\tilde{N}_c)^2} \; \frac{1}{1 + {B^2/B_0^2}}
  \, ,}
\end{eqnarray}
in the linear case $g=0$, which obviously satisfy $R^{{(0)}}+T^{{(0)}}=1$.

\subsection{{Contributions of first order in the nonlinearity}}

\begin{figure}[t]
  \begin{center}
    \includegraphics[width=0.6\linewidth]{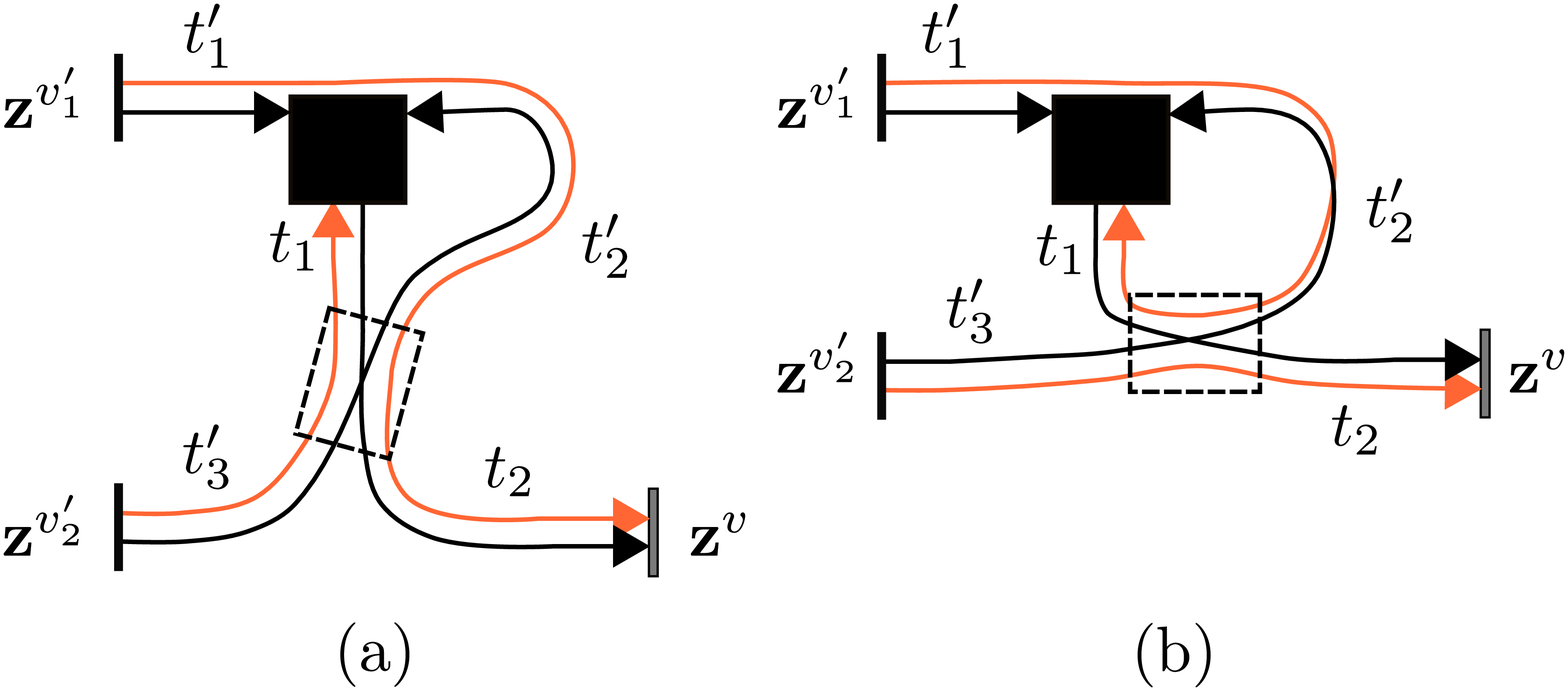}\\
    \includegraphics[width=0.8\linewidth]{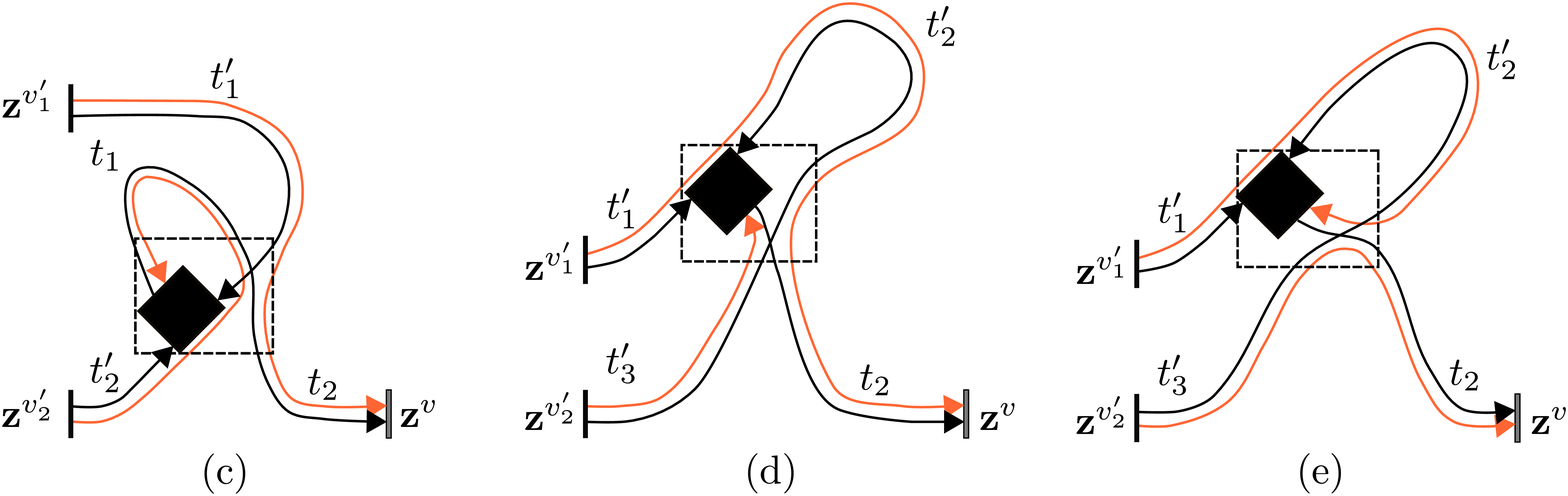}
  \end{center}
  \caption{\label{fig:firstorderloop}
    {Diagrammatic representation of the nonlinear loop contributions
      that arise in first order in the nonlinearity strength $g$.
      These diagrams involve encounters between different}
    trajectories rather than a self{-}encounter {of} one trajectory.
    {The upper and lower rows show, respectively, the two diagrams 
      (a) and (b) in which the nonlinearity block is located outside the 
      encounter region, as well as the three possibilities (c), (d), and
      (e) for} the nonlinearity event {to} enter the 
    encounter region.
    {The diagram (c), in which} the nonlinearity {block} moves 
    {along} a stretch through the whole encounter region{,}
    corresponds to the transition from diagram (a) to diagram (b).
    {Diagrams} {(d)} and {(e)}{, on the other hand,
      are obtained from diagrams (a) and (b), respectively, by pushing,
      in these latter two diagrams, the nonlinearity block along the 
      trajectory that starts at the block into the encounter region.}}
\end{figure}

In the case of nonvanishing interaction {between the atoms,} the
determination {of the loop contributions} to reflection and transmission
probabilities becomes more {involved due to} richer
possibilities {for associating correlated trajectories that exhibit small
action differences.}
{Loop contributions arise not only from}
self{-}encounters of single trajectories{,} but also {from} 
encounters of different trajectories in phase space. 
{This is illustrated} in Fig.~\ref{fig:firstorderloop} {which shows
the nonlinear diagrams that contribute to loop corrections of the
reflection and transmission probabilities in linear order in $g$.}
{As} it is quite {instructive, we begin} 
{our} {analysis of} loop corrections in the nonlinear case 
with the calculation of the contributions of {these diagrams. 
We shall first focus on diagrams (a) and (b) of 
Fig.~\ref{fig:firstorderloop} in which} the 
nonlinearity event can only move along parts of trajectories 
{that} are outside {the} encounter region{.} 

\begin{figure}[t]
  \begin{center}
    \includegraphics[width=0.8\linewidth]{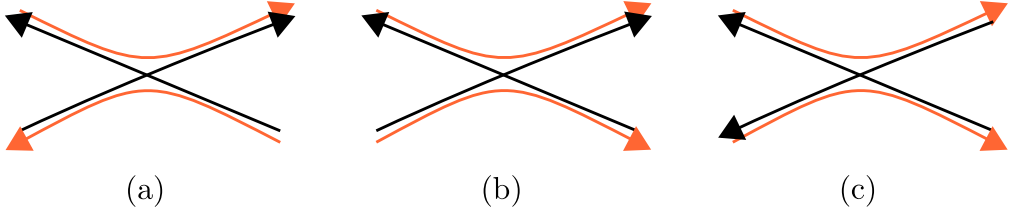}
  \end{center}
  \caption{\label{fig:encgauge} 
    Three different constellations for the stretches {involved} in an 
    encounter region. 
    {In the case of constellation (a), which arises also in the linear
      case discussed in Section \ref{sec:loop_linear}, the encounter region 
      itself does not provide a significant contribution to the dephasing
      in the presence of an external magnetic field, since}
    the flux integrals in the phases {associated with the individual
      stretches} cancel each other. 
    {This is different for the constellations} {(b)} and {(c)} 
    {in which} twice {(b)} and {four times (c)} the flux integral 
    from a stretch of duration $t_\mathrm{enc}$ remains, 
    {yielding {the phase factors} $\exp(-t_\mathrm{enc}/\tau_B)$ 
      {for constellation (b)} and 
      $\exp(-2 t_\mathrm{enc}/\tau_B)$ {for constellation (c)}.}}
\end{figure}

{As a starting point, we have to define the probability density $\tilde{w}$
for having a pair of trajectories $\gamma$ and $\gamma'$ that exhibit a
near-encounter in phase space{. 
This near-encounter results in the existence of an additional
pair of partner trajectories$\tilde{\gamma}$, $\tilde{\gamma'}$.}
In configuration space, the trajectories cross each other under a small angle
in {one of these two pairs, ($\gamma$, $\gamma'$) or 
($\tilde{\gamma}$, $\tilde{\gamma'}$){,} whereas the other pair
avoids this crossing.}}

{The probability density $\tilde{w}$ is specified for a given action 
difference $\Delta S$
between the trajectories $\gamma,\gamma'$ and the pair of partner
trajectories $\tilde{\gamma},\tilde{\gamma}'$, 
as well as for given partial propagation times $t$, $t'$ of the 
trajectories $\gamma$ and $\gamma'$, respectively, before 
(in the case of trajectory $\gamma$) or after 
(in the case of trajectory $\gamma'$) the {encounter region},
which may become relevant for the evaluation of magnetic dephasing.
It furthermore depends parametrically on the total propagation times
$T$ and $T'$ of the two trajectories $\gamma,\gamma'$ as well as on
the orientations of the individual trajectory stretches within the encounter 
region, as these orientations might give rise to additional contributions to
the magnetic dephasing.
Figure \ref{fig:encgauge} displays three different possibilities for 
orienting the trajectory stretches within the encounter region.
While the encounter region in constellation (a) does not contribute to
the dephasing in the presence of a magnetic field, the constellations (b)
and (c) contribute with phase factors $\exp(-t_\mathrm{enc}/\tau_B)$ and
$\exp(-2 t_\mathrm{enc}/\tau_B)$, respectively,
as there are one (b) and two (c) trajectory stretches 
\includegraphics[height=2ex]{lin_GF.eps} of the linear Green
function that are not balanced by the complex conjugate counterparts
\includegraphics[height=2ex]{lin_GF_conj.eps}.}

{Denoting by $n\in\{0,1,2\}$ the number of imbalanced pairs of stretches 
within the encounter region, we define}
\begin{eqnarray}
 {\tilde{w} (\Delta S,t,t'{;}T,T',n)}&=&
  {\int_0^\infty  \int_0^\infty  d \tilde{t} d \tilde{t}'
    \int_{-c}^c\int_{-c}^c ds du
  \frac{\delta\left(\Delta S-su\right)}{\Sigma(\mu)t_\mathrm{enc}(su)}
  \exp\left[-n\frac{t_\mathrm{enc}{(su)}}{\tau_B}\right]}\nonumber\\
  &&{\times\delta\left(t+\tilde{t}+t_\mathrm{enc}{(su)}-T\right)
    \delta\left(t'+\tilde{t}'+t_\mathrm{enc}{(su)}-T'\right)}
 \label{eq:partnerdensity2}
\end{eqnarray}
as the density of trajectory pairs $\gamma$, $\gamma'$
{that come close to} each other in {phase space} 
{and thus {have} partner trajectories 
with a combined action difference $\Delta S$.
In this} {expression, the integration variables $\tilde{t}$ and 
$\tilde{t}'$ correspond to the propagation times of the final and 
initial parts of the trajectories $\gamma$ and $\gamma'$, respectively, 
after leaving ($\tilde{t}$) and before entering ($\tilde{t}'$) 
the encounter region.
We have $n=0\equiv n_a$ for diagram (a) and $n=1 \equiv n_b$ 
for diagram (b) in Fig.~\ref{fig:firstorderloop}.}

{Calculating now the contributions of the diagrams shown in
the upper row of Fig.~\ref{fig:firstorderloop} 
(which are multiplied by a combinatorial factor 2 as there are 
two possibilities to construct these diagrams), we obtain}
\begin{eqnarray}
  \lefteqn{\left\langle
    \psi_{\tilde{n}}^{(0)*}\left(\delta\psi_{\tilde{n}}^{(1)}\right)
    \right\rangle_{\textrm{loop, {(a{/b})}}} =
    2 \frac{\hbar^2}{2m} \frac{\left|S_0\right|^4\left(\pi\hbar\right)^3}
    {\tilde{W} W^2} \left\langle \sum_{\nu_1',\nu_2',\nu=\pm 1}
    \AnfangEnde{G}^{*}\left(\mathbf{z}_{\tilde{n}}^{\nu}, \mathbf{z}_{i}^{\nu_1'},\mu\right)
    \right.} \nonumber\\
  &&\times\left.
  \int_\Omega d^2 r'
  \Ende{G}\left(\mathbf{z}_{\tilde{n}}^{\nu},\mathbf{r}',\mu\right)
  \Anfang{G}\left(\mathbf{r}',\mathbf{z}_{i}^{\nu_1'},\mu\right)
  \Anfang{G}\left(\mathbf{r}',\mathbf{z}_{i}^{\nu_2'},\mu\right)
  \Anfang{G}^{*}\left(\mathbf{r}',\mathbf{z}_{i}^{\nu_2'},\mu\right) 
  \right\rangle_{\textrm{loop, {(a{/b})}}} \label{eq:1_o_loop_a1} 
  \nonumber \\ 
  &=& 2 \frac{\hbar^2}{2m} 
  \frac{\left|S_0\right|^4\left(\pi\hbar\right)^3}{\tilde{W}W^2} 
  {\int_{-\infty}^\infty d(\Delta S)\int_0^\infty dt_1 
      \int_0^\infty d{t_2'}
      \sum_{\nu_1',\nu_2',\nu=\pm 1} \sum_{\gamma'}
      \left\langle\left|\AnfangEnde{A}_{\gamma'}
      \left(\mathbf{z}_{i}^{\nu_2'},\mathbf{z}_{i}^{\nu_1'},\mu \right)\right|^2\right\rangle}
      \nonumber\\
    && {\times \left(-\frac{i}{\hbar}\right)\int_0^{{\infty}} dt'_1
      \sum_{\gamma} \left\langle\left| \Ende{A}_{\gamma}
      \left[\mathbf{z}_{\tilde{n}}^{\nu},\mathbf{r}_{\gamma'}\left(t'_1\right),\mu\right]
      \right|^2\right\rangle}
    \nonumber\\
    &&{\times \tilde{w}
      \left(\Delta S,t_1,{t_2';}T_{\gamma},T_{\gamma'}{-t_1'},
      {n_{a/b}}\right)
      \exp\left(\frac{i}{\hbar}\Delta S\right)
      \exp\left(-{\frac{t_1 + t_2'}{\tau_B}}\right)}
  \label{eq:1_o_loop_a2} \nonumber \\
   &=&-i\frac{m}{p_{\tilde{n}}^\mathrm{l}(\mu)}
     \left(\frac{m\left|S_0\right|^2}{\hbar^2 p_i^\mathrm{l}(\mu)}\right)^2 
     \frac{1}{\tau_H^2} \int_0^\infty dt_1' \int_0^\infty dt_2' 
     \int_0^\infty dt_3' \int_0^\infty dt_1 \int_0^\infty dt_2\nonumber\\
   &&\times \exp\left(-\frac{t_1'+t_2'+t_3'+t_1+t_2}{\tau_D}
     -\frac{t_1+t_2'}{\tau_B}\right)\nonumber\\
   &&\times\int_{-c}^c ds\int_{-c}^c du
     \frac{1}{\Sigma(\mu)t_\mathrm{enc}{(su)}}
     \exp\left(\frac{i}{\hbar}su\right)
     \exp\left[-\frac{t_\mathrm{enc}{(su)}}{\tau_D}
       {- n_{a/b}\frac{t_\mathrm{enc}{(su)}}{\tau_B}} 
       \right] \, {,} \label{eq:1_o_loop_a3}
\end{eqnarray}
{where we applied the sum rules (\ref{eq:Gzr}, \ref{eq:Gzz})} 
to convert the sums over classical trajectories $\gamma$, $\gamma'$ into
integrations over trajectory durations $T_\gamma$, $T_{\gamma'}${,
taking into account that} we have to use a reduced effective time
$T_\gamma+T_{\gamma'}-t_\mathrm{enc}$ for the classical survival probability.
Gauge invariance of the result is ensured by the fact that the
encounter region closes the overall flux integral.
The integration over $s$ and $u$ is calculated in
\ref{sec:encint} and yields $-1/(\tau_D\tau_H)$ for $n=n_a=0$
as well as $-\left(1+\tau_D/\tau_B\right)/(\tau_D\tau_H)$ for $n=n_b=1$.
We therefore obtain
\begin{eqnarray}
  \left\langle
  \psi_{\tilde{n}}^{(0)*}\left(\delta\psi_{\tilde{n}}^{(1)}\right)
  \right\rangle_{\textrm{loop, {(a)}}} & = &
  i \frac{m}{p_{\tilde{n}}^\mathrm{l}(\mu)} \left(j^\mathrm{i}\right)^2
  \left(\frac{\tau_D}{\tau_H}\right)^3 \tau_D
  \left(\frac{1}{1+\tau_D/\tau_B}\right)^2\,{,}
  \label{eq:1_o_loop_a4} \\
  \left\langle \psi_{\tilde{n}}^{(0)*}\left(\delta\psi_{\tilde{n}}^{(1)}\right)
    \right\rangle_{\textrm{loop, {(b)}}}
  & = & i\frac{m}{p_{\tilde{n}}^\mathrm{l}(\mu)}\left(j^\mathrm{i}\right)^2
    \left(\frac{\tau_D}{\tau_H}\right)^3 \tau_D
    \left(\frac{1}{1+\tau_D/\tau_B}\right)
    \label{eq:1_o_loop_b4} 
 \end{eqnarray}
{for the cases where the nonlinearity block is located outside the
encounter region.}

As shown in the lower row of Fig.~\ref{fig:firstorderloop}, there are
two qualitatively different possibilities for the nonlinearity event to enter
the encounter region. 
In the first {scenario}{, depicted in 
Fig.~\ref{fig:firstorderloop}(c),} the nonlinearity event moves along a 
trajectory {that} provides a stretch {within} the encounter region.
{This scenario corresponds to the transition from diagram (a) to
  diagram (b).}
{Its contribution is calculated as}
\begin{eqnarray}
  &&\lefteqn{\left\langle \psi_{\tilde{n}}^{(0)*}\left(\delta\psi_{\tilde{n}}^{(1)}\right) 
  \right\rangle_{\textrm{loop, {(c)}}} =} \nonumber\\
  &=& 2 \frac{\hbar^2}{2m} 
  \frac{\left|S_0\right|^4\left(\pi\hbar\right)^3}{\tilde{W}W^2} 
  {\int_{-\infty}^\infty d(\Delta S)\int_0^\infty dt_1 \int_0^\infty dt'_1
      \sum_{\nu_1',\nu_2',\nu=\pm 1} 
      \sum_{\gamma'}
      \left\langle\left|\AnfangEnde{A}_{\gamma'}\left(\mathbf{z}_{i}^{\nu_2'},\mathbf{z}_{i}^{\nu_1'},\mu
      \right)\right|^2\right\rangle
      }\nonumber\\
    && {\times \left(-\frac{i}{\hbar}\right)
      \int_{{0}}^{{t_\mathrm{enc}(\Delta S)}} dt'\sum_{\gamma}
      \left\langle\left| \Ende{A}_{\gamma}
      \left[\mathbf{z}_{\tilde{n}}^{\nu},\mathbf{r}_{\gamma'}(t_1'{+t'}),\mu\right]
      \right|^2\right\rangle}
    \nonumber\\
    &&{\times \tilde{w}
      \left(\Delta S,t_1,t'_1{;}T_{\gamma},T_{\gamma'},0\right)
      \exp\left(\frac{i}{\hbar}\Delta S\right)
      \exp\left(-{\frac{t_1+t'}{\tau_B}}\right)
      } \label{eq:1_o_loop_c1} \nonumber \\
  &=&-i
    \frac{m}{p_{\tilde{n}}^\mathrm{l}(\mu)} \left(j^\mathrm{i}\right)^2 
    \frac{1}{\tau_H^2} \int_0^\infty dt_1' \int_0^\infty dt_2'
    \int_0^\infty dt_1 \int_0^\infty dt_2\nonumber\\
  &&\times\exp\left(-\frac{t_1'+t_2'+t_1+t_2}{\tau_D}
    -\frac{t_1+t_2'}{\tau_B}\right) \nonumber\\
  &&\times\int_{-c}^c ds \int_{-c}^c du
    \frac{1}{\Sigma(\mu)t_{\mathrm{enc}}{(su)}}
    \exp\left(\frac{i}{\hbar}\right)
    \exp\left(-\frac{t_\mathrm{enc}{(su)}}{\tau_D}\right)
      \int_0^{t_{\mathrm{enc}}{(su)}}dt'
      \exp\left(-\frac{t'}{\tau_B}\right) \nonumber\\
  &=& - i \frac{m}{p_{\tilde{n}}^\mathrm{l}(\mu)} \left(j^\mathrm{i}\right)^2
    \left(\frac{\tau_D}{\tau_H}\right)^3 \tau_D
    \left(\frac{1}{1+\tau_D/\tau_B}\right) {= -
    \left\langle \psi_{\tilde{n}}^{(0)*}\left(\delta\psi_{\tilde{n}}^{(1)}\right)
    \right\rangle_{\textrm{loop, (b)}}}
  \label{eq:1_o_loop_c2}
\end{eqnarray}
[see Fig.~\ref{fig:firstorderloop}(c) for the signification of 
$t_1,t_2,t'_1,t_2'$], where 
the integrations over $s$, $u$ and over the propagation time
$t'$ within the encounter region (whose gauge field dependence is taken
into account in the integration) yield, as shown in
\ref{sec:enc_nonlinint}, $1/\tau_H$.

{In the other scenario, depicted in Fig.~\ref{fig:firstorderloop}(d) 
and (e), a trajectory pair leaving the encounter and ending at a nonlinearity 
event becomes arbitrarily small until finally the nonlinearity event enters 
the encounter region but does not traverse it. 
This case} requires a modification of the {probability}
density of suitable partner trajectories as some stretches do not leave 
the encounter {region} any more but {terminate} at a certain point 
{within} it. 
{Following Refs.~\cite{BroRah06PRB,WalGutGouRic08PRL,GutWalKuiRic09PRE},
we define} a reduced encounter region {with the duration
\begin{equation}
  \bar{t}_\mathrm{enc}({\tilde{t}'},u) \equiv 
  {\tilde{t}'}+\frac{1}{\lambda}\ln\left(\frac{c}{|u|}\right)
\end{equation}
where}
${\tilde{t}'}\in[0,(1/\lambda)\ln(c/|s|)]$ {is} the time 
{interval} between the nonlinearity {event} and the 
{Poincar\'e surface of section} $\mathcal{P}$ {that is optimally
chosen to be located in the center of the encounter region.}
As a consequence{,} we have to extend the integration
over $s$ and $u$, associated with the possible action differences $su$, 
by an integration over all possible {time spans ${\tilde{t}'}$
defining the location of the nonlinearity event with respect to 
$\mathcal{P}$, which} substitutes one of the integrations over time in
{Eq.}~(\ref{eq:partnerdensity2}){.
This yields} {the modified {probability} density}
\begin{eqnarray}
 {\bar{w}(\Delta S,t,t'{;}T,T',n)}&=&
  { \int_{-c}^c\int_{-c}^c ds du \int_0^\infty  d\tilde{t} \int_0^{(1/\lambda)\ln(c/|s|)} \hspace{-2ex}d\tilde{t}'
  \frac{\delta\left(\Delta S-su\right)}{\Sigma(\mu)\bar{t}_\mathrm{enc}(\tilde{t}',u)}\exp\left(-n\frac{\bar{t}_\mathrm{enc}(\tilde{t}',u)}{\tau_B}\right)}\nonumber\\
  &&{\times\delta\left[t+\tilde{t}+\bar{t}_\mathrm{enc}(\tilde{t}',u)-T\right]\delta\left[t'+\bar{t}_\mathrm{enc}(\tilde{t}',u)-T'\right]}\,.
 \label{eq:partnerdensity_3}
\end{eqnarray}
Using this density for the calculation of diagram (d) in 
Fig.~\ref{fig:firstorderloop}, we  obtain
\begin{eqnarray}
  \lefteqn{\left\langle
      \psi_{\tilde{n}}^{(0)*}\left(\delta\psi_{\tilde{n}}^{(1)}\right)
    \right\rangle_{\textrm{loop, {(d)}}} =} \nonumber\\
 &=& 2 \frac{\hbar^2}{2m} 
    \frac{\left|S_0\right|^4\left(\pi\hbar\right)^3}{\tilde{W}W^2} 
     {\int_{-\infty}^\infty d(\Delta S)\int_0^\infty dt_2 \int_0^\infty d
       {t_2'} \sum_{\nu_1',\nu_2',\nu=\pm 1} \sum_{\gamma'}
    \left\langle\left|\AnfangEnde{A}_{\gamma'}
    \left(\mathbf{z}_{i}^{\nu_2'},\mathbf{z}_{i}^{\nu_1'},\mu\right)\right|^2\right\rangle}
     \nonumber\\
  && {\times \left(-\frac{i}{\hbar}\right)\int_0^{{\infty}} dt'_1
       \sum_{\gamma} \left\langle\left|\Ende{A}_{\gamma}
       \left[\mathbf{z}_{\tilde{n}}^{\nu},\mathbf{r}_{\gamma'}\left(t'_1\right),\mu\right]
       \right|^2\right\rangle}
    \nonumber\\
  &&{\times \bar{w}
      \left(\Delta S,t_2,{t_2'}{;}T_{\gamma},T_{\gamma'}{-t_1'},0\right)
      \exp\left(\frac{i}{\hbar}\Delta S\right)
      \exp\left(-\frac{{t_2'}}{\tau_B}\right)} \label{eq:1_o_loop_d1} 
    \nonumber \\
    &=& -i \frac{m}{p_{\tilde{n}}^\mathrm{l}(\mu)} \left(j^\mathrm{i}\right)^2
    \frac{1}{\tau_H^2} 
    \int_0^\infty dt_2 \int_0^\infty dt_2'\int_0^\infty dt_1' \int_0^\infty  dt_3'
    \exp\left(-\frac{t_1'+t_2'+t_3'+t_2}{\tau_D}-\frac{t_2'}{\tau_B}\right)
    \nonumber\\
  &&\times\int_{-c}^c ds \int_{-c}^c du
    \int_0^{(1/\lambda)\ln\left(c/|s|\right)} dt_1
    \frac{1}{\Sigma(\mu)\bar{t}_\mathrm{enc}(t_1,u)} \exp\left(\frac{i}{\hbar}su\right)
    \exp\left(-\frac{\bar{t}_\mathrm{enc}(t_1,u)}{\tau_D}\right)
    \nonumber\\
  &=&- i \frac{m}{p_{\tilde{n}}^\mathrm{l}(\mu)} \left(j^\mathrm{i}\right)^2
    \left(\frac{\tau_D}{\tau_H}\right)^3 \tau_D
    \left(\frac{1}{1+\tau_D/\tau_B}\right)
    {= \left\langle \psi_{\tilde{n}}^{(0)*}\left(\delta\psi_{\tilde{n}}^{(1)}\right)
    \right\rangle_{\textrm{loop, (c)}}}
  \label{eq:1_o_loop_d2}
\end{eqnarray}
where we evaluate the integration over $s$, $u$ and $t_1$ in 
\ref{sec:enc_nonlinint} yielding $1/\tau_H$.
The calculation of the contribution of diagram (e) in 
Fig.~\ref{fig:firstorderloop} proceeds in perfect analogy with the
one presented for diagram (d) and yields the same result 
\begin{equation}
 \left\langle \psi_{\tilde{n}}^{(0)*}\left(\delta\psi_{\tilde{n}}^{(1)}\right)
  \right\rangle_{\textrm{loop, {(e)}}}
  =- i \frac{m}{p_{\tilde{n}}^\mathrm{l}(\mu)} \left(j^\mathrm{i}\right)^2
\left(\frac{\tau_D}{\tau_H}\right)^3 \tau_D
\left(\frac{1}{1+\tau_D/\tau_B}\right)
{= \left\langle \psi_{\tilde{n}}^{(0)*}\left(\delta\psi_{\tilde{n}}^{(1)}\right)
    \right\rangle_{\textrm{loop, (c)}} \, .}
\end{equation}

The overall {nonlinear loop} contribution to 
$\langle \psi_{\tilde{n}}^{(0)*}(\delta\psi_{\tilde{n}}^{(1)})\rangle$ 
originating from {the} diagrams  shown in 
Fig.~\ref{fig:firstorderloop} {therefore} sums up to
\begin{equation}
 \left\langle \psi_{\tilde{n}}^{(0)*}\left(\delta\psi_{\tilde{n}}^{(1)}\right)
  \right\rangle_{\textrm{loop}}
 = i \frac{m}{p_{\tilde{n}}^\mathrm{l}(\mu)} \left(j^\mathrm{i}\right)^2
  \left(\frac{\tau_D}{\tau_H}\right)^3 \tau_D\left[
  \left(\frac{1}{1+\tau_D/\tau_B}\right)^2-2\left(\frac{1}{1+\tau_D/\tau_B}\right)
  \right]\, {.}
\end{equation}
{As this expression} is purely imaginary{, no modifications of} 
transmission and reflection probabilities {are expected in linear order
in the interaction strength $g$, which is in perfect accordance 
with the discussion in Section \ref{sec:crossed}}
[{see Eq.~(\ref{eq:back_refl_o1})}].

\subsection{Contributions of arbitrary order in the nonlinearity}

As {we can see from the above calculations,
the presence of an encounter region perfectly fits to
the picture of diagrams consisting of separated parts, 
which was developed in Section \ref{sec:alt_approach}},
{since we can perform the integrations corresponding to an encounter 
region independently from the remaining integrations over link durations.
{Encounter regions can, in the spirit of Section \ref{sec:alt_approach},
be interpreted as extended ``blocks'' which are connected via four links
to other (nonlinearity or encounter) blocks as well as to the leads of the 
system.
Care must be taken, though, if a nonlinearity block enters the encounter 
region, as is the case in the diagrams depicted in the lower row of 
Fig.~\ref{fig:firstorderloop}.
Under} consideration of the calculations performed in 
\ref{sec:loopappendix}, we can extend our diagrammatic rules listed in 
Section \ref{sec:alt_approach} by the following ones:
\begin{enumerate}[(1)]
 \setcounter{enumi}{\value{mycounter}}
 \item each encounter region {containing no} nonlinearity event 
   contributes a factor 
   $-(1+n\tau_B/\tau_D)/(\tau_D\tau_H)$ where $n=0$, $1$, $2$ 
   {counts the number of trajectory pairs with imbalanced stretches
   within the encounter region} (see Fig.~\ref{fig:encgauge}){;}
 \item each encounter region including a nonlinearity event contributes a 
   factor $1/\tau_H$.
\end{enumerate}
{As was already argued in Section} \ref{sec:alt_approach}, 
{diagrams containing a nonlinearity block with ladder-type input
(as the ones shown in the left and central column of 
Fig.~\ref{fig:basicblocks}) will not contribute to the reflection and
transmission probabilities as they are canceled by counterparts in which
this nonlinearity block is attached to the complex conjugate trajectory
stretch.
We can therefore restrict ourselves to ``crossed type'' nonlinearity blocks
shown in} {the right column in Fig.~\ref{fig:basicblocks}.}}

\begin{figure}[p]
  \begin{center}
    \includegraphics[width=0.8\linewidth]{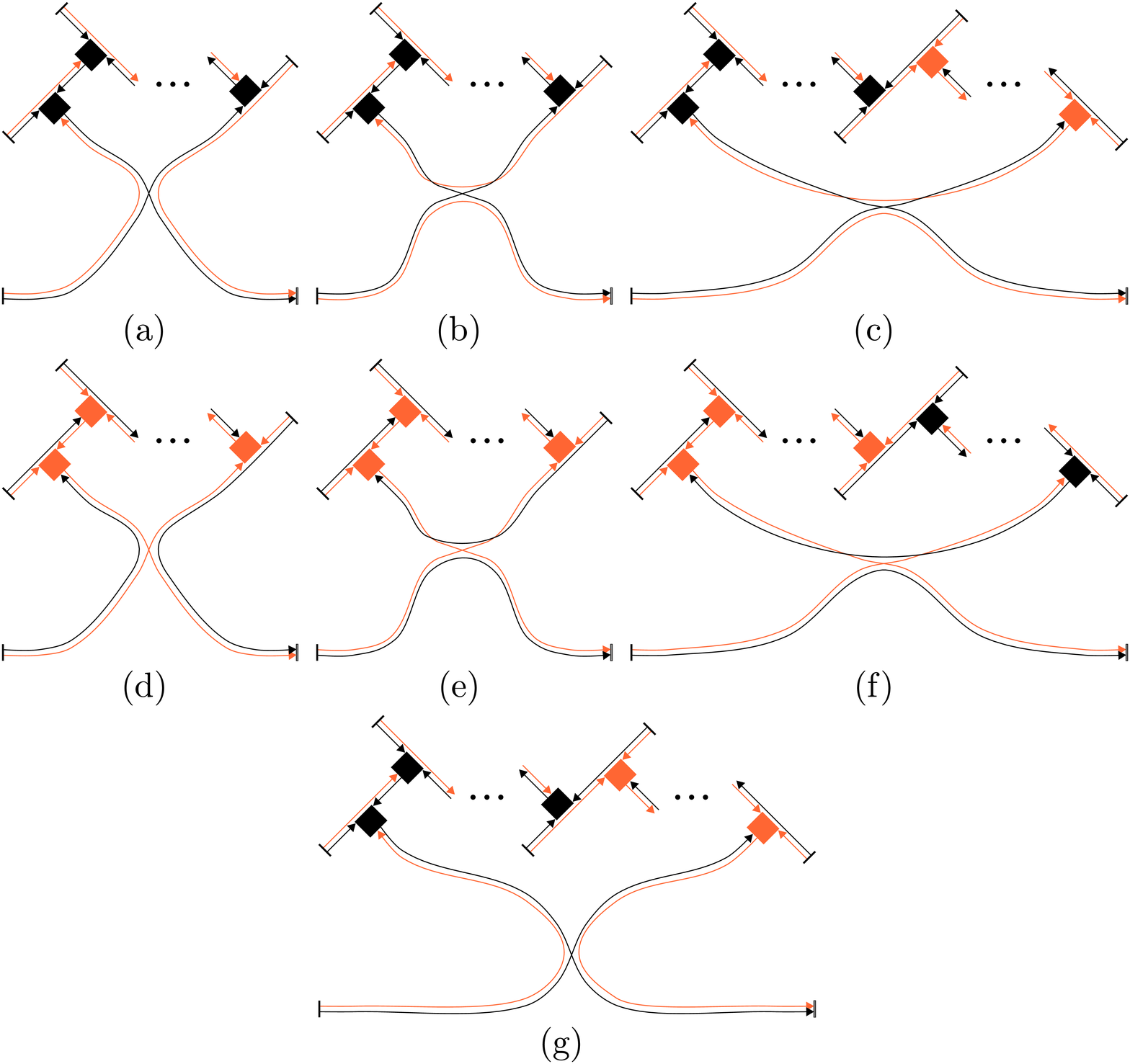}
  \end{center}
  \caption{\label{fig:kthorderloops} 
    Complete set of diagrams of $k$-th order in $g$
    containing one encounter region and $k$ {nonlinearity} 
    blocks of {the ``crossed'' type shown in} the right column of 
    Fig.~\ref{fig:basicblocks} {which are located outside the 
      encounter region.}
    The diagrams {(a)} - {(f)} are contributions coming from
    $\langle\psi_{{\tilde{n}}}^{(0)*}(\delta\psi_{{\tilde{n}}}^{(k)})\rangle$ and
    $\langle(\delta\psi_{{\tilde{n}}}^{(k)*})\psi_{{\tilde{n}}}^{(0)}\rangle$, 
    respectively, where (c) and (f) only exist for $k\geq2$.
    Contrary to the {other diagrams,} (g) is contained in
    $\langle(\delta\psi_{{\tilde{n}}}^{(k_1)*})
    (\delta\psi_{{\tilde{n}}}^{(k_2)})\rangle$ with $k_1+k_2=k$ {and}
    $k_1,k_2\geq1$.
    {As is shown in Eq.~(\ref{eq:bcef}), the contributions of the
      diagrams (b), (c), (e), and (f) exactly cancel each other.}
  }
\end{figure}

{Figure \ref{fig:kthorderloops} shows the complete set of nonlinear diagrams
contributing to reflection and transmission probabilities in arbitrary order 
in the nonlinearity strength $g$.
These diagrams consist of the same chains of nonlinearity blocks as the ones
shown in Fig.~\ref{fig:kth_refl}, with the main difference that these chains 
cannot be directly attached to a lead as we are calculating the scattering 
amplitude to an arbitrary (i.e., not necessarily the incident) channel in 
the reflected or transmitted lead;
instead, they have to be connected to ladder links via an encounter region.
In perfect analogy with Fig.~\ref{fig:firstorderloop}, the nonlinearity 
blocks at the right ends of the chains in the diagrams (a), (b), (d), and (e)
can be moved all the way through the entire encounter region, thereby giving 
rise to a transition from diagram (a) to diagram (b) as well as from
diagram (d) to diagram (e). 
The other blocks located at the ends of the chains in the diagrams (a) -- (g)
can, in analogy with the diagrams (d) and (e) of 
Fig.~\ref{fig:firstorderloop}, be pushed into the encounter region by moving 
them along the trajectories that start or end at those blocks.}

{Using the diagrammatic rules {(1) -- (8) stated in Section 
\ref{sec:alt_approach} and above,}
the contributions of these relevant diagrams as well as their corrections
{due to nonlinearity events entering the encounter region}
are {straightforwardly} evaluated. 
Defining by
\begin{equation}
  {g_0 \equiv \left(j^i\tau_Dc_{{ii}}^{(0)}\right)^{-1}
  = \left(j^\mathrm{i}\tau_D\frac{\tau_D}{\tau_H}\frac{1}{1+\tau_D/\tau_B}\right)^{-1}}
\end{equation}
the {relevant scale for the nonlinearity strength appearing} in
Eq.~(\ref{eq:c11gendresult}), we obtain for the diagrams (a) {--} (g)}
\begin{eqnarray}
  \left\langle\psi^{(0)*}_{{\tilde{n}}} \left(\delta\psi^{(k)}_{{\tilde{n}}}\right)
    \right\rangle_{\textrm{loop, {(a)}}}
  &=&\left[\left\langle \psi^{(0)}_{{\tilde{n}}} \left(\delta\psi^{(k)}_{{\tilde{n}}}\right)^{*}
    \right\rangle_{\textrm{loop,  {(d)}}}\right]^{*}\nonumber\\
  &=&-\frac{m}{p_{\tilde{n}}^\mathrm{l}(\mu)} j^\mathrm{i} \left(\frac{\tau_D}{\tau_H}\right)^2
    \frac{1}{1+\tau_D/\tau_B} \left(-\frac{i}{g_0}\right)^k\, ,\\
  \left\langle\psi^{(0)*}_{{\tilde{n}}} \left(\delta\psi^{(k)}_{{\tilde{n}}}\right)
    \right\rangle_{\textrm{loop, {(b)}}} 
  &=&\left[\left\langle \psi^{(0)}_{{\tilde{n}}} \left(\delta\psi^{(k)}_{{\tilde{n}}}\right)^{*}
    \right\rangle_{\textrm{loop, {(e)}}}\right]^{*}\nonumber\\
  &=&-\frac{m}{p_{\tilde{n}}^\mathrm{l}(\mu)} j^\mathrm{i} \left(\frac{\tau_D}{\tau_H}\right)^2
    \left(-\frac{i}{g_0}\right)^k\, ,\\
  \left\langle\psi^{(0)*}_{{\tilde{n}}} \left(\delta\psi^{(k_1+k_2)}_{{\tilde{n}}}\right)
    \right\rangle_{\textrm{loop, {(c)}}} 
  &=&\left[\left\langle \psi^{(0)}_{{\tilde{n}}} \left(\delta\psi^{(k_1+k_2)}_{{\tilde{n}}}\right)^{*}
    \right\rangle_{\textrm{loop, {(f)}}}\right]^{*}\nonumber\\
  &=&-\frac{m}{p_{\tilde{n}}^\mathrm{l}(\mu)} j^\mathrm{i} \left(\frac{\tau_D}{\tau_H}\right)^2
   {\left(-\frac{i}{g_0}\right)^{k_1}\left(\frac{i}{g_0}\right)^{k_2}}\,,\\
  {\left\langle\left(\delta\psi_{{\tilde{n}}}^{(k_1)*}\right)
    \left(\delta\psi_{{\tilde{n}}}^{(k_2)}\right)
    \right\rangle_\textrm{loop, (g)}}
  &=&{- \frac{m}{p_{\tilde{n}}^\mathrm{l}(\mu)} j^\mathrm{i} \left(\frac{\tau_D}{\tau_H}\right)^2
    \frac{1}{1+\tau_D/\tau_B}
    {\left(-\frac{i}{g_0}\right)^{k_1}\left(\frac{i}{g_0}\right)^{k_2}}}
\end{eqnarray}
{where $\delta\psi^{(k_1+k_2)}_{{\tilde{n}}}$ 
denotes the $k$-th order contribution to $\psi_{{\tilde{n}}}$, 
whose diagrammatic representations contain $k_1$ 
{nonlinearity blocks of the type}
\includegraphics[height=2ex]{vertex.eps}
and $k_2$ {complex conjugate nonlinearity blocks}
\includegraphics[height=2ex]{complex_vertex.eps} .
The corrections to these contributions} 
{due to nonlinearity events entering the encounter region}
are calculated as
\begin{eqnarray}
  \left\langle\psi^{(0)*}_{{\tilde{n}}} \left(\delta\psi^{(k)}_{{\tilde{n}}}\right)
    \right\rangle_{\textrm{loop, {(a)}, correct}} 
  &=& \left[\left\langle \psi^{(0)}_{{\tilde{n}}} \left(\delta\psi^{(k)}_{{\tilde{n}}}\right)^{*} 
    \right\rangle_{\textrm{loop,  {(d)}, correct}}\right]^{*} 
    \nonumber\\ 
  &&  \nonumber \\
  {= 2} \left\langle\psi^{(0)*}_{{\tilde{n}}} \left(\delta\psi^{(k)}_{{\tilde{n}}}\right)
    \right\rangle_{\textrm{loop, {(b)}, correct}}
  &=&{2} \left[\left\langle \psi^{(0)}_{{\tilde{n}}}\left(\delta\psi^{(k)}_{{\tilde{n}}}\right)^{*}
    \right\rangle_{\textrm{loop, {(e)}, correct}}\right]^{*} \nonumber\\
  &=&  {2}\frac{m}{p_{\tilde{n}}^\mathrm{l}(\mu)}j^\mathrm{i}\left(\frac{\tau_D}{\tau_H}\right)^2
    \left(-\frac{i}{g_0}\right)^k\, ,\\
  \left\langle\psi^{(0)*}_{{\tilde{n}}} \left(\delta\psi^{(k_1+k_2)}_{{\tilde{n}}}\right)
    \right\rangle_{\textrm{loop, {(c)}, correct}} 
  &=& \left[\left\langle \psi^{(0)}_{{\tilde{n}}} 
    \left(\delta\psi^{(k_1+k_2)}_{{\tilde{n}}}\right)^{*} 
    \right\rangle_{\textrm{loop, {(f)}, correct}}\right]^{*} \nonumber\\
{=\left\langle\left(\delta\psi_{{\tilde{n}}}^{(k_1)*}\right)
    \left(\delta\psi_{{\tilde{n}}}^{(k_2)}\right) 
    \right\rangle_\textrm{loop, (g), correct}}
  &=& 2 \frac{m}{p_{\tilde{n}}^\mathrm{l}(\mu)} j^\mathrm{i}
    \left(\frac{\tau_D}{\tau_H}\right)^2
    {\left(-\frac{i}{g_0}\right)^{k_1}\left(\frac{i}{g_0}\right)^{k_2}}
    \,{.}\\
    &&  \nonumber
\end{eqnarray}

By multiplying these {individual contributions} with the corresponding 
powers of $g$ and then summing over all orders $k$, $k_1$ and $k_2$
{in analogy with Eq.~(\ref{eq:c11alt})}, 
we finally obtain 
\begin{equation}
  {\left\langle|\psi_{{\tilde{n}}}|^2\right\rangle_{\textrm{loop, (a) $+$ (d)}}
    = - 2 \left\langle|\psi_{{\tilde{n}}}|^2\right\rangle_\textrm{loop, (g)}
    = 2\frac{m j^\mathrm{i}}{p_{\tilde{n}}^\mathrm{l}(\mu)}
    \frac{(\tau_D/\tau_H)^2}{1+\tau_D/\tau_B} 
    \frac{(g/g_0)^2}{1+(g/g_0)^2}}
\end{equation}
{for the summed contributions of the diagrams (a) and (d) as well as for
the contributions of the diagrams (g), and}
\begin{equation}
  {\left\langle|\psi_{{\tilde{n}}}|^2\right\rangle_{\textrm{loop, (b) $+$ (e)}}
  = - \left\langle|\psi_{{\tilde{n}}}|^2\right\rangle_{\textrm{loop, (c) $+$ (f)}}
  = 2\frac{mj^\mathrm{i}}{p_{\tilde{n}}^\mathrm{l}(\mu)} 
    \left(\frac{\tau_D}{\tau_H}\right)^2 
    \frac{(g/g_0)^2}{1+(g/g_0)^2}} \label{eq:bcef}
\end{equation}
{for the summed contributions of the diagrams (b) and (e) as well as of
(c) and (f), which implies that the contributions of these latter four
diagrams exactly cancel each other.}
The summation of the associated corrections due to nonlinearity blocks
entering the encounter region  yields {the contributions}
\begin{equation}
{- \left\langle|\psi_{{\tilde{n}}}|^2\right\rangle_{\textrm{loop, (a) $+$(d),
      correct}} = \left\langle|\psi_{{\tilde{n}}}|^2\right\rangle_{\textrm{loop, 
      (c) $+$ (f), correct}} = \frac{4 mj^\mathrm{i}}{p_{\tilde{n}}^\mathrm{l}(\mu)} 
    \left(\frac{\tau_D}{\tau_H}\right)^2
    \frac{(g/g_0)^2}{1+(g/g_0)^2} \,,}
\end{equation}
\begin{equation}
{-\left\langle|\psi_{{\tilde{n}}}|^2\right\rangle_{\textrm{loop, (b) $+$ (e), 
      correct}} = \left\langle|\psi_{{\tilde{n}}}|^2\right\rangle_\textrm{loop, 
    (g), correct} = \frac{2mj^\mathrm{i}}{p_{\tilde{n}}^\mathrm{l}(\mu)} 
    \left(\frac{\tau_D}{\tau_H}\right)^2
    \frac{(g/g_0)^2}{1+(g/g_0)^2} \,,}
\end{equation}
{which exactly cancel each other as well.
In effect, therefore, only the diagrams (a), (d), and (g) provide nonvanishing
contributions to the reflection or transmission probabilities, which are
summed up as}
\begin{eqnarray}
  \left\langle|\psi_{{\tilde{n}}}|^2\right\rangle_\textrm{loop, {{(a)}$+${(d)}$+$ {(g)}}}
  &=&\frac{m}{p_{\tilde{n}}^\mathrm{l}(\mu)} j^\mathrm{i} 
    {\left(\frac{\tau_D}{\tau_H}\right)^2}\frac{1}{1+\tau_D/\tau_B}
    \left(\frac{g}{g_0}\right)^2 \frac{1}{1+(g/g_0)^2}\, .
\end{eqnarray}

Together with the result obtained in the linear case{,} 
the overall loop {contribution} to $\langle|\psi_{{\tilde{n}}}|^2\rangle$ 
{reads}
\begin{eqnarray}
  \left\langle|\psi_{{\tilde{n}}}|^2\right\rangle_{\textrm{loop}}
  &=&\left\langle|\psi_{{\tilde{n}}}^{(0)}|^2\right\rangle_{\textrm{loop}}
    +\left\langle|\psi_{{\tilde{n}}}|^2\right\rangle_{\textrm{loop, {{(a)} $+${(d)}$+$ {(g)}}}}
    \nonumber\\
  &=&-\frac{m}{p_{\tilde{n}}^\mathrm{l}(\mu)} j^\mathrm{i}
  {\left(\frac{\tau_D}{\tau_H}\right)^2} \frac{1}{1+\tau_D/\tau_B}
  \left[1-\left(\frac{g}{g_0}\right)^2
  \frac{1}{1+(g/g_0)^2}\right]\nonumber\\
  &=&-\frac{m}{p_{\tilde{n}}^\mathrm{l}(\mu)} j^\mathrm{i} 
  {\left(\frac{\tau_D}{\tau_H}\right)^2}
  \frac{1}{1+\tau_D/\tau_B}\frac{1}{1+(g/g_0)^2}\, ,
\end{eqnarray}
which yields as the correction to the transmission and reflection probabilities
\begin{equation}
\label{eq:loops}
  \delta(t_{\tilde{n}i})^\textrm{loop}
  =\frac{p_{\tilde{n}}^\mathrm{l}(\mu)}{m j^\mathrm{i}}
    \left\langle|\psi_{{\tilde{n}}}|^2\right\rangle_{\textrm{loop}}
  =-{\left(\frac{\tau_D}{\tau_H}\right)^2}\frac{1}{1+\tau_D/\tau_B}\frac{1}{1+(g/g_0)^2}\, .
\end{equation}
As in the linear case [see Eq.~(\ref{eq:loopg0})], this correction
restores current conservation in the presence of the nonlinearity.
With Eq.~(\ref{eq:c11gendresult}) we obtain
\begin{equation}
{
  r_{ii} = \frac{1}{N_c+\tilde{N}_c} +
  \frac{(N_c+\tilde{N}_c - 1)\left( 1 + \tau_D/\tau_B \right)}
  {(N_c+\tilde{N}_c)^2 \left( 1 + \tau_D/\tau_B \right)^2 + 
    \left( g j^{\mathrm{i}} \tau_D \right)^2} \label{eq:r11wlg}
}
\end{equation}
for the probability of retro-reflection into the incident channel $n=i$,
as well as
\begin{equation}
{
  r_{ni} = t_{\tilde{n}i} = \frac{1}{N_c+\tilde{N}_c} -
  \frac{1 + \tau_D/\tau_B}
  {(N_c+\tilde{N}_c)^2\left( 1 + \tau_D/\tau_B \right)^2 + 
    \left( g j^{\mathrm{i}} \tau_D \right)^2} \label{eq:rn1wlg}
}
\end{equation}
{for the probability $r_{ni}$ of reflection into a different channel 
$n \neq i$ of the incident lead and for the probability $t_{\tilde{n}i}$ 
of transmission into channel $\tilde{n}$.
This yields the total reflection and transmission probabilities}
\begin{eqnarray}
  {R} & {=} & {\frac{N_c}{N_c+\tilde{N}_c} + 
    \frac{\tilde{N}_c \left( 1 + \tau_D/\tau_B \right)}
         {(N_c+\tilde{N}_c)^2\left( 1 + \tau_D/\tau_B \right)^2 + 
           \left( g j^{\mathrm{i}} \tau_D \right)^2}
         \, ,} \label{eq:Rwlg} \\
  {T} & {=} & {\frac{\tilde{N}_c}{N_c+\tilde{N}_c} -
    \frac{\tilde{N}_c \left( 1 + \tau_D/\tau_B \right)}
         {(N_c+\tilde{N}_c)^2\left( 1 + \tau_D/\tau_B \right)^2 + 
           \left( g j^{\mathrm{i}} \tau_D \right)^2}
  \, ,} \label{eq:Twlg}
\end{eqnarray}
{which obviously satisfy $R+T=1$.}

\section{{Comparison with numerical results}}

\label{sec:comp}

\begin{figure}
  \begin{center}
    \includegraphics[width=\linewidth]{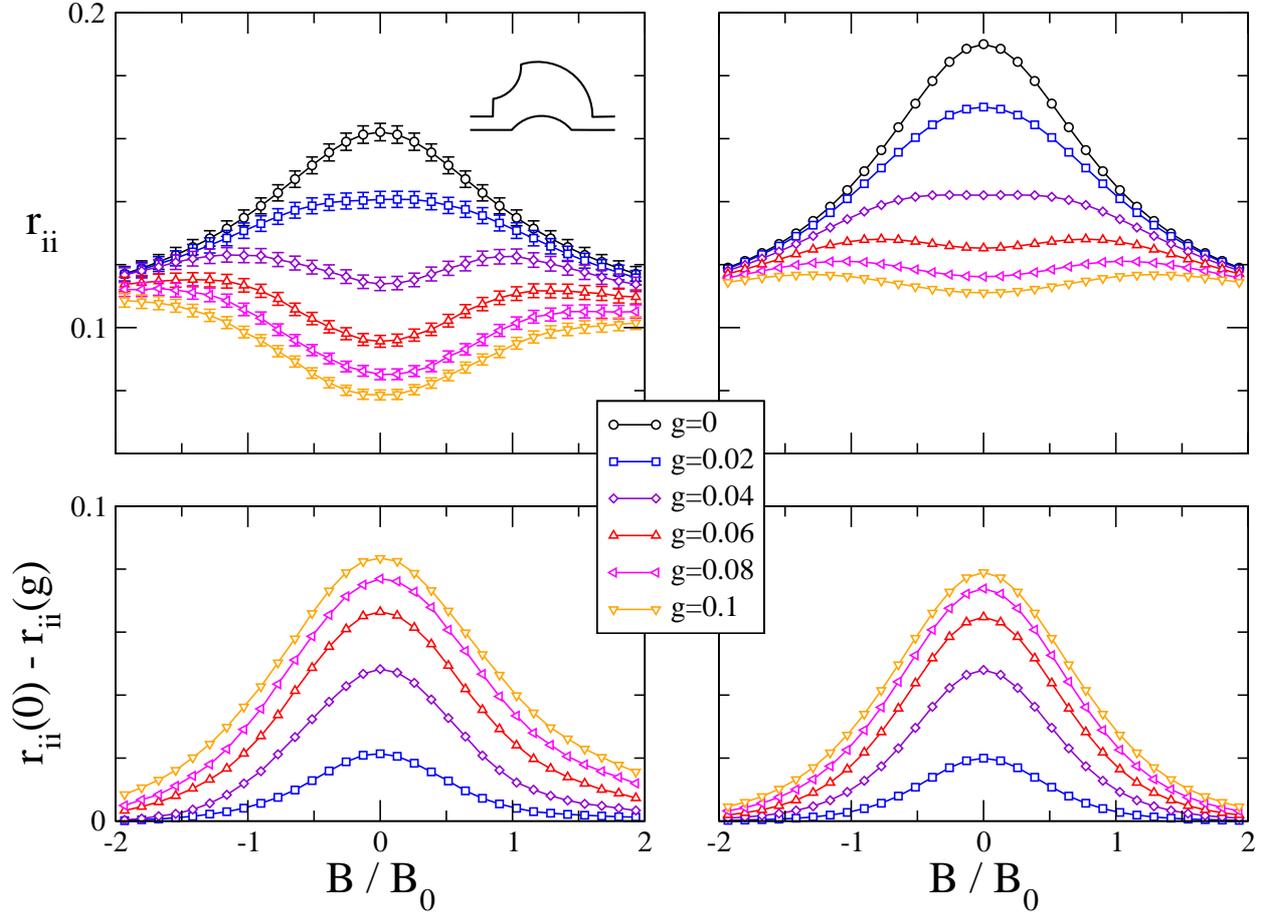}
    \caption{\label{fig:compg1}
      Nonlinearity-induced destruction of weak localization for the 
      billiard a displayed in the left panel of Fig.~\ref{fig:billiard}.
      Plotted are the numerically computed backscattering probabilities 
      (upper left panel) and their semiclassical prediction according to 
      Eq.~(\ref{eq:r11wlg}) (upper right panel) as a function of the 
      effective magnetic field for various values of the nonlinearity $g$. 
      We use the magnetic field scale 
      $B_0 \simeq {1.55\times10^{-3}\, m{\mu_0}/\hbar}$
      and the average population $j^{\mathrm{i}} \tau_D \simeq 267${,
      which were inferred from an analysis of the classical dynamics} 
      within the billiard.
      The lower panels display the differences of the backscattering 
      probabilities for finite $g$ with respect to the backscattering 
      probabilities of the linear system{. 
        Good agreement is found between} the numerical data 
      (lower left panel) and  the semiclassical prediction (right
      panel).
    }
  \end{center}
\end{figure}
    
\begin{figure}
  \begin{center}
    \includegraphics[width=\linewidth]{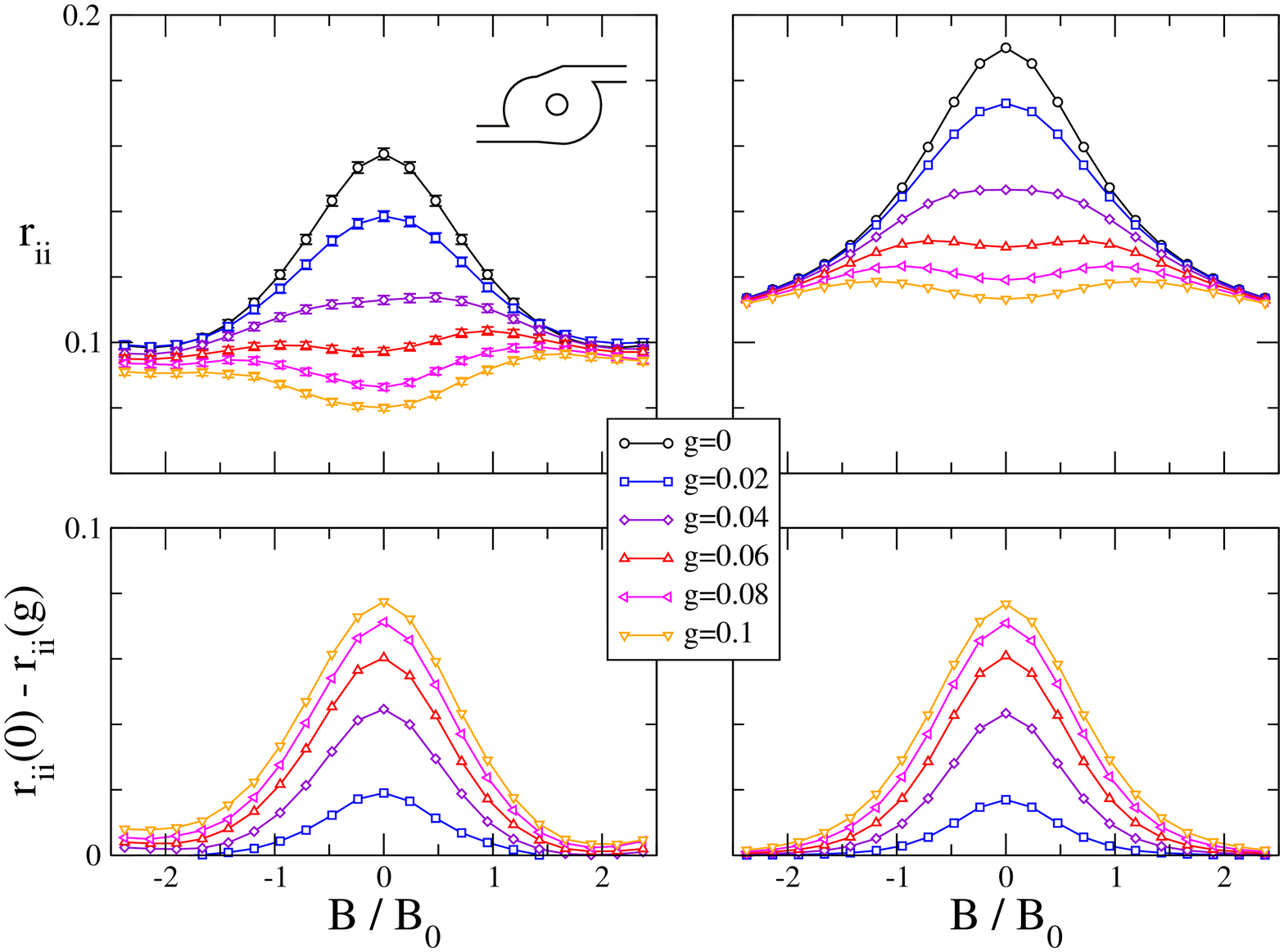}
    \caption{\label{fig:compg2}
      Same as Fig.~\ref{fig:compg1} for the billiard b 
      displayed in the right panel of Fig.~\ref{fig:billiard}.
      We use the magnetic field scale 
      $B_0 \simeq {4.21\times10^{-4}\, m{\mu_0}/\hbar}$ and the 
      average population $j^{\mathrm{i}} \tau_D \simeq 241${,
      which were inferred from an analysis of the classical dynamics} 
      within the billiard.} 
  \end{center}
\end{figure}

Figures \ref{fig:compg1} and \ref{fig:compg2} display (in their right columns)
the semiclassical prediction (\ref{eq:r11wlg}) for the probability of
retro-reflection into the incident channel as evaluated for the billiards
a and b, respectively, that are shown in Fig.~\ref{fig:billiard}.
The sizes of the two billiards are chosen such that both the
incident and the transmitted leads exhibit five open channels, i.e.\
$N_c = \tilde{N}_c = 5${, at the energy that corresponds to the 
chemical potential $\mu$ of the incident beam.
We specifically have the areas $\Omega \simeq 3.41 \times 10^3\,
\hbar^2/(m\mu_0)$ for billiard a and $\Omega \simeq 3.29 \times 10^3\, 
\hbar^2/(m\mu_0)$ for billiard b, where $\mu_0$ defines the characteristic 
energy scale for the chemical potential of the atomic beam
(i.e.\ we choose $\mu=\mu_0$ for the evaluation of the semiclassical
retro-reflection probability).
The incident current is chosen as $j^{\mathrm{i}} = 1.0\, \mu_0/\hbar$.
As described in \ref{sec:cladyn}, the}
dwell time $\tau_D$ and the characteristic scale $B_0$ of the effective
magnetic field were classically determined from the numerically computed 
length and area distributions within the two billiards, respectively; 
 we obtained
${j^{\mathrm{i}}}\tau_D \simeq {267}$ {and
$B_0 \simeq 1.55\times10^{-3}\, m\mu_0/\hbar
\simeq 0.844 \times 2\pi\hbar/\Omega$}
for billiard a {as well as}
${j^{\mathrm{i}}}\tau_D \simeq {241}$ {and
$B_0 \simeq 4.21\times10^{-4}\, m\mu_0/\hbar
\simeq 0.221 \times 2\pi\hbar/\Omega$} for billiard b.

In the linear case $g=0$, a Lorentzian peak is obtained for the 
retro-reflection probability as a function of the effective
magnetic field, on top of an incoherent background at 
$r_{ii} \simeq 1/(N_c+\tilde{N}_c) = 0.1$.
This is the characteristic signature of weak localization.
As is evident from Eq.~(\ref{eq:r11wlg}), the presence of a finite 
nonlinearity $g$ gives rise to a reduction of the coherent enhancement of
the backscattering probability, which ultimately approaches the incoherent
background $1 / (N_c + \tilde{N}_c)$ for $g\to\infty$.
This reduction, however, is more effective {at} the center of the 
backscattering peak than {in} its wings, such that for intermediate values
of $g$ a local {minimum} may be encountered in the reflection probability
around $B=0$.

\begin{figure}
  \begin{center}
    \includegraphics[width=\linewidth]{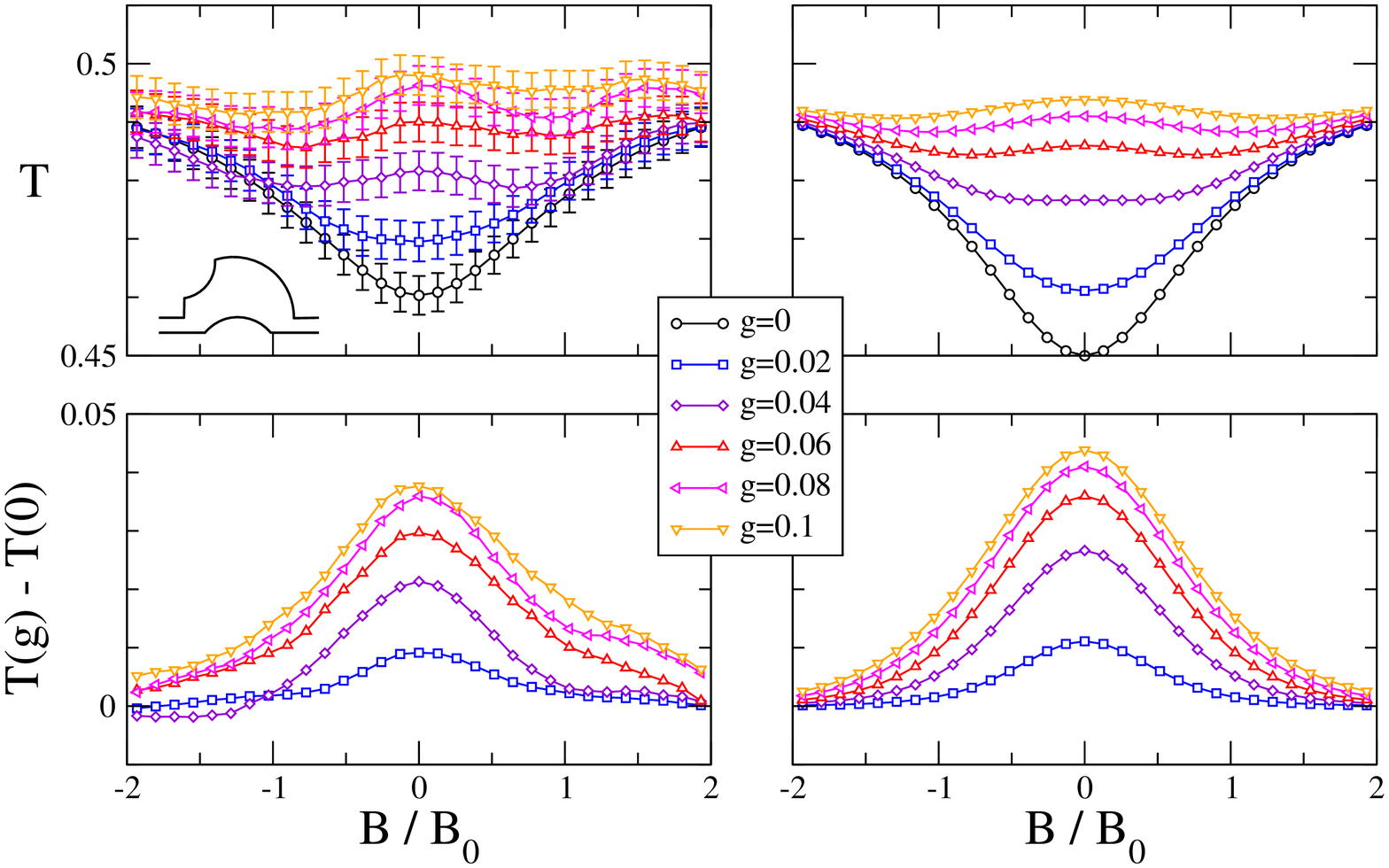}
    \caption{\label{fig:compg3}
     {Total transmission for the billiard a.
     Plotted are the numerically computed transmission probabilities 
     (upper left panel) and their semiclassical prediction according to 
     Eq.~(\ref{eq:Rwlg}) (upper right panel) as a function of the 
     effective magnetic field for various values of the nonlinearity $g$,
     using the same parameters as for Fig.~\ref{fig:compg1}.
     The lower panels display the differences of the transmission 
     probabilities for finite $g$ with respect to the transmission 
     probabilities of the linear system{.
       Good agreement is found between} the numerical data 
     (lower left panel) and  the semiclassical prediction (right panel).}}
  \end{center}
\end{figure}

\begin{figure}
  \begin{center}
    \includegraphics[width=\linewidth]{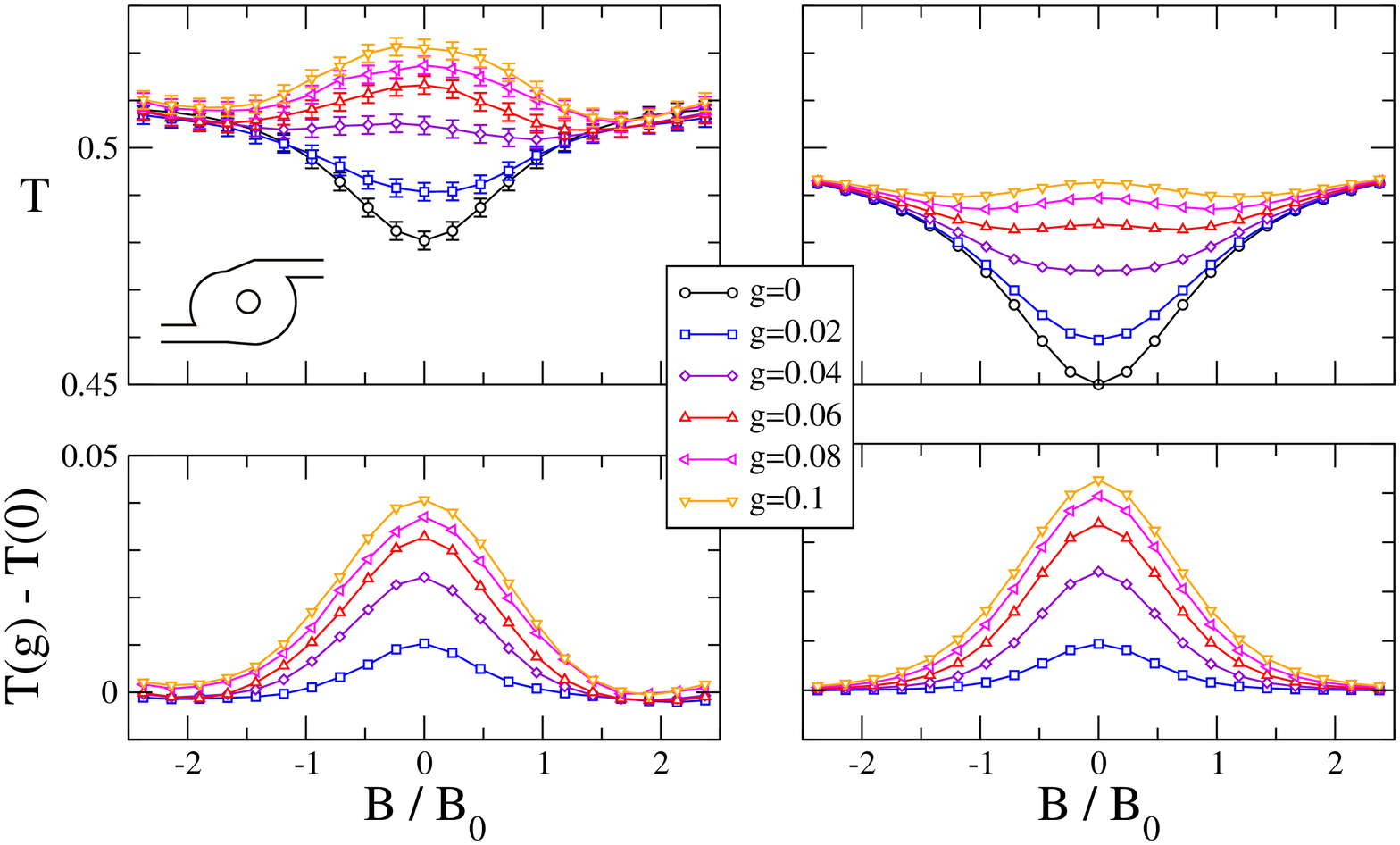}
    \caption{\label{fig:compg4}
      Same as Fig.~\ref{fig:compg3} for the billiard b.}
  \end{center}
\end{figure}

This prediction is indeed confirmed by numerical computations of the
quasi-stationary transport process within the two billiards under
consideration.
As explicit numerical propagations of the time-dependent
Gross-Pitaevskii equation (\ref{eq:GPEs}) are rather time-consuming, 
we use, in practice, a different approach in 
order to compute the scattering states of the system.
As explained in \ref{sec:numerics}, this
approach uses a Newton search algorithm in order to 
construct self-consistent scattering states of the nonlinear system{.}
Among all those scattering states that are identified at given chemical 
potential $\mu$ and given incident current 
$j^{\mathrm{i}}$
(there is only one such scattering state in the linear system, but several
of them generally exist in the presence of the nonlinearity, see, e.g.,
Ref.~\cite{PauRicSch05PRL}), we select the one that {would be first
encountered in the presence of an adiabatic increase of the nonlinearity
strength $g$.}
More technical details concerning this approach will be provided in a 
subsequent publication {\cite{HarO}}.

The left columns of Figs.~\ref{fig:compg1} and \ref{fig:compg2} display the
results that are obtained from these numerical computations.
To obtain good statistics, we did not only perform an energy average
of the reflection probability{, within the energy interval 
$0.93\, \mu_0 \leq \mu \leq 1.18\, \mu_0$ for which there are exactly 5 open 
channels within each lead}, but also averaged over different positions of 
the semicircular and circular obstacles in the case of billiard a and b, 
respectively.
{This additional configurational average is necessary as the above energy 
interval contains only a limited number of resonances within the billiard.}
Moreover, we averaged over different choices of the incident channel $i$,
even though only the choice $i=1$ appears realistic from the experimental
point of view.
The error bars attached to the data points consequently indicate the size 
of the statistical standard deviation {that results from these averages}.

As shown in the {upper} panels of Figs.~\ref{fig:compg1} and 
\ref{fig:compg2},  {the} relative height of the peak with 
respect to the incoherent background  significantly deviates 
from the universal semiclassical prediction, even in the linear case $g=0$.
This discrepancy may, on the one hand, be attributed to a limited
applicability of the semiclassical framework in our context.
Indeed, as is seen in Fig.~\ref{fig:billiard}, the wavelength of the
matter-wave beam is not sufficiently small to rule out the influence of
possible diffraction effects at the rounded corners of the billiard.
On the other hand, \emph{non-universal} scattering phenomena that explicitly 
depend on the shape of the billiard under consideration may play a role.
Specifically, among the backreflected trajectories
that start and end in a given channel, there is possibly a significant 
fraction of \emph{self-retracing} trajectories which are identical {to} 
their time-reversed counterpart.
As those self-retracing trajectories do not contribute to the crossed 
part of the coherent backscattering probability, their semiclassical 
contribution then should be subtracted from the sum-rule based expression 
(\ref{eq:psin2srcross}) of the crossed density.
Indeed, the presence of a prominent circular obstacle within billiard b
should allow for a number of rather short self-retracing trajectories with
a relatively small Lyapunov exponent (and therefore with a relatively large 
weight in the semiclassical Green function), namely those trajectories that 
directly head toward the obstacle, undergo a self-retracing 
reflection there, and subsequently exit the billiard in the incident channel.
Similarly relevant self-retracing trajectories bouncing off the 
semicircular obstacles should exist in billiard a.

\begin{figure}[t]
  \begin{center}
   \includegraphics[width=\textwidth]{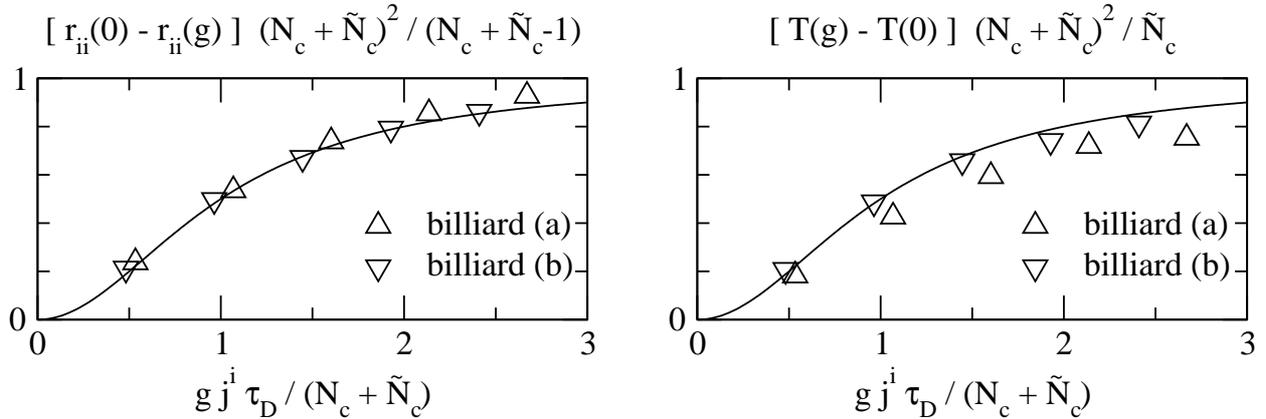}
   \caption{\label{fig:peak}
     {
       Change of height of the weak localization peak (or dip) at $B=0$
       as compared to the linear case.
       Plotted are, in the left panel, 
       $[r_{ii}(g=0) - r_{ii}(g)] (N_c + \tilde{N}_c)^2 / 
       (N_c + \tilde{N}_c + 1)$ and, in the right panel,
       $[T(g) - T(g=0)] (N_c + \tilde{N}_c)^2 / \tilde{N}_c$
       as a function of $g j^{\mathrm{i}} \tau_D / (N_c + \tilde{N}_c)$
       for the two billiards under consideration.
       A comparison with the semiclassical predictions (\ref{eq:r11wlg}) 
       and (\ref{eq:Twlg}) (solid lines), which are given by
       $1/(1+x^{-2})$ with 
       $x \equiv g j^{\mathrm{i}} \tau_D / (N_c + \tilde{N}_c)$ in both cases,
       shows good agreement.
     }
   }
  \end{center}
\end{figure}

{In view of the diagrammatic theory developed in Sec.~\ref{sec:semi}, 
we note that such self-retracing trajectories do not affect the 
nonlinearity-induced \emph{corrections} $c_{ii}^{(g)}-c_{ii}^{(0)}$ 
to the coherent backscattering probability.
Indeed, as is evident e.g.\ from Eq.~(\ref{eq:crosseddiags}), those 
corrections are distinctly different from ladder contributions
and will therefore not be overcounted if they happen to involve self-retracing
trajectories.
We consequently find, as shown in the lower panels of Figs.~\ref{fig:compg1} 
and \ref{fig:compg2}, rather good agreement between the numerical data
and the semiclassical prediction if we specifically compare those 
corrections, i.e.\ the reduction of the weak localization peak with respect
to the linear case $g=0$.
This is furthermore confirmed in the left panel of Fig.~\ref{fig:peak} 
which shows the reduction of the height of the weak localization peak at 
$B=0$ as a function of the nonlinearity strength $g$.
Renormalizing the horizontal and vertical axes in terms of the scales that
are suggested by the analytical prediction (\ref{eq:r11wlg}), we find rather
good agreement with this universal prediction for both billiards.
This underlines the validity of the approach developed in Sections 
\ref{sec:semi} and \ref{sec:loop}.
}

\begin{figure}[t]
  \begin{center}
   \includegraphics[width=0.7\textwidth]{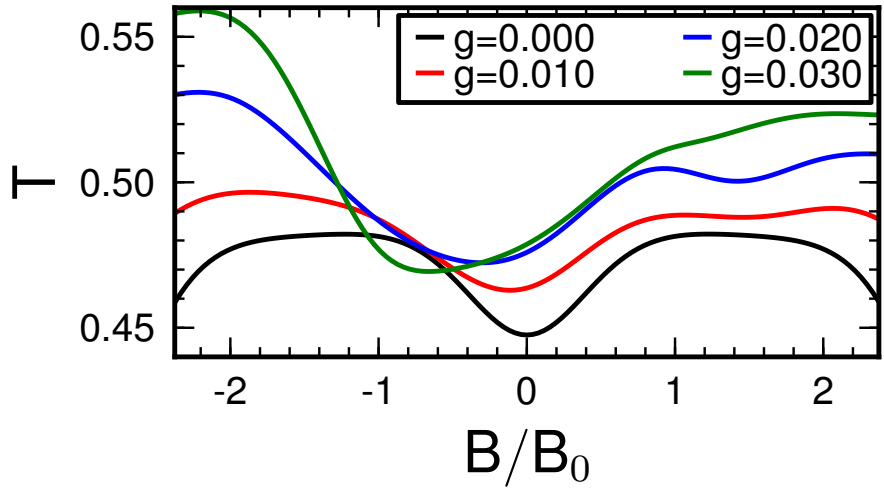}
   \caption{\label{fig:onsager} 
     {
       Total mesoscopic transmission 
       $\mathcal{T}=\frac{1}{N_c} \sum_i\sum_{\tilde{n}} |t_{\tilde{n}i}|^2$
       as a function of the magnetic field $B$ for billiard b at a fixed
       chemical potential $\mu = 1.08306\, \mu_0$ and at different 
       strengths of the nonlinearity $g$.
       In the absence of interaction (black curve), $\mathcal{T}$ is 
       symmetric in $B$, which is a consequence of Onsager's relations 
       \cite{FerGoo}.
       This symmetry in $B$ is, however, broken in the presence of the
       nonlinearity (red, blue, and green curves).
       A sufficiently large energetic and configurational average would 
       restore the symmetry in $B$ as is visible in Fig.~\ref{fig:compg4}.
     }}
  \end{center}
\end{figure}

Finally, in order to {demonstrate} the relevance of the loop
contributions, we show in {Figs.~\ref{fig:compg3}, \ref{fig:compg4}, and in
the right panel of Fig.~\ref{fig:peak}} 
the comparison of the numerical results for the 
transmission with our analytical prediction obtained from Eq.~(\ref{eq:Twlg}). 
Once again, excellent agreement is found after removing non-universal 
effects.
{Remarkably, Figs.~\ref{fig:compg3} and \ref{fig:compg4} display 
asymmetries in the transmission as a function of the magnetic field, 
i.e.\ we do not necessarily have $T(-B) = T(B)$.
This finding seems to constitute a violation of Onsager's relations
\cite{FerGoo} which state that the total mesoscopic transmission
$\mathcal{T} \equiv \sum_i\sum_{\tilde{n}} |t_{\tilde{n}i}|^2$, which
represents the incoherent sum over the individual transmissions that 
result from all available choices of the incident channel $i$ (and which
is implicitly computed in Figs.~\ref{fig:compg3} and \ref{fig:compg4}
due to the averaging over the incident channel), be symmetric in the
magnetic field $B$ for any given scattering geometry at any given
chemical potential.}
It should be noted, however, that Onsager's relations are based on the
unitarity of the scattering matrix $S$ and its symmetry property 
$S(-B) = S^T(B)$ \cite{Dat} and thereby implicitly rely on the linearity
of the scattering process.
Indeed, computing the total mesoscopic transmission $\mathcal{T}$ across
billiard b in the absence of interaction for a specific choice of the 
chemical potential, we obtain perfect symmetry of $\mathcal{T}$ in $B$ 
as shown in Fig.~\ref{fig:onsager}.
This symmetry is broken at finite values of the nonlinearity $g$.
Similar findings have been reported in electronic transport
through mesoscopic structures in the presence of strong bias voltages
\cite{SanBut04,GusKvoOlsPlo09}.

\section{{Conclusion}}
\label{sec:conc}

{
In summary, we studied, both analytically and numerically, weak localization 
of guided matter waves that originate from interacting Bose-Einstein
condensates and propagate through chaotic billiard geometries.
Our analytical approach is based on a nonlinear diagrammatic perturbation theory
\cite{WelGre08PRL,WelGre09PRA,Wel09AP} that originates from the
Gross-Pitaevskii equation, which is combined with a semiclassical expansion of
the  linear (single-particle) Green function within the billiard.
Summing all terms of this diagrammatic perturbation theory and utilizing
standard semiclassical sum rules in ergodic billiards, we obtain
analytical expressions for the retro-reflection probability 
[Eq.~(\ref{eq:r11wlg})] as well as for the total reflection and transmission 
[Eqs.~(\ref{eq:Rwlg}) and (\ref{eq:Twlg})] in dependence of the effective
interaction strength and of the strength of an artificial gauge field
that breaks time-reversal invariance and simulates the effect of a
magnetic field for charged particles.
These expressions also involve the analysis of loop corrections in leading 
order \cite{RicSie02PRL} which restore current conservation.}

Globally, we find that the peak of weak localization decreases with 
increasing nonlinearity strength $g$ and eventually disappears beyond a
characteristic scale of $g$ given by the inverse average population within 
the billiard.
This suggests that the presence of the nonlinearity introduces an 
additional dephasing mechanism that affects the constructive interference
between backscattered trajectories and their time-reversed counterparts.
The decrease of the peak height with $g$ is found to be stronger at the center
of the peak (i.e.\ for vanishing gauge field $B=0$) than in its wings,
which eventually gives rise to a tiny local dip in the backscattering 
probability around $B=0$.
While this dip, as it is predicted by the general semiclassical theory, is 
presumably too small to be  of experimental relevance, it can be more 
pronounced in individual billiard geometries, as the ones specifically 
studied in this work, in which the backscattering probability develops a 
global minimum, instead of a maximum, at $B=0$.
We thereby encounter a signature of \emph{weak antilocalization} in those 
billiards, which is of genuinely different nature than antilocalization in 
electronic transport processes involving spin-orbit interaction 
\cite{ZumO02PRL}.

{Comparisons of the numerically computed absolute and relative heights
 of the weak localization peaks with the semiclassical prediction seem to
suggest that this weak antilocalization{-type} phenomenon is caused by the 
occurrence of \emph{self-retracing} trajectories in the scattering system.
Indeed, the presence of such self-retracing trajectories reduces the 
probability for coherent backscattering as compared to the universal
semiclassical prediction in the linear case, as the application of the
standard sum rule would give rise to an overcounting of interference 
contributions between backscattered trajectories and their time-reversed 
counterparts.
It does, however, not affect the nonlinearity-induced corrections
to this coherent backscattering probability.
Consequently, the peak of weak localization can turn into a finite dip
in billiard geometries that exhibit prominent self-retracing trajectories
of short length and therefore of large weight in the semiclassical
backscattering amplitude.}

{This observation also sheds new light on the inversion of the 
coherent backscattering peak that was found in the coherent propagation
of Bose-Einstein condensates through two-dimensional disorder potentials
\cite{HarO08PRL}.
As a matter of fact, such disorder potentials also exhibit self-retracing 
trajectories, which essentially arise from a retro-reflection at the first 
impurity that the incident matter wave encounters within the disorder region.
Diagrammatic calculations within such disordered systems 
{\cite{Eck}}
do indeed suggest that short reflected paths are at the origin of the 
inversion of the coherent backscattering peak in disordered systems.}

{In this study, we considered a number of idealizations concerning the
setup for the matter-wave transport process.
For the sake of analytical tractability of the problem, we particularly
imposed hard-wall boundaries of the wave guides and the billiard
and assumed a continuous monochromatic flow of atoms through this scattering
region.
We are convinced, however, that the phenomenology studied in this work
should be sufficiently robust to manifest also in the case of harmonic 
waveguides and harmonic-like confinement geometries with chaotic
(or mixed regular-chaotic) dynamics, which could possibly be realized by 
combinations of red- and blue-detuned laser beams that are perpendicularly 
focused onto the waveguide {\cite{GatO11PRL}}, as well as in the case of 
atomic wave packet scattering processes which may be easier to realize than 
guided atom-laser beams.
Weak localization and antilocalization of interacting Bose-Einstein 
condensates should therefore be observable with present-day cold-atom 
technologies.}

\subsection*{Acknowledgements}

We would like to thank \.{I}nan\c{c} Adagideli, 
{David Gu\'ery-Odelin,} Michael Hartung, 
Jack Kuipers, and Daniel Waltner for helpful and inspiring discussions.
This work was supported by the DFG Forschergruppe FOR760 
``Scattering Systems with Complex Dynamics''.
C.P.\ acknowledges financial support by the Alexander von Humboldt
foundation and by CEA eurotalent.

\appendix

\section{Numerical computation of stationary scattering states}
\label{sec:numerics}

In this appendix{,} we {explain} how {we numerically compute} 
stationary solutions $\psi \equiv \psi(\mathbf{r})$ of the 
time{-}dependent inhomogeneous Gross-Pitaevskii equation~(\ref{eq:GPEs}).
{Such stationary solutions satisfy the nonlinear equation}
\begin{equation}
\label{eq:nlgreen}
\left[ \mu-H \right]\psi(\mathbf{r})
-g(\mathbf{r})\frac{\hbar^2}{2m}|\psi(\mathbf{r})|^2\psi(\mathbf{r}) - 
S(\mathbf{r}) = 0
\end{equation}
{with $S(\mathbf{r}) = S_0 \chi_i(y) \delta(x - x_L)$,}
which is equivalent to Eq.~(\ref{eq:scstate_g}).
This equation is discretized on a two{-}dimensional lattice where 
only points inside the cavity {and the leads are taken} 
into account.
The single{-}particle Hamiltonian $H$ given by Eq.~(\ref{eq:H})
can be approximated using a finite{-}difference scheme 
\cite{FerGoo} where we incorporate the vector potential using a 
Peierls phase \cite{Pei33ZP}.
We choose {the lattice spacing} $\Delta$ small enough that the 
approximation error {[}which scales as ${\mathcal{O}}(\Delta^2)${]} 
becomes negligible{,} which is the case for roughly 30 lattice points 
per wavelength.
The interaction strength $g(\mathbf{r})$ is{, as explained in 
{Ref.}~\cite{PauHarRicSch07PRA}, considered to be constant within
the scattering region and} adiabatically {ramped} off 
in the leads \cite{rem_gr}.
{The} effects of the infinite leads {can then be incorporated}
through self{-}energies as in the recursive Green function method
\cite{Dat,LeeFis81PRL}{, which allows one to}
restrict the {numerical computation} to a finite {spatial} region.

{The complex solution $\psi(\mathbf{r})$ of the nonlinear wave equation
(\ref{eq:nlgreen}) can be represented as a $2\mathcal{N}$-dimensional real
vector where $\mathcal{N}$} is the number of grid points.
Defining
\begin{equation}
{F: \mathbb{R}^{2{\mathcal{N}}} \longrightarrow 
  \mathbb{R}^{2{\mathcal{N}}},} \;
\psi(\mathbf{r}) \longmapsto \left[ \mu-H_0 \right] \psi(\mathbf{r})
-g(\mathbf{r}){\frac{\hbar^2}{2m}}|\psi(\mathbf{r})|^2\psi(\mathbf{r}) - S(\mathbf{r})\, ,
\end{equation}
we search now for a solution of $F(\psi)=0$. 
This is done with Newton's method \cite{NocWri}.
One selects a starting vector $\psi_0(\mathbf{r})$ and constructs a sequence 
of vectors $\left\{\psi_k\right\}_{k=1}^{\infty}$ 
(here $k$ is the iteration number) using the iteration
$\psi_{k+1}=\psi_{k}-\left(\mathcal{D} F\right)^{-1} F(\psi_k)$.
If the derivate $\mathcal{D} F$ at the solution is not singular,
this iteration is guaranteed to converge to a solution of 
the nonlinear equation {(\ref{eq:nlgreen}), provided the starting vector 
is chosen in a suitably close vicinity of this solution.}

{The efficiency of this method strongly depends on the
starting vector $\psi_0(\mathbf{r})$.
An obvious choice would be} the solution of the linear {wave equation
(for $g=0$). This choice, however, works out only} for very small 
{nonlinearities}.
In the general case{,} one has to use a continuation method 
\cite{NocWri,Sey}.
{To this end, we consider $g$, i.e.\ the constant value of the
interaction strength within the billiard \cite{rem_ggr},}
as an additional free parameter {and}
reinterpret {$F\equiv F[\psi(\mathbf{r});g]$} as a function 
$F:~\mathbb{R}^{{2\mathcal{N}}}{\times\mathbb{R}} \to 
\mathbb{R}^{2{\mathcal{N}}}$.
Now $F^{-1}(0)$ is a one{-}dimensional manifold \cite{rem_rank}
which can {be conveniently} parametrized by the arclength $s$
{through} the parametric curve 
$s \mapsto \left[ g(s),\psi(s) \right]$.
Starting from $g=0$, the numerical algorithm follows this curve 
until {the desired} value {$g$ of the nonlinearity strength is 
reached, and returns the wavefunction $\psi(\mathbf{r})$ that is obtained 
at the end of this curve-tracking process \cite{rem_sweep}.}

\begin{figure}[t]
  \begin{center}
   \includegraphics[width=0.6\textwidth]{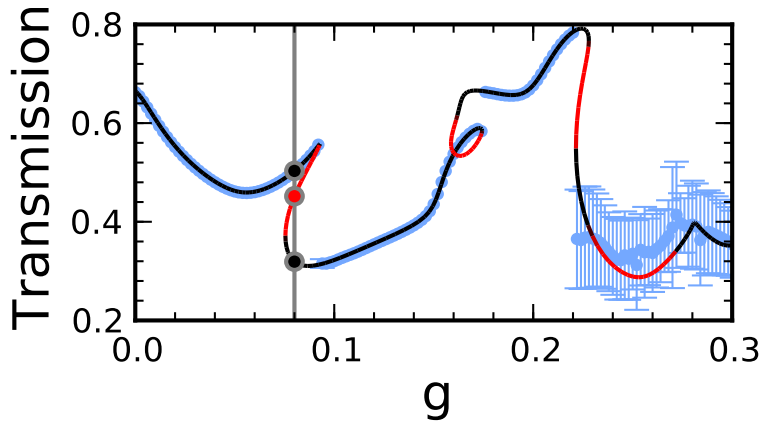}
    \caption{\label{fig:curvetrack} 
      {Transmission of the stationary scattering state versus the 
        interaction strength $g$ as obtained from the numerical 
        curve-tracking method.
        Black lines indicate dynamically stable branches and red lines 
        indicate dynamically unstable branches.
        Due to the nonlinearity in the wave equation, several scattering
        states with different transmissions may be encountered at a given 
        value of $g$.
        At $g=0.08$ for instance, there are three different scattering states,
        two of them being dynamically stable (black dots) and one being 
        unstable (red dot).
        The light-blue data points show the time-averaged transmissions 
        that are obtained from a numerical simulation of the time-dependent
        scattering process in the presence of an adiabatically slow ramping 
        of the source amplitude.
        As is visible from the error bars which indicate the associated 
        standard deviation of the transmission at given value of $g$, the
        atomic current through the billiard develops a pronounced 
        time dependence in the absence of stable stationary scattering 
        states (e.g.\ around $g=0.25$).
        The calculation is done for billiard b at $B=-0.001\, m \mu_0/\hbar$ 
        and $\mu = 0.935102\, \mu_0$, with the condensate being injected
        in the transverse ground mode ($i=1$) of the incident lead.
      }
    }
  \end{center}
\end{figure}

{Fig.~\ref{fig:curvetrack} shows, for a specific set of parameters,
a projection of this curve onto the two-dimensional parameter space
spanned by the nonlinearity strength $g$ and the total transmission 
that is associated with the stationary scattering state $\psi(\mathbf{r})$.
As is characteristic for nonlinear wave equations, several 
stationary solutions are found for some values of $g$, e.g. at $g=0.08$.
Some of these solutions may be dynamically unstable and can therefore not
be populated in the time-dependent propagation process.
At sufficently large values of the nonlinearity, no dynamically stable
scattering state is found any longer, which implies that the scattering
process becomes permanently time-dependent and develops turbulent-like
behaviour \cite{PauO05PRA}.}

{To determine the dynamical stability of the numerically computed 
stationary scattering state, we linearize} the time{-}dependent 
Gross-Pitaevskii equation around the stationary solution 
{$\psi(\mathbf{r})$ using} the {Bogoliubov} ansatz
\begin{equation}
  \tilde\psi({\mathbf{r},}t) = {\left[\psi(\mathbf{r}) + 
    u(\mathbf{r}) \exp\left(-i \xi t\right) + 
    v^{*}(\mathbf{r}) \exp\left( i \xi^{*} t\right) 
    \right] \exp\left(- \frac{i}{\hbar} \mu t \right)}
  \label{eq:Bog}
\end{equation}
{for the time-dependent scattering wavefunction 
$\tilde\psi(\mathbf{r},t)$}.
This leads to the Bogoliubov-de Gennes equation \cite{PetSmi}
\begin{equation}
 \begin{pmatrix}
   H-\mu+2g{(\mathbf{r})} {\frac{\hbar^2}{2m}} |\psi{(\mathbf{r})}|^2  & 
   g{(\mathbf{r}) \frac{\hbar^2}{2m} \left[\psi(\mathbf{r})\right]^2} \\
   g{(\mathbf{r}) \frac{\hbar^2}{2m} \left[\psi^*(\mathbf{r})\right]^2} & 
   H^*-\mu+2{g(\mathbf{r})\frac{\hbar^2}{2m}} |\psi{(\mathbf{r})}|^2
 \end{pmatrix}
 \begin{pmatrix}
   u{(\mathbf{r})} \\
   v{(\mathbf{r})}
 \end{pmatrix}
 {
 =
 \hbar \xi
 \begin{pmatrix}
   +1 & 0  \\
   0  & -1
 \end{pmatrix}
 \begin{pmatrix}
   u(\mathbf{r}) \\
   v(\mathbf{r})
 \end{pmatrix}
 } \, . \label{eq:BdG}
\end{equation}

{This generalized eigenvalue problem is numerically solved using the 
implicit restarted Arnoldi method as realized in the software library 
\texttt{ARPACK} \cite{LehSorYan97}.}
Special care must be taken in order to describe the infinite leads properly.
This is done using the method of smooth exterior complex scaling \cite{Moi98PR}
which exponential{ly} damps the outgoing waves {of the collective modes}
in the leads.
{As a consequence, the Bogoliubov eigenfrequencies $\xi$ become 
\emph{complex}.
If one of them is found to exhibit a positive imaginary part, 
i.e.\ Im$(\xi)>0$, we can infer from Eq.~(\ref{eq:Bog}) that the 
scattering state $\psi(\mathbf{r})$ under consideration is dynamically 
unstable.}

\begin{figure}[t]
  \begin{center}
   \includegraphics[width=0.49\textwidth]{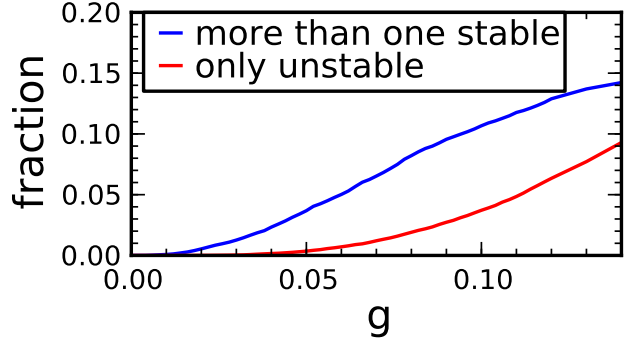}
    \caption{\label{fig:unstable} 
      This figure shows {as a function of $g$} the fraction of 
      configuration{s} of $B$, $\mu$ and {the incident} mode 
      {$i$ that exhibit} only dynamical{ly} 
      unstable solutions {(red curve)} or support more than one
      dynamical{ly} stable solution {(blue curve)}.
      {In the latter case, we select the one that is first encountered 
        in the curve-tracking algorithm starting from $g=0$.}}
  \end{center}
\end{figure}

{Fig.~\ref{fig:unstable} shows (red curve) the fraction of parameter 
configurations of the chemical potential $\mu$, the magnetic field $B$, and
the incident mode number $i$ for which no dynamically stable scattering state
is found.
This fraction of unstable configurations is found to increase with the
nonlinearity strength $g$, which} imposes restrictions on the shape of the 
cavities and the maximum value of $g$ one can use for {numerical} 
simulations. 
{In particular, we find that} the fraction of configurations {that} 
support no stable solution increase{s rather} rapidly
with $g$ {for small widths of the leads, i.e.\ for a very low number of
open channels.
This is attributed to the reduced spectral width that quasi-bound resonance
states within the billiard exhibit in that case, which in turn leads to an
enhanced interaction energy 
$\int d\mathbf{r} g(\mathbf{r}) |\psi(\mathbf{r})|^4$ at such resonances.
Rather wide leads with a large number of open channels, on the other hand,
compromise the effect of weak localization and reduce the visibility of
its signature in the reflection and transmission probabilities.}
{For the billiard sizes and geometries under consideration,} the fraction 
of configurations with unstable solutions {remains} below $0.02$ 
{until} $g=0.08$.

When {encountering} a configuration with only unstable solutions{,} 
we {select} the one {that exhibits} the smallest {Lyapunov exponent,
the latter being defined by the largest imaginary part of the eigenvalues 
$\xi$ of the Bogoliubov-de Gennes equations (\ref{eq:BdG}).}
This {choice is supported by} time{-}dependent {propagations} of 
{the inhomogeneous Gross-Pitaevskii equation} (\ref{eq:GPEs}){, 
which directly simulate, as in Ref.~\cite{HarO08PRL}, the 
time{-}dependent scattering process in the presence of an adiabatic 
increase of the source amplitude.
In such simulations, which were carried out for specific parameter
configurations, we find,} for not {too} large values of $g${, that
the time-dependent current in the transmitted lead displays regular 
oscillations} around the {current} of the stationary solution 
{that is the least unstable one and exhibits the smallest Lyapunov 
exponent.
The \emph{time-averaged} transmission, which is the main experimental
observable in such scattering processes, is then correctly reproduced by}
the transmission of the {unstable} stationary {scattering state.
For larger values of $g$, however, the time-dependent dynamics of the 
propagating wavefunction becomes chaotic, which means that the 
transmission associated with an unstable stationary scattering state 
loses its significance.}

\section{The eikonal approximation}
\label{sec:eikonal}

In this appendix, we {explicitly} derive the semiclassical expression for
the Green function in the presence of a weak perturbation of the Hamiltonian.
Considering the Hamiltonian $H = H_0 + \delta H$ with
$H_0 \equiv H_0(\mathbf{p},\mathbf{r})$ the unperturbed part and
$\delta H \equiv \delta H(\mathbf{p},\mathbf{r})$ a perturbation that is
slowly varying with $\mathbf{r}$, and defining
$G_0 \equiv (E - H_0 + i 0)^{-1}$, we can express the Green function of the
system by means of a Dyson equation of the form
\begin{equation}
  G \equiv (E - H + i 0)^{-1} = G_0 + G_0 \delta H G
  = G_0 \sum_{k=0}^\infty (\delta H G_0)^k \, . \label{eq:G1}
\end{equation}

Let us first evaluate the first-order term of this Born series,
\begin{equation}
  \delta G^{(1)}(\mathbf{r},\mathbf{r}',E) \equiv 
  \langle \mathbf{r} | G_0 \delta H G_0 | \mathbf{r}' \rangle
  = \int d^2r'' G_0(\mathbf{r},\mathbf{r}'',E) 
  \delta H(\hat{\mathbf{p}},\mathbf{r}'')
  G_0(\mathbf{r}'',\mathbf{r}',E) \, , \label{eq:G1born1}
\end{equation}
in the semiclassical approximation.
Using the semiclassical expression (\ref{eq:G0sc}) for the Green function,
\begin{equation}
  G_0(\mathbf{r},\mathbf{r}',E) = \sum_\gamma
  A_\gamma(\mathbf{r},\mathbf{r}',E) 
  {\exp\left[\frac{i}{\hbar} S_\gamma(\mathbf{r},\mathbf{r}',E)-i\frac{\pi}{2}\mu_\gamma \right]}
  \, . \label{eq:G0scApp}
\end{equation}
we see that the momentum operator $\hat{\mathbf{p}}$ in $\delta H$ acts, 
in leading semiclassical order (i.e., in lowest order in $\hbar$), 
only on the action integrals $S_\gamma(\mathbf{r}'',\mathbf{r}',E)$ 
in the exponents.
This means that $\hat{\mathbf{p}}$ can  
be replaced by the final momenta of the trajectories leading from 
$\mathbf{r}'$ to $\mathbf{r}''$.
We therefore obtain
\begin{eqnarray}
  \delta G^{(1)}(\mathbf{r},\mathbf{r}',E) & = &
  \int d^2r'' \sum_{\gamma_1, \gamma_2} A_{\gamma_2}(\mathbf{r},\mathbf{r}'',E) 
  A_{\gamma_1}(\mathbf{r}'',\mathbf{r}',E) \delta H(\mathbf{p}'',\mathbf{r}'')
  \nonumber \\
  & & \times \exp\left\{
  {\frac{i}{\hbar} [ S_{\gamma_2}(\mathbf{r},\mathbf{r}'',E) +
      S_{\gamma_1}(\mathbf{r}'',\mathbf{r}',E) ]} {-i\frac{\pi}{2}\left(\mu_{\gamma_1} + \mu_{\gamma_2}\right)}
   \right\}
  \label{eq:dG1}
\end{eqnarray}
with $\mathbf{p}'' \equiv \mathbf{p}_{\gamma_1}^{\mathrm{f}}
(\mathbf{r}'',\mathbf{r}',E)$,
where the indices $\gamma_1$ and $\gamma_2$ respectively represent the 
trajectories from $\mathbf{r}'$ to $\mathbf{r}''$ and from $\mathbf{r}''$
to $\mathbf{r}$.

Using now the fact that $A_{\gamma_1}$, $A_{\gamma_2}$, and 
$\delta H(\mathbf{p}'',\mathbf{r}'')$ are slowly varying functions of 
$\mathbf{r}''$ on the length scale of the de Broglie wavelength of the atoms,
we can apply the stationary phase approximation to evaluate the integral over
$\mathbf{r}''$.
The stationary phase condition yields
\begin{equation}
  \frac{\partial}{\partial \mathbf{r}''} 
  \left[ S_{\gamma_2}(\mathbf{r},\mathbf{r}'',E) 
    + S_{\gamma_1}(\mathbf{r}'',\mathbf{r}',E) \right] = 0 \, ,
\end{equation}
i.e., $\mathbf{p}_{\gamma_1}^{\mathrm{f}}(\mathbf{r}'',\mathbf{r}',E) = 
\mathbf{p}_{\gamma_2}^{\mathrm{i}}(\mathbf{r},\mathbf{r}'',E)$.
This condition is satisfied if and only if the trajectory $\gamma_2$ is the 
direct continuation of $\gamma_1$.
The double sum in Eq.~(\ref{eq:dG1}) can therefore be contracted to a single
sum over trajectories $\gamma$ that are going from $\mathbf{r}'$ to 
$\mathbf{r}$ at energy $E$.
Combining the prefactors that result from the spatial integration 
perpendicular to this trajectory as well as from the amplitudes 
$A_{\gamma_1},A_{\gamma_2}$, and transforming the spatial 
integration parallel to the trajectory into an integration along the 
propagation time, we finally obtain
\begin{equation}
  \delta G^{(1)}(\mathbf{r},\mathbf{r}',E) = -\frac{i}{\hbar} 
  \sum_{\gamma} A_\gamma(\mathbf{r},\mathbf{r}',E) 
  {\exp\left[\frac{i}{\hbar} S_\gamma(\mathbf{r},\mathbf{r}',E)-i\frac{\pi}{2}\mu_\gamma \right]}
  \int_0^{{T_\gamma}} \delta H[\mathbf{p}_\gamma(t),\mathbf{q}_\gamma(t)] dt
  \, . \label{eq:dG1sp}
\end{equation}

Similarly, higher order terms in the Born series (\ref{eq:G1}) can be
evaluated yielding
\begin{eqnarray}
  \delta G^{(k)}(\mathbf{r},\mathbf{r}',E) & \equiv & 
  \langle \mathbf{r} | G_0 (\delta H G_0)^k | \mathbf{r}' \rangle \nonumber \\
  & = & \sum_\gamma A_\gamma(\mathbf{r},\mathbf{r}',E) 
  {\exp\left[\frac{i}{\hbar} S_\gamma(\mathbf{r},\mathbf{r}',E)-i\frac{\pi}{2}\mu_\gamma \right]}\nonumber\\
  &&\times\frac{1}{k!} \left( -\frac{i}{\hbar} \int_0^{{T_\gamma}} 
  \delta H[\mathbf{p}_\gamma(t),\mathbf{q}_\gamma(t)] dt \right)^k \, .
\end{eqnarray}
This finally yields the modified Green function 
\begin{equation}
  G(\mathbf{r},\mathbf{r}',E) = \sum_\gamma A_\gamma(\mathbf{r},\mathbf{r}',E) 
  {\exp\left[\frac{i}{\hbar} S_\gamma(\mathbf{r},\mathbf{r}',E)-i\frac{\pi}{2}\mu_\gamma -\frac{i}{\hbar} \phi_\gamma(\mathbf{r},\mathbf{r}',E)\right]}
  \label{eq:G1scApp}
\end{equation}
where
\begin{equation}
  \phi_\gamma(\mathbf{r},\mathbf{r}',E) \equiv \int_0^{{T_\gamma}} 
  \delta H[\mathbf{p}_\gamma(t),\mathbf{q}_\gamma(t)] dt \, .
  \label{eq:Phi}
\end{equation}
represents an effective modification of the action integral $S_\gamma$ 
due to the presence of the perturbation.
{Eq.~(\ref{eq:G1scApp}) reflects a general result in classical mechanics 
that, to leading order in the perturbation, the action difference between
unperturbed and perturbed orbits is, for periodic orbits, given by 
Eq.~(\ref{eq:Phi}) \cite{BohO95Nl}.}

In the case of a perturbation due to a weak magnetic field, we have
\begin{equation}
  \delta H(\mathbf{p},\mathbf{r}) = 
  - \frac{1}{m} \mathbf{A}(\mathbf{r}) \cdot \mathbf{p}
  + \frac{1}{2m} \mathbf{A}^2(\mathbf{r}) \, .
\end{equation}
Hence, we can write $\phi_\gamma = - \varphi_\gamma - \tilde{\varphi_\gamma}$ 
with
\begin{eqnarray}
  \varphi_\gamma(\mathbf{r},\mathbf{r}',E) & \equiv &
  \frac{1}{m} \int_0^{{T_\gamma}} \mathbf{p}_\gamma(t) \cdot 
  \mathbf{A}[\mathbf{q}_\gamma(t)] dt \, , \\
  \tilde{\varphi_\gamma}(\mathbf{r},\mathbf{r}',E) & \equiv &
  - \frac{1}{2m} \int_0^{{T_\gamma}}\mathbf{A}^2[\mathbf{q}_\gamma(t)] dt \, 
\end{eqnarray}
corresponding, respectively, to a paramagnetic and a diamagnetic 
contribution to the effective action integral.

Also the presence of a weak nonlinearity within the scattering system 
can be accounted for in this framework, provided only ladder contributions
are considered.
Comparing Eq.~(\ref{eq:scst1sc}) in the cases (\ref{eq:l1cond}) and
(\ref{eq:l2cond}) with Eq.~(\ref{eq:G1born1}), we see that we have to set
\begin{equation}
  \delta H(\mathbf{r}) = 2 g \frac{\hbar^2}{2m} 
  |\psi^{(0)}(\mathbf{r})|_{\rm d}^2
\end{equation}
where $|\psi^{(0)}(\mathbf{r})|_{\rm d}^2$ represents, according to
Eq.~(\ref{eq:psi2diag}), the density at position $\mathbf{r}$ as
evaluated within the diagonal approximation.
We then obtain the effective modification of the action integral 
[defined by $\chi_\gamma(\mathbf{r},\mathbf{r}',\mu)$ in 
Sec.~\ref{sec:ladder}] as
\begin{equation}
  \phi_\gamma(\mathbf{r},\mathbf{r}',\mu) = g \frac{\hbar^2}{m} 
  \int_0^{{T_\gamma}} |\psi^{(0)}[\mathbf{q}_\gamma(t)]|_{\rm d}^2 \, dt \, .
\end{equation}

\section{Sum rules}
\label{sec:sumrules}

In this appendix, we derive the generalized Hannay-Ozorio de Almeida 
sum rules \cite{HanOzo84JPA,Sie99JPA} that we need in order to evaluate
energy averages of squares of the Green function in the diagonal 
approximation.
To keep the derivation as general as possible, we introduce a new 
parametrization of the initial and final phase space points according to
$(\mathbf{p},\mathbf{r}) \equiv (\xi_1,\xi_2,\xi_3,\xi_4)$ and
$(\mathbf{p}',\mathbf{r}') \equiv (\eta_1',\eta_2',\eta_3',\eta_4')$
where the sets $(\xi_1,\xi_2,\xi_3,\xi_4)$ and 
$(\eta_1',\eta_2',\eta_3',\eta_4')$ contain the components $(p_x,p_y,x,y)$ 
and $(p_x',p_y',x',y')$ of the final and initial phase space points, 
respectively, in some arbitrary order.
We shall now be interested in the Green function from the coordinates
$(\eta_1',\eta_2')$ to the coordinates $(\xi_1,\xi_2)$.
In the diagonal approximation, the energy average of the modulus square 
of this Green function reads
\begin{eqnarray}
  \left\langle \left| G\left[ (\xi_1,\xi_2),(\eta_1',\eta_2'),E 
    \right] \right|^2 \right\rangle_{\mathrm{d}} & = & 
  \sum_\gamma \left\langle \left| A_\gamma\left[ (\xi_1,\xi_2),
    (\eta_1',\eta_2'),E \right] \right|^2 \right\rangle \\
  & = & \frac{1}{2\pi\hbar^3} \sum_\gamma \left\langle \left|
  \det \frac{\partial\left[(p_x',p_y'),(x',y'),{T}\right]}
  {\partial\left[(\xi_1,\xi_2),(\eta_1',\eta_2'),E\right]} \right| 
  \right\rangle \, .
\end{eqnarray}
We furthermore need the corresponding expression for the crossed average
which includes, in addition, a magnetic dephasing.
This yields
\begin{eqnarray}
  \lefteqn{\left\langle G^*\left[ (\eta_1',\eta_2'),(\xi_1,\xi_2),E \right]
  G\left[ (\xi_1,\xi_2),(\eta_1',\eta_2'),E \right] 
  \right\rangle_{\mathrm{c}} =} \nonumber \\
  & = & \sum_\gamma \left\langle \left| A_\gamma\left[ (\xi_1,\xi_2),
    (\eta_1',\eta_2'),E \right] \right|^2 \right\rangle %e^{-t_\gamma/\tau_B}
  {\exp\left(-\frac{T_\gamma}{\tau_B}\right)} \\
  & = & \frac{1}{2\pi\hbar^3} \sum_\gamma \left\langle \left|
  \det \frac{\partial\left[(p_x',p_y'),(x',y'),{T}\right]}
  {\partial\left[(\xi_1,\xi_2),(\eta_1',\eta_2'),E\right]} \right| 
  \right\rangle
  {\exp\left(-\frac{T_\gamma}{\tau_B}\right)}
\end{eqnarray}
where $\tau_B \sim B^{-2}$ [see Eq.~(\ref{eq:tB})] is the dephasing time.

Applying standard rules for multidimensional integrations over 
$\delta$-distributions, we can derive
\begin{eqnarray}
  \sum_\gamma \left|\det \frac{\partial[(p_x',p_y'),(x',y'),{T}]}
    {\partial[(\xi_1,\xi_2),(\eta_1',\eta_2'),E]}\right| f({T_\gamma}) & = &
    \int d\xi_3 \int d\xi_4 \int d\eta_3' \int d\eta_4' \int_0^\infty d{T}
     \delta\left[ E - H_0(\mathbf{p}',\mathbf{r}')\right] \nonumber \\
    && \times
    \delta\left[ \mathbf{r} - \mathbf{q}(\mathbf{p}',\mathbf{r}',{T}) \right]
    \delta\left[ \mathbf{p} - \mathbf{p}(\mathbf{p}',\mathbf{r}',{T}) \right]
     f({T})
    \label{eq:sc_amplitude_transform}
\end{eqnarray}
{for any $f({T})$}, where we define
$\mathbf{q}(\mathbf{p}',\mathbf{r}',{T}) \equiv 
(q_x,q_y)(\mathbf{p}',\mathbf{r}',{T})$
and
$\mathbf{p}(\mathbf{p}',\mathbf{r}',{T}) \equiv 
(p_x,p_y)(\mathbf{p}',\mathbf{r}',{T})$
as the position and momentum variables that result from the propagation of a
classical trajectory over time ${T}$ with the initial values 
$\mathbf{p}'\equiv(p_x',p_y')$ and $\mathbf{r}'\equiv(x',y')$.
Furthermore, assuming classical ergodicity, which is valid if the dynamics
within the billiard is fully chaotic, we can state that each phase-space point
$(\mathbf{p},\mathbf{r})$ within the billiard has equal probability to be hit 
by a given trajectory after a given evolution time ${T}$, provided it lies 
within the shell of constant energy $E$.
This probability, however, decreases exponentially with the evolution time,
due to the possibility for escape from the billiard via the openings. 
We therefore obtain
\begin{equation}
  \left\langle 
  \delta\left[ \mathbf{r} - \mathbf{q}(\mathbf{p}',\mathbf{r}',{T}) \right]
  \delta\left[ \mathbf{p} - \mathbf{p}(\mathbf{p}',\mathbf{r}',{T}) \right]
  \right\rangle = \frac{\delta\left[H_0(\mathbf{p}',\mathbf{r}') - 
      H_0(\mathbf{p},\mathbf{r})\right]}
  {\int d^2p \int d^2q \delta\left[H_0(\mathbf{p}',\mathbf{r}') - 
      H_0(\mathbf{p},\mathbf{q})\right]}
  {\exp\left(-\frac{T}{\tau_D}\right)}
  \label{eq:ergod}
\end{equation}
where the ``dwell time'' $\tau_D$ corresponds to the mean evolution time
that a classical trajectory spends within the billiard before escaping to
one of the waveguides.
This altogether yields
\begin{equation}
  \left\langle\left|G\left[(\xi_1,\xi_2),(\eta_1',\eta_2'),E\right]
  \right|^2\right\rangle_{\mathrm{d}} =
  \frac{\tau_D}{2\pi\hbar^3} \, \frac{
    \int d\xi_3 \int d\xi_4 \delta\left[E - H_0(\mathbf{p},\mathbf{r})\right]
    \int d\eta_3' \int d\eta_4'
    \delta\left[E - H_0(\mathbf{p}',\mathbf{r}')\right]}
  {\int d^2p \int d^2q \delta\left[E - H_0(\mathbf{p},\mathbf{q})\right]}
  \label{eq:G2gen}
\end{equation}
and
\begin{equation}
\label{eq:G2cl}
\left\langle G^*\left[ (\eta_1',\eta_2'),(\xi_1,\xi_2),E \right]
  G\left[ (\xi_1,\xi_2),(\eta_1',\eta_2'),E \right] \right\rangle_{\mathrm{c}} 
  = \frac{1}{1+\tau_D/\tau_B} 
  \left\langle\left|G\left[(\xi_1,\xi_2),(\eta_1',\eta_2'),E\right] \right|^2\right\rangle_{\mathrm{d}} \, .
\end{equation}

The phase space integrals appearing in Eq.~(\ref{eq:G2gen}) can be 
straightforwardly computed.
We obtain
\begin{eqnarray}
  \int d^2q \delta\left[ E - H_0(\mathbf{p},\mathbf{q}) \right] & = & 
  \Omega \delta\left( E - \frac{p^2}{2m} \right) \, , \\
  \int d^2p \delta\left[ E - H_0(\mathbf{p},\mathbf{q}) \right] & = & 
  2 \pi m \chi_\Omega(\mathbf{q}) \, , \\
  \int d^2p \int d^2q \delta\left[ E - H_0(\mathbf{p},\mathbf{q}) \right] 
  & = & 2 \pi m \Omega  \, ,
\end{eqnarray}
where $\chi_\Omega(\mathbf{q})$ represents the characteristic function of the
scattering system and $\Omega$ denotes the area of the billiard.
Furthermore, for the case of ``mixed'' initial or final conditions 
$\mathbf{z} \equiv (x_L,p_y)$ specified within the incident lead, we calculate
\begin{equation}
  \int dy \int dp_x \delta\left[ E - H_0(\mathbf{p},\mathbf{q}) \right] =
  \int_0^W dy \int_0^\infty dp_x \delta\left( E - \frac{p_x^2 + p_y^2}{2} 
  \right) = \frac{mW}{\sqrt{2mE - p_y^2}} \, .
\end{equation}
Here the longitudinal momentum $p_x$ is restricted to positive (or negative)
values corresponding to an initial (or final) condition with an incoming
(or outgoing) trajectory.
The width $W$ of the waveguide is to be replaced by $\tilde{W}$ in the case
of a final condition within the transmitted lead.

Putting these ingredients together and specifying the choice of the
phase space variables $(\xi_1,\xi_2)$ and $(\eta_1,\eta_2)$ that appear as
arguments in the Green function, we finally obtain
\begin{eqnarray}
  \left\langle\left|{\Anfang{G}}\left(\mathbf{r},\mathbf{z}',E\right)\right|^2
  \right\rangle_{\mathrm{d}} & = &
  \frac{\tau_D}{\tau_H} \left( \frac{m}{\hbar^2} \right)^2
  \chi_\Omega(\mathbf{r}) \frac{W}{2\pi} \frac{1}{\sqrt{2mE - p_y'^2}} \, , 
  \label{eq:Grz} \\
  {\left\langle\left|\Ende{G}\left(\mathbf{z},\mathbf{r},E\right)\right|^2%
  \right\rangle_{\mathrm{d}}} & {=} &%
   {\frac{\tau_D}{\tau_H} \left( \frac{m}{\hbar^2} \right)^2%
   \chi_\Omega(\mathbf{r}) \frac{\tilde{W}}{2\pi} \frac{1}{\sqrt{2mE - p_y^2}}}\, , 
  \label{eq:Gzr} \\
  \left\langle\left|{\AnfangEnde{G}}\left(\mathbf{z},\mathbf{z}',E\right)\right|^2
  \right\rangle_{\mathrm{d}} & = &
  \frac{\tau_D}{\tau_H} \left( \frac{m}{\hbar^2} \right)^2
  { \frac{W\tilde{W}}{(2\pi)^2}} \frac{1}{\sqrt{2mE - p_y^2}} 
  \frac{1}{\sqrt{2mE - p_y'^2}} \label{eq:Gzz}
\end{eqnarray}
for $\mathbf{z}$, $\mathbf{z}'$ being defined in the {transmitted and}
incident lead, {respectively,}
where $\tau_H \equiv m \Omega / \hbar$ denotes the Heisenberg time of the 
billiard.
The corresponding energy-averaged crossed densities are obtained
by a multiplication with the prefactor $(1+\tau_D / \tau_B)^{-1}$,
as is seen from Eq.~(\ref{eq:G2cl}).

\section{Analysis of the classical dynamics}
\label{sec:cladyn}

{
The aim of this section is to explain how we numerically determine the
classical dwell time $\tau_{D}$ and the dimensionless scaling parameter
$\eta$ appearing in Eq.~(\ref{eq:B0}) that characterizes magnetic dephasing.
To this end, we compute, with a ray-tracing algorithm, an ensemble of 
classical trajectories that enter the cavity from the left lead.
The initial conditions $(x^{(0)},y^{(0)},p_x^{(0)},p_y^{(0)})$ 
of these trajectories are randomly selected from the intervals 
$y^{(0)}\in[0,W]$ and $p_y^{(0)}\in[-p,p]$ in a uniform manner, 
while we fix $x^{(0)} = x_L$ and $p_x^{0} = \sqrt{p^2 - [p_y^{(0)}]^2}$ 
(with $p \equiv \sqrt{2m\mu}$ the total momentum of the classical particle).
The propagation of a trajectory is continued until it exits the billiard via
one of the leads.}

\begin{figure}[t]
  \begin{center}
   \includegraphics[width=0.45\textwidth]{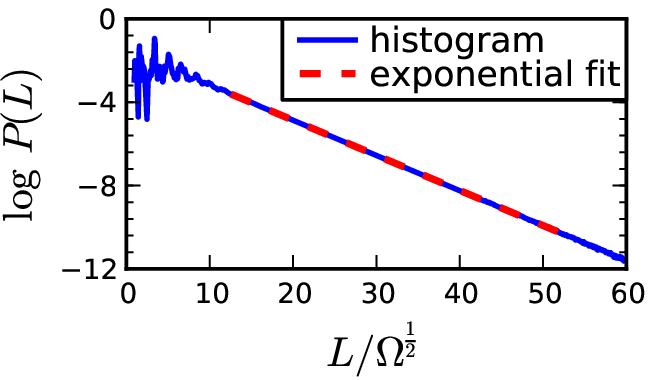}
   \hspace{0.04\textwidth}
   \includegraphics[width=0.45\textwidth]{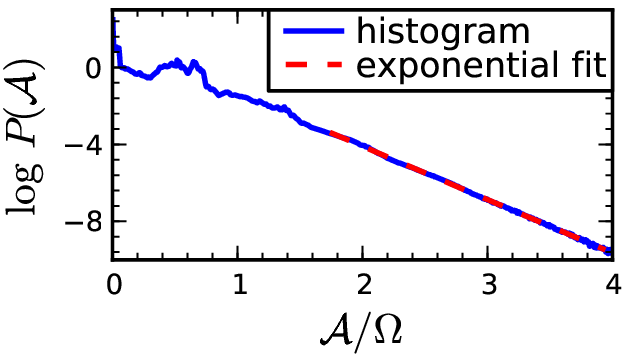}
    \caption{\label{fig:alhisto} 
    {
    Probability distributions (blue lines) for the path length $L$
    (left panel) and for the directed area ${\mathcal{A}}$ 
    (right panel) for the case of the annular stadium billiard
    shown in Fig.~\ref{fig:billiard}({(b)}.
    Neglecting a short transient region, an exponential decay 
    (red dashed lines) is fitted to these distributions.
      }
    }
  \end{center}
\end{figure}

{
Fig.~\ref{fig:alhisto} shows, for the case of the annular stadium billiard
shown in Fig.~\ref{fig:billiard}({(b)}, the numerically obtained probability 
distributions for the path length $L$ and for the modulus of the directed 
area $\mathcal{A}$ that is accumulated along the trajectories according 
to Eq.~(\ref{eq:phipara2}).
As it is expected for chaotic motion \cite{adist1,Jen91}, both probability 
distributions follow an exponential law after a short transient region.
Fitting an exponential decay $P(L) \propto \exp(-L/L_0)$ to the asymptotic 
behaviour of the probability distribution for the trajectory lengths, we
obtain the dwell time via $\tau_{D}= L_0 / v$ with $v \equiv p/m$ the
velocity of the particle.}

{
The distribution of directed areas $P(\mathcal{A})$ can be determined from 
the distribution $P(t,\mathcal{A})$ [see Eq.~(\ref{eq:tadistri})] via
\begin{equation}
  P({\mathcal{A}})=\tau_D^{{-1}} \int_0^{\infty} 
  P(t,{\mathcal{A}}) \exp(-t/\tau_D) dt
  =\frac{1}{\sqrt{{\eta\Omega^{3/2} v \tau_D}}} 
  \exp\left[-\frac{2\left|\mathcal{A}\right|}
      {\sqrt{\eta \Omega^{3/2} v \tau_D}}\right]
\end{equation}
and is also predicted to decrease exponentially with $|\mathcal{A}|$
\cite{adist1,Jen91}.
The exponential decay 
$P(\mathcal{A}) \propto \exp\left(-|\mathcal{A}| / \mathcal{A}_0\right)$
that is numerically encountered after a short transient region allows
us to determine the characteristic scaling parameter via
$\eta = 4 \mathcal{A}_0^2 / (\Omega^{3/2} v \tau_D)$,
using the dwell time $\tau_D$ that is obtained from the length distribution
as explained above.}

Comparing the numerically computed dwell time $\tau_D^{(\text{num.})}$ with 
the ``universal'' prediction (\ref{eq:tauD}), we obtain 
$\tau_D^{(\text{num.})}\simeq 0.79 (\pi \Omega)/[(W + \tilde{W})v]$.
This deviation is attributed to the finite width of the leads, which 
effectively compromises the assumption of ergodic motion that underlies 
the derivation of Eq.~(\ref{eq:tauD}).
Regular islands of appreciable size, which might also give rise to a 
deviation of the dwell time from the universal prediction, could not be 
identified in the phase space of the two billiards.

\section{Frequent integrals in the calculations of loop corrections}

\label{sec:loopappendix}

\subsection{The standard encounter integral}

\label{sec:encint}
In this appendix, we calculate the contribution of the encounter region.
We first consider the absence of a nonlinearity. 
The corresponding integral is given by
\begin{equation}
  I_1^n(E)=\left\langle\int_{-c}^c ds \int_{-c}^c du 
  \frac{1}{\Sigma(E) t_\mathrm{enc}({su})}
  \exp\left(\frac{i}{\hbar}su\right)
           {\exp\left(-\frac{t_\mathrm{enc}({su})}{\tau_n}
             \right)}\right\rangle\,,
           \label{eq:enc_int_std}
\end{equation}
where $1/\tau_n{\equiv}1/\tau_D+n/\tau_B$ accounts for the {fact that}
we can have $n\in\{0,1,2\}$ stretches with a gauge field dependence
{within} the encounter region{, and}
$\Sigma(E)=2\pi m \Omega$ is the volume of
the energy shell in phase space. 
Following {Ref.}~\cite{BroRah06PRB}, we first split
the integration over $u$ in two parts,
\begin{equation}
  \int_{-c}^c du \, \ldots= \int_{-c}^0 du \,\ldots +\int_0^c du\,\ldots,
  \label{eq:usplit} 
\end{equation}
{and make} the variable transformation {$(s,u) \mapsto (S,y)$ with}
{$S \equiv su/c^2 \in [-1,1]$ and $y \equiv c / |u| = \mp c / u\in [1,1/S]$,
with the associated Jacobian determinant $c^2/y$,}
where the {sign} in the {definition of $y$ refers} to the first and the 
second integral {on the right-hand side of Eq.}~(\ref{eq:usplit}), 
respectively. 
{In physical terms,} we transform here from {the} phase space coordinates 
$s,u$ to the action difference $su$ measured in terms of $c^2${,} and 
{to} a coordinate {$y$} related to the time $t_u=(1/\lambda)\ln(c/u)$ 
needed for the unstable {phase space} coordinate to 
{evolve from the Poincar\'e surface 
of section $\mathcal{P}$ to} the limiting value ${\mp} c$.

As $t_\mathrm{enc}({su})=(1/\lambda)\ln(c^2/|su|) = 
t_\mathrm{enc}(|S|)$ {does not depend on $y$, we obtain}
\begin{equation}
  \int_1^{1/|S|}dy\frac{1}{y}=\ln\left(\frac{1}{|S|}\right)
  =\lambda t_\mathrm{enc}(|S|)
\end{equation}
{for the integration over $y$.}
We {then have}
\begin{eqnarray}
  I_1^n(E)&=&\left\langle\frac{2c^2\lambda}{\Sigma(E)}\int_{-1}^1 dS 
  \exp\left(\frac{i}{\hbar}Sc^2\right)
  \exp\left(-\frac{t_\mathrm{enc}(|S|)}{\tau_n}\right)\right\rangle\nonumber\\
  &=&\left\langle
  \frac{4c^2\lambda}{\Sigma(E)} \int_0^1 dS 
  \cos\left(\frac{Sc^2}{\hbar}\right)
  {S^{1/(\lambda\tau_n)}}\right\rangle\nonumber\\
  &=& \left\langle \frac{4\hbar\lambda}{\Sigma(E)} 
  \sin\left(\frac{c^2}{\hbar}\right)\right\rangle
  -\left\langle{\frac{4\hbar}{\Sigma(E)\tau_n}}
  \int_0^1 dS \frac{\sin(Sc^2/\hbar)}{S} 
  {S^{1/(\lambda\tau_n)}}\right\rangle\,.
  \label{eq:I1nfull}
\end{eqnarray}
As {the limiting scale} $c$ {for the coordinates $s$ and $u$
(i.e., the scale until which the linerization around the reference trajectory 
is still valid) generally} depends on the energy $E${,} the first term 
{in Eq.~(\ref{eq:I1nfull}) is expected to strongly oscillate} when varying
$E$ and {would thus vanish} when performing the energy average. 
For the second term, we {obtain after} the transformation 
$S\mapsto S'\equiv Sc^2/\hbar$
\begin{equation}
  I_1^n(E)  =-\left\langle{\frac{4\hbar}{\Sigma(E)\tau_n}}
  \int_0^{c^2/\hbar} dS' \frac{\sin\left(S'\right)}{S'}
  {\left(\frac{S'\hbar} {c^2}\right)^{1/(\lambda\tau_n)}}
  \right\rangle\, . \label{eq:I1nunapprox}
\end{equation}
{Assuming that the Ehrenfest time $\tau_E \equiv (1/\lambda)\ln(c^2/\hbar)$
is much smaller than the dwell time $\tau_D$ and the magnetic dephasing time
scale $\tau_B$, we have $\tau_E \ll \tau_n$ as well as $\lambda\tau_n\gg 1$ 
and can approximate}
\begin{equation}
  {\left(\frac{S' \hbar}{c^2}\right)^{1/(\lambda\tau_n)}
  = (S')^{1/(\lambda\tau_n)}\exp\left[-\frac{\tau_E}{\tau_n}\right]
  \simeq 1 \, .}
\end{equation}
{The} remaining integral can then be evaluated 
{in the semiclassical limit $\hbar \to 0$}
by sending the upper limit of the integration (\ref{eq:I1nunapprox}) 
to infinity, which {finally} yields
\begin{eqnarray}
  I_1^n(E)&=&-{\frac{4{\hbar}}{\Sigma(E)\tau_n}}
  \int_0^\infty dS \frac{\sin\left(S\right)}{S}
  =-{\frac{2\pi\hbar}{\Sigma(E)\tau_n}}
  = -\frac{1}{\tau_D\tau_H}\left(1+n\frac{\tau_D}{\tau_B}\right) \,.
\end{eqnarray}

\subsection{The encounter integral with {an embedded nonlinearity event}}

\begin{figure}[t]
  \begin{center}
    \includegraphics[width=0.25\textwidth]{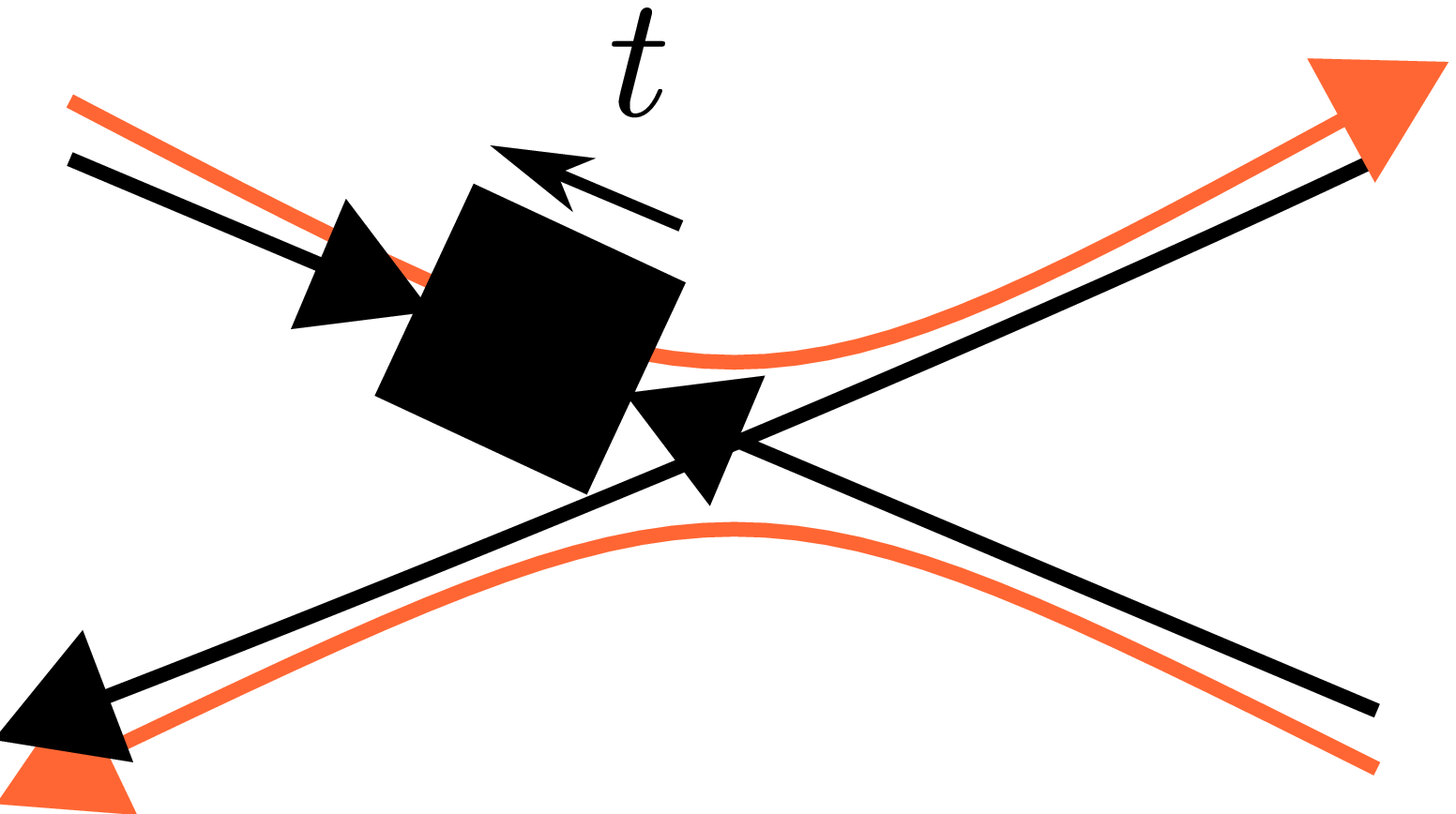}
    \hspace{0.25\textwidth}
    \includegraphics[width=0.25\textwidth]{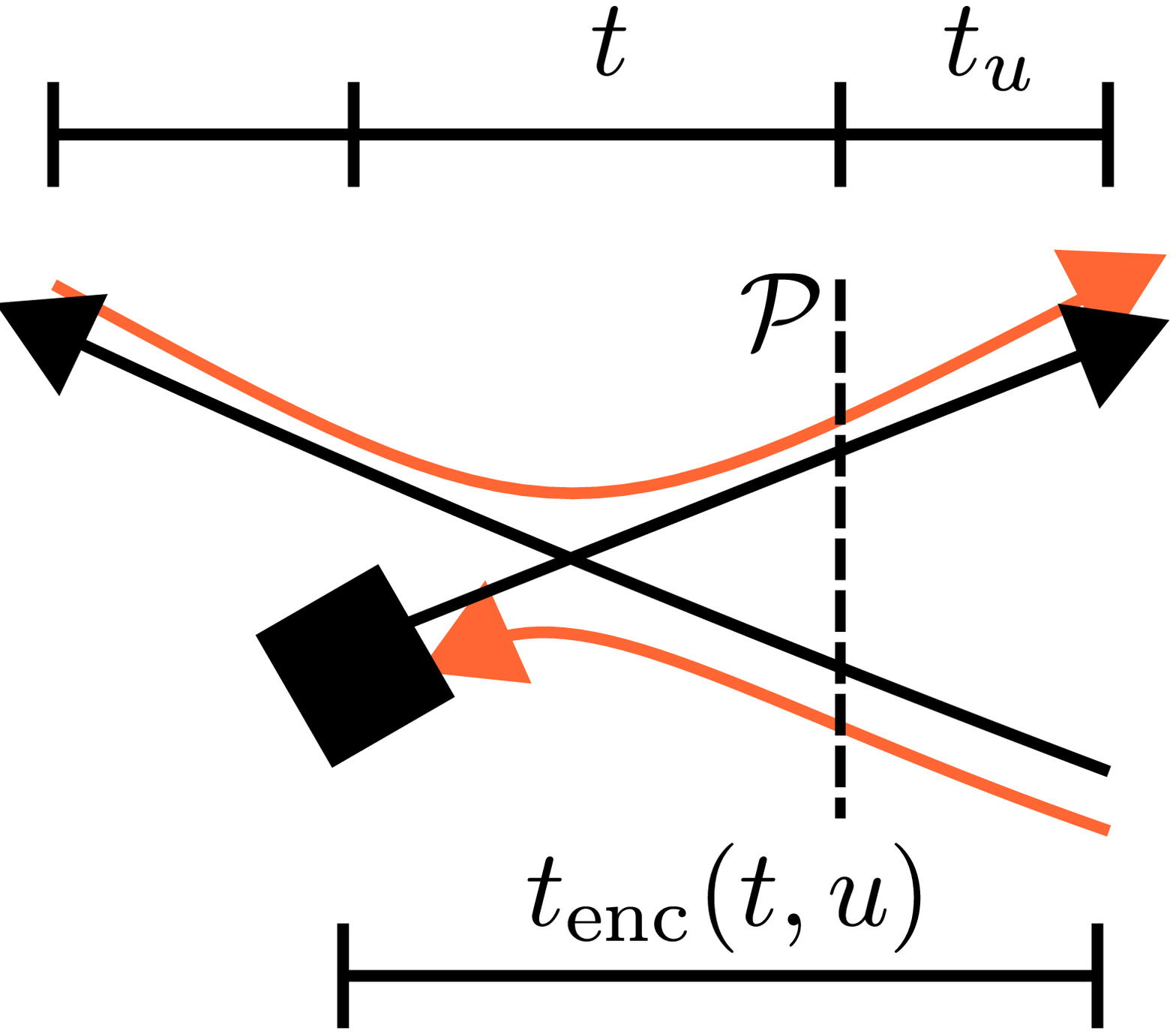}
    \caption{\label{fig:nonlinenc}
      {Sketch of the two different possibilities for a nonlinearity event
        to enter the encounter region.
        In the scenario depicted on the left-hand side, the nonlinearity 
        event moves along a stretch all the way through the encounter region.
        On the right-hand side, two of the four encounter stretches end at the
        nonlinearity block.
        This gives rise to a reduced encounter duration 
        $\bar{t}_\mathrm{enc}(t,u) = t + t_u$
        which depends on the time interval $t$
        between the nonlinearity event and the Poincar\'e surface of section
        $\mathcal{P}$, and on the time $t_u =(1/\lambda)\ln(c/|u|)$
        between $\mathcal{P}$ and the borders of the encounter region.}
      } 
  \end{center}
\end{figure}

\label{sec:enc_nonlinint}

Now we {consider the presence} of a nonlinearity event 
{within the encounter region.
We first focus on the case that the nonlinearity event is}
moving {along a stretch all the way} through {the}
encounter region{,} as depicted {on the left-hand side of} 
Fig.~\ref{fig:nonlinenc}. 
{For this case}, we have to evaluate {the integral}
\begin{equation}
  I_2(E)=\left\langle{\int\limits_{-c}^c\int\limits_{-c}^c  ds du} 
  \frac{1}{\Sigma(E) t_\mathrm{enc}{(su)}}
  \exp\left(\frac{i}{\hbar}su\right)
  \exp\left[-\frac{t_\mathrm{enc}{(su)}}{\tau_D}\right]
  \int_0^{t_\mathrm{enc}{(su)}} dt 
  \exp\left(-\frac{t}{\tau_B}\right)\right\rangle\,,
  \label{eq:encnonlinint}
\end{equation}
where the additional integration {variable} $t$ represents the 
{location} of the nonlinearity on a stretch within
the encounter region. 
The gauge field dependence{, manifested by the dephasing factor 
$\exp(-t/\tau_B)$,} emerges from the stretch {along which the 
nonlinearity moves}. 
We {have}
\begin{equation}
  \int_0^{t_\mathrm{enc}}\hspace{-1ex}dt 
  \exp\left(-\frac{t}{\tau_B}\right)
  =\tau_B\left[1-\exp\left(-\frac{t_\mathrm{enc}}{\tau_B}\right)\right]\,,
\end{equation}
which would also be {obtained} if the integrand in 
Eq.~(\ref{eq:encnonlinint}) {was}
$\exp[-\left(t_\mathrm{enc}-t\right)/\tau_B]$, corresponding to the
case that the other part of the stretch {guiding} the nonlinearity event 
would {provide} the gauge field dependence. 
{Using} the results from section \ref{sec:encint}{, we} obtain
\begin{equation}
  I_2(E)=\tau_B\left[I_1^0(E)-I_1^1(E)\right]= \frac{1}{\tau_H}\,.
\end{equation}

We now analyze the {second scenario, shown on the right-hand side of 
Fig.~\ref{fig:nonlinenc},} where stretches of the encounter
region terminate at a nonlinearity{.}
The integral {that} has to be {evaluated in this case} is given by
\begin{equation}
  I_3^n(E)=\left\langle{\int\limits_{-c}^c\int\limits_{-c}^c ds du}
  \int_0^{(1/\lambda)\ln(c/|s|)}dt
  \frac{1}{\Sigma(E) \bar{t}_\mathrm{enc}(t,u)}
  \exp\left(\frac{i}{\hbar}su\right)
  \exp\left(-\frac{\bar{t}_\mathrm{enc}(t,u)}{\tau_n}\right)
  \right\rangle, \label{eq:enc_nonlinint_2}
\end{equation}
where {we define} $\bar{t}_\mathrm{enc}(t,u){\equiv}t+(1/\lambda)\ln(c/|u|)$
{as} the reduced encounter time {and $1/\tau_n\equiv 1/\tau_D+n/\tau_B$,
with $n=0,1,2$ the number of pairs of imbalanced stretches that give rise to
a gauge field dependence.
As indicated in Fig.~\ref{fig:nonlinenc}, the integration variable} $t$ 
{represents} the time between the nonlinearity and the 
{Poincar\'e surface of section} $\mathcal{P}$ {within} which the 
stable and unstable coordinates are defined{.}

{Following Refs.}~\cite{BroRah06PRB,WalGutGouRic08PRL,GutWalKuiRic09PRE}{,
we split,} as in {S}ection \ref{sec:encint}{, the integration over $u$}
according to {Eq.}~(\ref{eq:usplit}) {and make the variable 
transformation $(s,u,t) \mapsto (S,y,\bar{t})$ with
$S \equiv su/c^2 \in[-1,1]$, 
$\bar{t} \equiv \bar{t}_\mathrm{enc}(t,u) =
t+(1/\lambda)\ln(c/|u|) \in[0,(1/\lambda)\ln(1/|S|)]$, and
$y \equiv c/|u| \in [1,\exp(\lambda {\bar{t}})]$, with the associated
Jacobian determinant $c^2/y$}. 
The integration over $y$  yields
\begin{equation}
  \int_1^{\exp(\lambda {\bar{t}})}dy\frac{1}{y}= \lambda {\bar{t}}\, {.}
\end{equation}
{We then evaluate}
\begin{eqnarray}
  I_3^n(E)&=& \left\langle\frac{2c^2\lambda}{\Sigma(E)}  \int_{-1}^1 dS
  \int_0^{(1/\lambda)\ln\left(1/|S|\right)}d{\bar{t}} 
  \exp\left(\frac{i}{\hbar}Sc^2\right)
  \exp\left(-\frac{{\bar{t}}}{\tau_n}\right)
  \right\rangle\nonumber\\
  &=&\left\langle{\frac{2c^2\lambda \tau_n}{\Sigma(E)}}
  \int_{-1}^1dS 
  \left[1-{|S|^{1/(\tau_n\lambda)}}\right]
  \exp\left(\frac{i}{\hbar}Sc^2\right) \right\rangle\nonumber\\
  &=&\left\langle{\frac{4c^2\lambda \tau_n}{\Sigma(E)}}
   \left[\frac{\hbar}{c^2}\sin\left(\frac{c^2}{\hbar}\right)
   -\int_0^1dS \cos\left(\frac{Sc^2}{\hbar}\right) 
   {S^{1/(\tau_n\lambda)}}\right] \right\rangle\,. \label{eq:last}
\end{eqnarray}
The first {contribution in the last line of Eq.~(\ref{eq:last})}
vanishes when performing the energy average, {whereas}
the second {term} yields $[-(\hbar\pi)/(2\lambda c^2\tau_n)]$, 
as seen in {S}ection \ref{sec:encint}. 
We {thus} obtain
\begin{equation}
  I_3^n(E)={\frac{4c^2\lambda\tau_n}{\Sigma(E)}
    \frac{\hbar\pi}{2\lambda c^2\tau_n}}
  =\frac{2\pi\hbar}{\Sigma(E)}=\frac{1}{\tau_H}\,{.}
\end{equation}


\begin{thebibliography}{83}
\bibitem[{rem(????)}]{rem_stable}
\bibinfo{note}{At the largest value $g = 0.1$ of the nonlinearity that we
  consider, we find stable solutions for about 95 percent of all choices of the
  billiard geometry, the chemical potential, and the effective magnetic field.}
\bibitem[{rem(????{\natexlab{b}})}]{rem_omega}
\bibinfo{note}{In the case that the point $\mathbf{r}$ cannot be connected to
  $\mathbf{r}_L$ by a straight-line trajectory due to geometric reasons (e.g.\
  because such a trajectory would cross boundaries of the billiard), one can
  alternatively define another trajectory for the point $\mathbf{r}$ that leads
  from there to $\mathbf{r}_L$ within a few bounces.}
\bibitem[{hyp(????)}]{hyperbolicity}
\bibinfo{note}{Note that our results are also valid for the general hyperbolic
  case. For this case one then has to use the asymptotic Lyapunov exponent for
  long trajectories, as e.g.~done in \cite{TurRic03JPA}.}
\bibitem[{rem(????)}]{rem_gr}
\bibinfo{note}{In a physical realization of this transport setup, this can be
  accomplished e.g.\ by varying the lateral confinement of the leads.}
\bibitem[{rem(????{\natexlab{a}})}]{rem_ggr}
\bibinfo{note}{Formally, we set $g(\mathbf{r})\equiv g g_0(\mathbf{r})$ and
  vary $g$ while keeping $g_0(\mathbf{r})$ fixed.}
\bibitem[{rem(????{\natexlab{b}})}]{rem_rank}
\bibinfo{note}{We assume that the derivative ${\mathcal{D} F}$ has full rank on
  all solutions of $F(\psi;g)=0$. Cases in which this assumption fails are very
  rare. This can happen at bifurcations or cusps.}
\bibitem[{rem(????{\natexlab{c}})}]{rem_sweep}
\bibinfo{note}{In practice, a single sweep of the nonlinearity strength is
  sufficent to compute all relevant scattering states within a finite interval
  $0 \leq g \leq g_{\mathrm{max}}$.}
\end{thebibliography}
\end{document}